\documentclass[prapplied, aps,
 twocolumn,
 superscriptaddress,
 amsmath,amssymb,
 10pt
]{revtex4-2}

\usepackage{times}
\usepackage{color}
\usepackage{graphicx}
\usepackage[dvipsnames]{xcolor}
\usepackage{physics}
\usepackage{bm}
\usepackage{mathtools}
\usepackage{upgreek}
\usepackage[colorlinks=true,allcolors=blue]{hyperref}
\usepackage{footmisc}
\usepackage{makecell}
\usepackage{mathtools}
\usepackage{soul}
\usepackage{xcolor}
\usepackage{lipsum}
\usepackage{wasysym}

\soulregister{\cite}{7}
\soulregister{\ref}{7}
\soulregister{\eqref}{7}
\soulregister{\label}{7}
\soulregister{\onlinecite}{7}
\definecolor{mycolor}{HTML}{ffffd4}
\sethlcolor{mycolor}

\begin{document}

\title{Crosstalk-Robust Quantum Control in Multimode Bosonic Systems}

\author{Xinyuan You}
\email[xinyuan@fnal.gov]{}
\affiliation{Superconducting Quantum Materials and Systems Center, Fermi National Accelerator Laboratory (FNAL), Batavia, IL 60510, USA}

\author{Yunwei Lu}
\affiliation{Department of Physics and Astronomy, Northwestern University, Evanston, IL 60208, USA}

\author{Taeyoon Kim}
\affiliation{Superconducting Quantum Materials and Systems Center, Fermi National Accelerator Laboratory (FNAL), Batavia, IL 60510, USA}
\affiliation{Department of Physics and Astronomy, Northwestern University, Evanston, IL 60208, USA}
\affiliation{Center for Applied Physics and Superconducting Technologies, Northwestern University, Evanston, IL 60208, USA}

\author{Do\~ga Murat K\"urk\c{c}\"uo\~glu}
\affiliation{Superconducting Quantum Materials and Systems Center, Fermi National Accelerator Laboratory (FNAL), Batavia, IL 60510, USA}

\author{Shaojiang Zhu}
\affiliation{Superconducting Quantum Materials and Systems Center, Fermi National Accelerator Laboratory (FNAL), Batavia, IL 60510, USA} 

\author{David van Zanten}
\affiliation{Superconducting Quantum Materials and Systems Center, Fermi National Accelerator Laboratory (FNAL), Batavia, IL 60510, USA}

\author{\\Tanay Roy}
\affiliation{Superconducting Quantum Materials and Systems Center, Fermi National Accelerator Laboratory (FNAL), Batavia, IL 60510, USA}

\author{Yao Lu}
\affiliation{Superconducting Quantum Materials and Systems Center, Fermi National Accelerator Laboratory (FNAL), Batavia, IL 60510, USA}

\author{Srivatsan Chakram}
\affiliation{Department of Physics and Astronomy, Rutgers University, Piscataway, NJ 08854, USA}

\author{Anna Grassellino}
\affiliation{Superconducting Quantum Materials and Systems Center, Fermi National Accelerator Laboratory (FNAL), Batavia, IL 60510, USA}

\author{Alexander Romanenko}
\affiliation{Superconducting Quantum Materials and Systems Center, Fermi National Accelerator Laboratory (FNAL), Batavia, IL 60510, USA}

\author{Jens Koch}
\affiliation{Department of Physics and Astronomy, Northwestern University, Evanston, IL 60208, USA}
\affiliation{Center for Applied Physics and Superconducting Technologies, Northwestern University, Evanston, IL 60208, USA}

\author{Silvia Zorzetti}
\email[zorzetti@fnal.gov]{}
\affiliation{Superconducting Quantum Materials and Systems Center, Fermi National Accelerator Laboratory (FNAL), Batavia, IL 60510, USA}

\begin{abstract}

High-coherence superconducting cavities offer a hardware-efficient platform for quantum information processing. 
To achieve universal operations of these bosonic modes, the requisite nonlinearity is realized by coupling them to a transmon ancilla. 
However, this configuration is susceptible to crosstalk errors in the dispersive regime, where the ancilla frequency is Stark-shifted by the state of each coupled bosonic mode. 
This leads to a frequency mismatch of the ancilla drive, lowering the gate fidelities. 
To mitigate such coherent errors, we employ quantum optimal control to engineer ancilla pulses that are robust to the frequency shifts. These optimized pulses are subsequently integrated into a recently developed echoed conditional displacement (ECD) protocol for executing single- and two-mode operations. 
Through numerical simulations, we examine two representative scenarios: the preparation of single-mode Fock states in the presence of spectator modes and the generation of two-mode entangled Bell-cat states. Our approach markedly suppresses crosstalk errors, outperforming conventional ancilla control methods by orders of magnitude. 
These results provide guidance for experimentally achieving high-fidelity multimode operations and pave the way for developing high-performance bosonic quantum information processors.

\end{abstract}

\maketitle

\section{Introduction}\label{sec:intro}
Building quantum information processors capable of executing advanced quantum algorithms in the NISQ era~\cite{NISQ}, and ultimately achieving quantum error correction~\cite{QEC1,QEC2}, necessitates control of a large Hilbert space, configurable connectivity, and high-fidelity operations.
One approach to meet these demands is to physically interconnect a large number of qubits. 
Among the various implementations of qubits, superconducting circuits have shown promise, with successful integration of over 1000 qubits into a superconducting processor~\cite{ibm}. Despite this progress, challenges persist in further scaling up the number of qubits. 
In particular, connectivity between qubits is often limited due to complexity in on-chip wiring and feedline layouts~\cite{Rosenberg2017}. Furthermore, maintaining coherence times, especially when dealing with a large number of qubits, remains a significant hurdle to overcome~\cite{McEwen2022}.

An alternative to increasing the size of the accessible Hilbert space is to expand the Hilbert space utilized within a single physical entity, such as a quantum harmonic oscillator, which ideally offers an infinite Hilbert space. Recent advancements in superconducting cavities, featuring ultra-high quality factors on the order of \(Q \approx 10^{10}\)~\cite{Romanenko2020,Milul2023}, suggest their capability to support thousands of Fock levels with coherence times in the millisecond range. This large, high-coherence subspace has been leveraged in two distinct encoding schemes.
The first approach involves encoding quantum information directly into \(d>2\) computational states of a system, known as qudits~\cite{Wang2020, Wu2020,Ringbauer2022,Chi2022}. This form of encoding is versatile and efficient for a wide range of applications, including those for quantum chemistry~\cite{MacDonell2021} and quantum simulation~\cite{Rico2018}.
The second approach focuses on the redundant encoding of qubit information within this large Hilbert space, enabling the detection and correction of specific types of errors~\cite{QEC_review}. This has led to the development of various bosonic error-correction codes, such as cat codes~\cite{cat1,cat2,Touzard2018}, binomial codes~\cite{binomial,Hu2019,Ni2023}, and Gottesman-Kitaev-Preskill (GKP) codes~\cite{GKP,Campagne-Ibarcq2019,gkp_review,Sivak2022}.

\begin{figure}
    \centering
    \includegraphics[width=\columnwidth]{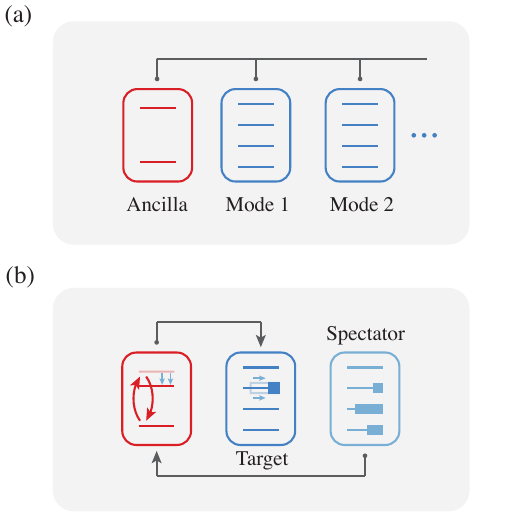}
    \caption{(a) Schematic of a multimode architecture. A transmon ancilla (red) is directly linked to all cavity modes (blue). 
    (b) Illustration of the crosstalk within this architecture. It occurs when two (or more) modes are coupled to the same ancilla. The quantum state of the spectator mode alters the resonant condition of the ancilla, leading to an error in the operation of the target mode.}
    \label{fig:fig_architecture} 
\end{figure}

With multiple bosonic modes, the accessible Hilbert space experiences a dramatic expansion. In the context of qudit encoding, the dimension of computational subspace scales as \(d^N\) for \(N\) qudits. For quantum error correction, the integration of multiple bosonic modes facilitates universal quantum computation among protected logical qubits.
In most hardware setups designed to realize such multimode systems, a transmon ancilla is directly integrated with various cavity modes~\cite{Naik2017a,Chakram2020,Chakram2022,Alessandro}, as depicted in Fig.~\ref{fig:fig_architecture}(a). This architecture allows the ancilla to couple simultaneously with all bosonic modes, thereby facilitating the execution of both single and multimode operations across arbitrary modes. 
However, this ``all-to-one'' configuration introduces a significant challenge: crosstalk among the different bosonic modes, as illustrated in Fig.~\ref{fig:fig_architecture}(b). Specifically, in the widely-used dispersive regime, the photon population in each cavity mode induces a dispersive shift in the frequency of the transmon ancilla~\cite{blais2021}. 
This becomes a critical issue because standard cavity control protocols depend on driving the ancilla at specific frequencies, which are typically resonant.
Consequently, a change in the population of one mode (unknown in a general circuit compilation) can adversely affect the frequency of the ancilla, thereby compromising the fidelity of its controls. This, in turn, can introduce errors in the control of other cavity modes coupled to the ancilla, or even the mode itself. For the purposes of this discussion, we refer to both of these scenarios as crosstalk~\footnote{It's crucial to differentiate this form of crosstalk from that caused by the cross-Kerr effect, which is generally negligible in comparison to the Stark shift-induced crosstalk.}.

The susceptibility to crosstalk varies according to the nature of the cavity control protocols.
The ones that involve a selective ancilla drive with a narrow pulse bandwidth, such as the selective number-dependent arbitrary phase (SNAP) gate~\cite{Heeres2015b,Krastanov2015b,Wang2021}, are particularly susceptible to crosstalk. This susceptibility arises from their sensitivity to the ancilla frequency shift, even with the help of optimal control~\cite{Kudra2022}.
On the other hand, protocols that involve unselective ancilla control, such as the echoed conditional displacement (ECD)~\cite{Eickbusch2021} and conditional not displacement (CNOD)~\cite{Diringer2023}, exhibit reduced sensitivity but still remain susceptible to frequency shifts.
In this study, we address the challenge of crosstalk by harnessing the power of quantum optimal control~\cite{Kelly2014,Machnes2018,Werninghaus2021,Goerz2017}. Our approach involves the development and implementation of robust control, a concept initially introduced in the field of nuclear magnetic resonance~\cite{hahn1950spin,dridi2020robust}.
We demonstrate that gates realized by our robust pulse are insensitive to a broad range of frequency variations in the ancilla. 
We then integrate these robust pulses into a multimode generalization of the ECD protocol, resulting in a notable reduction of crosstalk error, as compared to the standard derivative removal by adiabatic gate (DRAG) pulse~\cite{Motzoi2009,Chow2010,Gambetta2011,Motzoi2013c,Chen2016a,Theis2018a}. 
It is crucial to highlight that the use of robust ancilla control is not exclusively applicable to the ECD protocol. Indeed, it can be widely applied to a variety of other cavity control protocols that involve unselective control of the ancilla.

This paper is organized as follows: In Sec.~\ref{sec:ecd_intro}, we detail the implementation of the multimode ECD protocol, with an emphasis on the spurious phase correction due to cavity-ancilla coupling.
Following this, in Sec.~\ref{sec:qoc}, we employ quantum optimal control to engineer robust ancilla control pulses. Our key findings, which illustrate the suppression of crosstalk, are showcased in Sec.~\ref{sec:main}. Specifically, we simulate the effect of crosstalk in two representative experimental scenarios: state preparation with spectator modes and a two-mode entanglement operation.
We discuss and compare our multimode architecture with other existing ones in the literature in Sec.~\ref{sec:discussion}, before sharing our conclusions in Sec.~\ref{sec:conclusions}. Supplementary details are included in the appendices.

\section{Two-mode ECD protocol with correction of spurious phase accumulation}\label{sec:ecd_intro}

\begin{figure*}\label{fig:fig_ecd} 
    \centering
    \includegraphics[width=\textwidth]{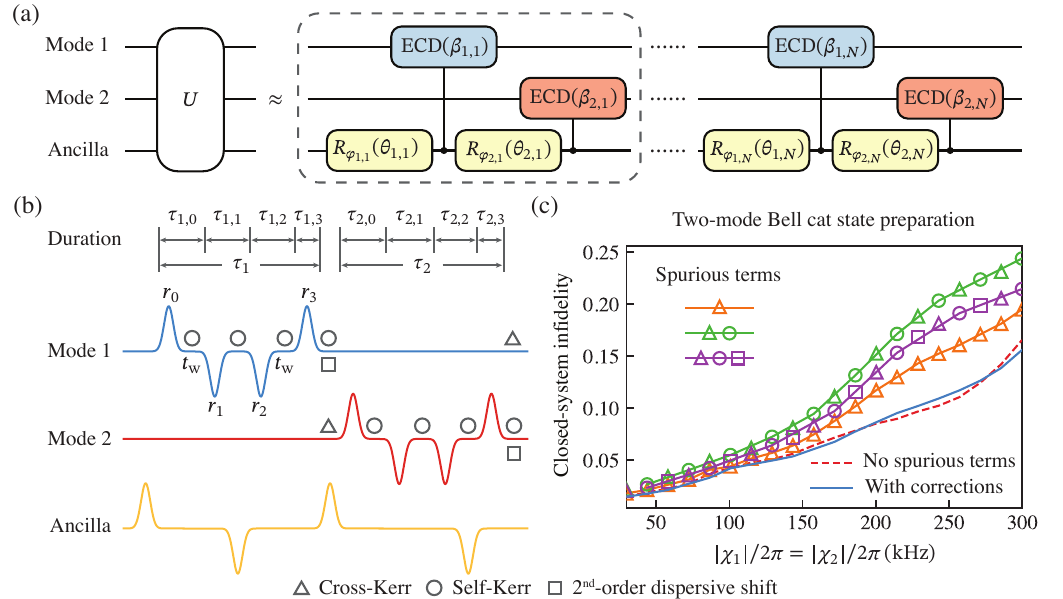}
    \caption{Universal two-mode ECD protocol with spurious phase corrections. 
    (a) Decomposition of an arbitrary control of two cavity modes with a modular sequence, wherein each unit block (enclosed with dashed lines) comprises a pair of ECD gates acting on each mode, interspersed with ancilla rotations. 
    (b) Pulse construction of a unit block in (a). Each single-mode ECD gate consists of an ancilla echo pulse and four Gaussian displacement pulses, parameterized by their amplitudes $(r_0,r_1,r_2,r_3)$ and separation $(t_\text{w})$. Corrections for spurious effects from cross-Kerr, self-Kerr, and second-order dispersive shift are applied as phase shifts in the upcoming pulse sequences for both cavity modes, represented by a triangle $\triangle$, circle $\Circle$, and square $\square$, respectively. 
    The durations used for estimating the spurious phase are detailed above the pulse sequences. 
    The ancilla rotations can be realized using a variety of pulse schemes, including Gaussian pulses, DRAG pulses, or optimal control pulses in general. 
    (c) Closed-system infidelity in the generation of entangled Bell-cat states, \(\mathcal{N}(|\alpha\rangle|\alpha\rangle + |{-}\alpha\rangle|{-}\alpha\rangle)\), with \(\mathcal{N}\approx 1/\sqrt{2}\) and  $\alpha = 4$. This is plotted as a function of identical dispersive shifts \( \chi_{1} \) and \( \chi_{2} \). The red dashed line represents the ideal scenario, free from spurious terms as mentioned in (b), and is constrained solely by crosstalk resulting from the dispersive shift. Solid lines with distinct markers represent situations incorporating selected spurious terms. Notably, the blue solid line indicates the outcome with spurious phase correction, demonstrating a significant reduction in error.
    [Parameters: $K_\text{q}/2\pi = -200\,\text{MHz}$, $\Delta_{1}/2\pi = \Delta_{2}/2\pi = 2\,\text{GHz}$, with other nonlinearities derived from Eq.~\eqref{eq:dispersive_relation}. Note that the frequencies of the cavity modes are chosen to be equal for convenience.]}
\end{figure*}

Universal control of a bosonic mode necessitates an entanglement operation between the linear cavity mode and an additional nonlinear ancilla. One notable example of such an operation is the ECD gate, which was introduced in Ref.~\onlinecite{Eickbusch2021,Sivak2022}, and is defined as follows:
\begin{equation}\label{eq:ecd}
\hat{D}(\beta/2) \otimes|\text{e}\rangle \langle \text{g}| + \hat{D}(-\beta/2) \otimes|\text{g}\rangle \langle \text{e}|,
\end{equation}
where $\hat{D}(\alpha) = \exp(\alpha \hat{a}^\dag - \alpha^* \hat{a})$ is the displacement operator and $|g,e\rangle$ represent the qubit ground and excited states. This entangling gate performs different displacement operations on the cavity mode, depending on the state of the ancilla qubit.
Any unitary on a single cavity mode can be approximated with a sequence of ECD gates and ancilla rotations $\hat{R}_{\varphi}(\theta)=\exp[-i(\theta/2)(\hat{\sigma}_x \cos\varphi +\hat{\sigma}_y \sin \varphi)]$.

To extend the universal control to two or more cavity modes coupled to the ancilla qubit, we implement a strategy involving alternating ECD gates on each mode~\cite{eesh}, as depicted in Fig.~\ref{fig:fig_ecd}(a).
Each unit block of the gate sequence (enclosed with dashed lines) consists of a pair of ECD gates acting on each mode, separated by two ancilla rotations.
This configuration facilitates indirect entanglement between cavity modes through a common ancilla. Given the established universality of the single-mode ECD gate~\cite{Eickbusch2021}, confirming multimode universality simply requires the ability to construct the beamsplitter unitary. Detailed procedures for the construction of such a unitary are elaborated in Appendix~\ref{app:beamsplitter}.

In practice, to realize a specific target gate or state transfer, we generalize the optimization procedure outlined in Ref.~\onlinecite{Eickbusch2021} for a single mode to account for the case of two cavity modes. Specifically, we start with a chosen number of unit blocks $N$. The gate sequence consists of $6N$ parameters $\{\beta_{1,n}, \beta_{2,n}, \varphi_{1,n}, \varphi_{2,n}, \theta_{1,n}, \theta_{2,n}\}$, whose roles are elucidated in Fig.~\ref{fig:fig_ecd}(a). Here, the first index in the subscript identifies the cavity mode, while the second index $n\in [1,N]$ refers to the $n$th block in the gate sequence.
Subsequently, we optimize these parameters to achieve the requisite circuit-level fidelity. If the optimization fails within a specified number of iterations, we increase the number of unit blocks and repeat the optimization until the desired fidelity is reached.
The numerical implementation is similar to the case for a single cavity mode. See Appendix S7 in Ref.~\onlinecite{Eickbusch2021} for more details.

The above discussion outlines the process of generating the gate sequence for a given task. However, translating this abstract sequence into concrete control pulses requires a thorough understanding of the underlying physical systems. In the following, we provide a brief overview of simulating a driven cavity-ancilla system (with detailed derivations in Appendix~\ref{app:theory}), and then outline the procedure for constructing the corresponding pulse sequence.

\subsection{Simulation of the driven coupled cavity-ancilla system}
Within the dispersive regime, the coupled cavity-ancilla system can be well described by the following Hamiltonian:
\begin{align*}
\hat{H}_\text{disp} &= \sum_{i=1,2} \left[\omega_i\hat{a}_i^\dag \hat{a}_i
+ (\chi_i \hat{a}_i^\dag \hat{a}_i + \frac{\chi_i'}{2} \hat{a}_i^{\dag 2} \hat{a}_i^2) \hat{q}^\dag \hat{q} 
+ \frac{K_{i}}{2} \hat{a}_i^{\dag 2} \hat{a}_i^2 \right] \nonumber\\ 
& + K_{12} \hat{a}_1^\dag \hat{a}_1 \hat{a}_2^\dag \hat{a}_2  + \omega_\text{q} \hat{q}^\dag \hat{q} + \frac{K_\text{q}}{2}\hat{q}^{\dag 2}\hat{q}^{2},
\end{align*}
where $\omega_\text{q}$ and $\omega_\text{i}$ denote the frequencies of the ancilla and cavity modes. 
The nonlinearities are defined as follows. 
The self- and cross-Kerr interaction strengths for the two cavity modes are described by \(K_{i}\) and  \(K_{12}\), and the first- and second-order dispersive shifts between the cavity modes and the ancilla are symbolized by $\chi_i$ and $\chi_{i}'$, respectively.
Note that the above nonlinearities are inherited from the transmon anharmonicity, and are interconnected through the relations in Eq.~\eqref{eq:dispersive_relation}.
The effects of higher-order transmon nonlinearities are analyzed in Appendix~\ref{app:six_order}.
Although our protocol does not necessitate identical cavity parameters, we take them to be identical for simplicity in the numerical simulations throughout this paper.

To execute ECD gates, microwave drives are applied on both the ancilla and the two cavity modes. In particular, the ancilla is driven resonantly with temporal profile $\varepsilon(t)$, while the two cavity modes are driven off-resonantly with detunings $\omega_i - \omega_{\text{d},i} = -\chi_{i}/2$ and temporal profiles $\Omega_i(t)$. 
To simulate the driven dynamics efficiently, it is advantageous to first go into the rotating frame defined by both the cavity and ancilla drive frequencies, and then the displaced frame, characterized by the unitary transformation $\hat{U}_\alpha(t) = \Pi_{i=1,2}\hat{D}_i^\dag[\alpha_i(t)]$. Here, the displacement $\alpha_i(t)$ is obtained by solving the trajectory equation in Eq.~\eqref{eq:ode}. 
The resulting displaced frame Hamiltonian is detailed in Appendix~\ref{app:displaced_frame}. 
In addition to those static terms, including the dispersive interaction $\chi_i \hat{a}_i^\dag \hat{a}_i$ leading to the conditional displacement in Eq.~\eqref{eq:ecd}, there exists time-dependent diagonal terms, which depend on the displacement $\alpha_i(t)$:
\begin{align}
     \hat{H}_\text{diag}(t) &= \sum_{i=1,2} 
     \bigg[ (2K_i|\alpha_i|^2 + K_{12}|\alpha_{\bar{i}}|^2 )\hat{a}_i^\dag \hat{a}_i  \label{eq:diag_ham}\\ 
     & + 2\chi_i'|\alpha_i|^2 \hat{a}_i^\dag \hat{a}_i \hat{q}^\dag \hat{q} 
     + (\chi_i|\alpha_i|^2 - \frac{\chi'_i}{2}|\alpha_i|^4)\hat{q}^\dag \hat{q}\bigg]. \nonumber
\end{align}
In the presence of strong cavity nonlinearities, the above terms introduce significant corrections to the system dynamics, necessitating careful consideration in the design of high-fidelity cavity control pulses.

\subsection{Construction of pulses mitigating spurious phase}
Our focus now shifts to the formulation of control pulses realizing universal control of two cavity modes. As illustrated in Fig.~\ref{fig:fig_ecd}(a), each unit block (enclosed with dashed lines) comprises paired single-mode ECD gates for every cavity mode, interleaved by ancilla rotations.
In this section, we employ standard DRAG pulses for ancilla control, displayed for simplicity as Gaussian shapes in Fig.~\ref{fig:fig_ecd}(b). (These pulses can be replaced by the ones obtained from optimal control to improve ancilla gate fidelity, as will be discussed in later sections.)
For single-mode ECD gates, we adopt the pulse ansatz introduced in Ref.~\onlinecite{Eickbusch2021}. This ansatz consists of four cavity Gaussian displacement pulses and an ancilla echo pulse in the middle, as depicted in Fig.~\ref{fig:fig_ecd}(b). 
The ansatz is parametrized by the amplitudes \( r_0, r_1, r_2, r_3 \) and the temporal separations \( t_\text{w} \) of these displacement pulses. 
We then follow the same routine in Ref.~\onlinecite{Eickbusch2021} to optimize these parameters for desired ECD gates.
This optimization procedure generates control pulses that yield high-fidelity operations for the small cavity nonlinearity parameters taken in Ref.~\onlinecite{Eickbusch2021}. 

\subsubsection{Characterization of the spurious phase}
In scenarios with strong cavity-ancilla coupling and significant cavity displacement, the effects from cavity nonlinearities are non-negligible and need proper treatment. One major consequence of these nonlinearities (self- and cross-Kerrs) is that they result in a time-dependent cavity frequency shift in the form of \(\delta_\text{NL}(t)\hat{a}_i^\dag \hat{a}_i\).
This can be inspected in the displaced frame, as shown in \(\hat{H}_\text{diag}(t)\) in Eq.~\eqref{eq:diag_ham}.
During the time interval \(\tau\), these terms alone lead to relative phase accumulations following \(\exp[-i\int_\tau \mathrm{d}t' n_i\delta_\text{NL}(t')]|n_i\rangle\), where \(|n_i\rangle\) denotes the \(n\)th Fock state of the $i$th cavity mode.
However, when off-diagonal terms are present [see Eq.~\eqref{eq:off_diag}], describing the spurious dynamics solely by phase accumulation is an approximation. These off-diagonal terms induce additional spurious dynamics, such as ancilla-state dependent displacement (e.g., $\hat{a}_i{\hat{q}}^\dag\hat{q}$) and squeezing (e.g., $\hat{a}_i^2\hat{q}^\dag\hat{q}$), which do not commute with the diagonal terms. Together, these effects make the phase accumulation description approximate.
In the following, we first characterize the approximated phase accumulations from various nonlinearities in a unit block [enclosed by dashed lines in Fig.~\ref{fig:fig_ecd}(a)], and then detail the protocol to mitigate them.

\paragraph{Spurious phase from cavity self-Kerr:}
The first half of the unit block  involves four displacement pulses on mode 1. From Eq.~\eqref{eq:diag_ham}, the phase accumulation due to the self-Kerr for the \( k \)th displacement (\( k \in \{0,1,2,3\} \)) is estimated as:
\begin{equation}\label{eq:self_kerr}
    \Phi_{1,k}^{\text{self-Kerr}} = \int_{\tau_{1,k}} \mathrm{d}t\,2K_{1}|\alpha_{1}(t)|^2.
\end{equation}
Here, the integration interval \( \tau_{1,k} \) denotes the duration of the \( k \)th displacement on mode 1 and is depicted in Fig.~\ref{fig:fig_ecd}(b). During the first half of the unit block, there is no displacement on mode 2 and thus no phase accumulation due to self-Kerr.

\paragraph{Spurious phase from cavity cross-Kerr:}
The displacement on mode 1 induces phase accumulation on mode 2 through the term proportional to cross-Kerr in Eq.~\eqref{eq:diag_ham}. This phase shift is estimated as:
\begin{equation}\label{eq:cross_kerr}
    \Phi_{2}^{\text{cross-Kerr}} = \int_{\tau_{1}} \mathrm{d}t\,K_{12}|\alpha_{1}(t)|^2,
\end{equation}
where \( \tau_1 \) denotes the time interval of the first half of the unit block, as shown in Fig.~\ref{fig:fig_ecd}(b).

\paragraph{Spurious phase from second-order dispersive shift:}
The dispersive shift term in Eq.~\eqref{eq:diag_ham} shifts the cavity frequency depending on the state of the ancilla. The presence of the echo pulse in the middle of the duration \( \tau_1 \) ensures that the ancilla spends approximately~\footnote{This is not strictly valid in an open system where the ancilla can decay. However, considering that the gate duration is much shorter than a typical transmon lifetime, this is still a good approximation.} an equal amount of time in both the ground and excited states. Since phase accumulation only occurs when the ancilla has population in the excited state, the estimated phase from this process is:
\begin{equation}\label{eq:2nd_disp}
    \Phi_{1}^{\text{2nd-disp}} = \int_{\tau_{1}} \mathrm{d}t\,\chi_{1}'|\alpha_{1}(t)|^2,
\end{equation}
where the factor of 2 in Eq.~\eqref{eq:diag_ham} is removed to account for the effect of the echo pulse.

The above discussion details the spurious phase accumulated during the first half of the unit block. Similar results can be derived for the second half, with the roles of modes 1 and 2 interchanged.

\subsubsection{Mitigation of the spurious phase}
To counteract the spurious phase accumulation, we implement virtual phase gates on the cavity modes (analogous to virtual \(\hat{Z}\) gates for qubits~\cite{McKay2017}). 
These gates are realized by introducing compensatory phase shifts into the upcoming cavity displacement pulses. The procedure for these phase corrections is depicted in Fig.~\ref{fig:fig_ecd}(b), and we elaborate our correction scheme on a single unit block as follows.

First, to correct the self-Kerr-induced spurious phase in Eq.~\eqref{eq:self_kerr}, we apply phase shifts to the individual cavity displacement pulses. The locations of these corrections are marked with circles $\Circle$ in Fig.~\ref{fig:fig_ecd}(b). The resulting gate sequence on mode 1 is (from right to left):
\begin{equation}
    \hat{\mathcal{Z}}_{1,3}^\text{self-Kerr}\hat{D}_{1,3} \hat{\mathcal{Z}}_{1,2}^\text{self-Kerr}\hat{D}_{1,2} \hat{\mathcal{Z}}_{1,1}^\text{self-Kerr}\hat{D}_{1,1} \hat{\mathcal{Z}}_{1,0}^\text{self-Kerr}\hat{D}_{1,0},
\end{equation}
where \(\hat{D}_{1,k}\) denotes the \(k\)th displacement on cavity mode 1. The phase shifts are implemented as virtual phase rotations:
\begin{equation}
    \hat{\mathcal{Z}}_{1,k}^\text{self-Kerr} = \exp\left[-i\left(\sum_{m=0}^k \Phi_{1,m}^{\text{self-Kerr}} +\Phi_\text{1,acc}\right) \hat{a}^\dag_1 \hat{a}_1\right],
\end{equation}
where \(\Phi_\text{1,acc}\) accounts for the accumulated phase correction on mode 1 prior to the considered unit block. 
At the end of the ECD gate for mode 1, we introduce a phase correction to mitigate the effects of the second-order dispersive shift, which is represented by a square $\square$ in Fig.~\ref{fig:fig_ecd}(b). The corresponding phase rotation is:
\begin{equation}
    \hat{\mathcal{Z}}_{1}^\text{2nd-disp} = \hat{\mathcal{Z}}_{1,3}^\text{self-Kerr} \exp\left(-i\Phi_{1}^{\text{2nd-disp}} \hat{a}^\dag_1 \hat{a}_1\right),
\end{equation}
where the factor \(\hat{\mathcal{Z}}_{1,3}^\text{self-Kerr}\) is used to track the applied phase corrections.
Simultaneously, we cancel the phase accumulation on mode 2, induced by the cross-Kerr interaction, shown as a triangle $\triangle$ in Fig.~\ref{fig:fig_ecd}(b). The corresponding virtual phase gate is:
\begin{equation}
    \hat{\mathcal{Z}}_{2}^\text{cross-Kerr} = \exp\left[-i\left(\Phi_{2}^{\text{cross-Kerr}} + \Phi_\text{2,acc}\right) \hat{a}^\dag_2 \hat{a}_2\right],
\end{equation}
where \(\Phi_\text{2,acc}\) accounts for the accumulated phase correction on mode 2, prior to the considered unit block. 
Finally, we repeat this process with switched roles of the two cavity modes, completing the second half of the gates in one unit block.

\subsubsection{Benchmarking the effectiveness of mitigating the spurious phase}
We demonstrate the effectiveness of our scheme by considering the generation of entangled Bell-cat states. The target state has the form of \(\mathcal{N}(|\alpha\rangle|\alpha\rangle + |{-}\alpha\rangle|{-}\alpha\rangle)\), with the normalization constant \(\mathcal{N}\approx 1/\sqrt{2}\) and the cat size $\alpha = 4$.
We conduct pulse-level simulations across a range of cavity-ancilla coupling strengths, such that the dispersive shift extends the range
$|\chi_{1}|/2\pi=|\chi_{2}|/2\pi\in[30, 300]$ kHz. According to Eq.~\eqref{eq:dispersive_relation}, this also adjusts other relevant nonlinearities in a consistent manner, including self-Kerrs, cross-Kerrs, and second-order dispersive shifts. Consequently, the spurious phase accumulations associated with each nonlinearity also scale up correspondingly, generally reducing fidelities.

In the following, we study infidelities arising from various spurious sources as a function of the dispersive shifts. 
We begin by presenting a reference infidelity benchmark where all those spurious cavity-related nonlinearities are deliberately removed, by setting  \(K_{12}=K_i=\chi_i'=0\). The resulting infidelity is depicted as the red dashed line in Fig.~\ref{fig:fig_ecd}(c). 
We then incrementally incorporate these nonlinearities. The orange line with triangles shows the infidelity with cross-Kerr \(K_{12}\). Adding self-Kerr \(K_i\) changes this to the green line marked with triangles and circles. Including the second-order dispersive shift \(\chi_i'\) results in the purple line, marked with triangles, circles, and squares.
As the dispersive shift increases, the infidelities in the presence of these nonlinearities rise markedly. Interestingly, the result including all three nonlinearities exhibits lower infidelity compared to the one with only self-Kerr and cross-Kerr. This is due to $\chi_{i}'$ possessing an opposite sign to $K_{12}$ and $K_{i}$, which partially offsets the spurious phase accumulation from the latter two.

Once we implement the phase corrections introduced earlier, the infidelity is shown as the blue solid line. Remarkably, this almost fully eliminates the infidelity arising from spurious phase accumulation due to self-Kerr, cross-Kerr, and second-order dispersive shift. For certain values of $\chi_i$, it even marginally outperforms the reference case, thereby validating our phase-correction strategy.

\section{Crosstalk error suppression based on quantum optimal control}\label{sec:qoc}
While the phase correction method detailed in Sec.~\ref{sec:ecd_intro} significantly reduces infidelity, coherent errors still persist, particularly in cases of strong ancilla-cavity coupling.
The primary source of the remaining coherent error in operating multiple cavity modes originates from the crosstalk. This crosstalk is mediated by the coupled transmon ancilla, and arises due to the dispersive shift. Generally, the fidelity of ancilla operations is influenced by the quantum states of the cavity modes. For instance, when a cavity mode is in a Fock state \( |n\rangle \), it induces a Stark shift \( n\chi \) on the ancilla frequency. Without accounting for this frequency change, the ancilla operations become off-resonant, thereby lowering the fidelity. This error can then propagate through the system, leading to crosstalk between different cavity modes.
While integrating tunable couplers into the multimode architecture could potentially mitigate this error~\cite{manipulator}, our approach instead utilizes resonant drives at the bare ancilla frequency combined with modulation of the pulse envelopes.
This approach ensures that the fidelity is insensitive to the detuning between the drive and the Stark-shifted ancilla frequency. Although the concept of Stark shift becomes less definitive when the cavity modes are in superposition or entangled states, robust ancilla control remains a sufficient protocol to mitigate the crosstalk error. 
In what follows, we first detail our method for optimizing these detuning-robust pulse envelopes for an isolated ancilla. 
Specifically, we focus on the optimization of $\hat{X}_{\pi/2}$ and $\hat{X}_{\pi}$ gates, which are sufficient for universal control and efficient in experimental calibration (see Appendix~\ref{app:decomposition}).
Subsequently, we integrate these optimized pulses into a composite ancilla-cavity system with minimal modifications, proven effective in reducing crosstalk through numerical simulations.

\subsection{Robust control of an isolated transmon ancilla}
\begin{figure}
    \centering
    \includegraphics[width=\columnwidth]{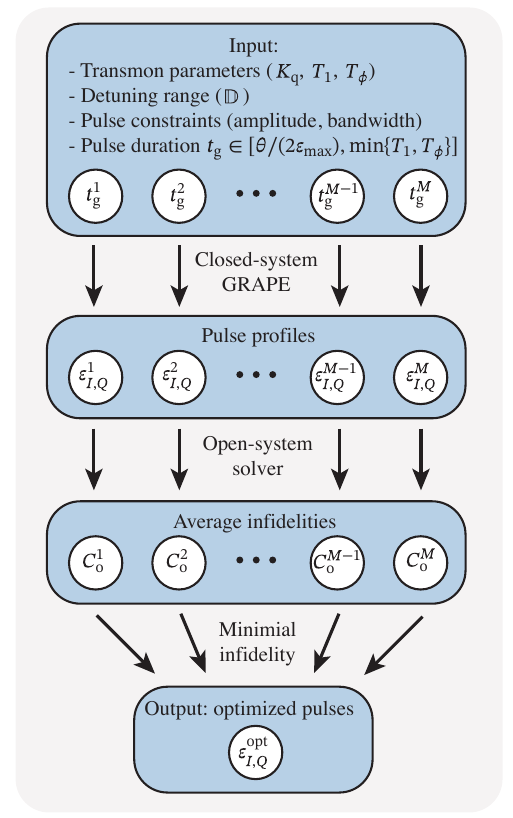}
    \caption{
    Optimization procedure to obtain detuning-robust $\hat{X}_\theta$ gates in transmons. Input includes pulse constraints, transmon parameters, and a set of $M$ pulse durations. The choice of duration should be shorter than decay time $T_1$ and longer than $1/\epsilon_\text{max}$, where $\epsilon_\text{max}$ is the maximum pulse amplitude. Closed-system GRAPE is performed for each duration to obtain the corresponding pulse profiles. Subsequently, an open-system solver evaluates the cost function $C_\text{o}$ for pulses with varying durations, selecting the pulse that minimizes $C_\text{o}$.}
    \label{fig:optflow}
\end{figure}

Detuning-robust control for two-level systems has been extensively investigated using various approaches, including reverse engineering~\cite{PhysRevLett.111.050404inverseengi,PhysRevLett.125.250403inverseengi}, space curve quantum control~\cite{barnes2022dynamicallyspacecurve}, pulse engineering~\cite{barnes2022dynamicallyspacecurve,wang2012compositepulse,PhysRevA.83.053420pulse,PhysRevA.86.022315pulse,tycko1983broadbandpulse,PhysRevA.107.023103pulse,ball2021softwarepulseengi,wimperis1994broadbandpulse,pasini2009optimizedinitialpulse}, and the collocation method~\cite{propson2022robustcollocation,Trowbridge2023}. However, protocols developed for two-level systems cannot be directly applied to weakly-anharmonic transmons due to leakage out of the qubit subspace. One promising solution is to apply the DRAG scheme to robust single-quadrature pulses developed for two-level systems~\cite{hai2023universalcompare}. 
Additionally, Ref.~\cite{qctrl} directly optimizes two-quadrature pulses for transmons.
In this work, we introduce an alternative optimization-based procedure to design detuning-robust control for transmon $\hat{X}_{\theta}$ gates and benchmark its performance. A detailed numerical comparison with selected literature is presented in Appendix~\ref{app:robust_pulse}.

\subsubsection{Metrics of robust control}
\label{metric}
During the gate operation, the Hamiltonian of interest may depend on an unknown quantity $\delta$. This uncertainty can cause the realized gate to deviate from the target gate, thus generally lowering gate fidelity $\mathcal{I}(\delta)$. To quantify the robustness of the infidelity against $\delta$, we consider both the mean infidelity and its standard deviation as key metrics,
\begin{equation*}
\mathcal{I}_\mu = \int_\mathbb{D} \mathrm{d}\delta\, p(\delta)\mathcal{I}(\delta), \quad
\mathcal{I}_\sigma =\sqrt{\int_\mathbb{D} \mathrm{d}\delta \, p(\delta)\big[\mathcal{I}(\delta)-I_\mu\big]^2 }.
\end{equation*}
Here, the probability distribution $p(\delta)$ satisfies $\int_\mathbb{D}\mathrm{d}\delta \, p(\delta)=1$ for a specified range $\mathbb{D}$. 

In our case of interest, the Hamiltonian governing the transmon qubit in the co-rotating frame of the drive frequency is 
\begin{equation*}
    \hat{H}_\text{trans}(t) = K_\text{q} \hat{q}^{\dagger 2} \hat{q}^2/2 + \varepsilon_I(t)(\hat{q}^\dagger+\hat{q})+i\varepsilon_Q(t)(\hat{q}^\dagger-\hat{q})+\delta \hat{q}^\dagger \hat{q}.
\end{equation*}
Here, $\varepsilon_I(t)$ and $\varepsilon_Q(t)$ describe control envelopes in the $I$ and $Q$ quadratures, and $\delta$ denotes the difference between the drive frequency and unknown shifted qubit frequency. The closed-system gate infidelity follows 
\begin{equation}\label{eq:fid_single}
\mathcal{I}_\text{c}(\delta) = 1-
\frac{1}{4}\left|\Tr\left[\hat{P}_{0,1}\hat{U}_\text{target}^\dagger \hat{U}_\delta(t_\text{g})\right]\right|^2,
\end{equation}
where $\hat{P}_{0,1}$ is a projector onto the computational subspace formed by the lowest two eigenstates. We define $\hat{U}_\text{target}$ as the target unitary on the transmon, and $\hat{U}_\delta(t_\text{g})$ $=\mathcal{T}\exp[-i\int_0^{t_\text{g}} \mathrm{d}\tau \, \hat{H}_\text{trans}(\tau)]$ as the actual unitary achieved over a gate duration $t_\text{g}$, where $\mathcal{T}$ is the time-ordering operator. 

In the presence of decoherence, the open-system infidelity follows~\cite{abdelhafez2019quantumthesis}
\begin{equation}
\mathcal{I}_\text{o}(\delta) = 1-\frac{1}{4}\Tr[\mathcal{P}_{0,1}\mathcal{L}_\text{target}^\dagger \mathcal{L}_\delta(t_\text{g})].
\end{equation}
Here, the density matrix is vectorized by stacking the elements row-by-row. $\mathcal{P}_{0,1}=\hat{P}_{0,1}\otimes\hat{P}_{0,1}$ is the superprojector onto the computational subspace, $\mathcal{L}_\text{target}=\hat{U}_\text{target}\otimes \hat{U}^*_\text{target}$ is the target superoperator, and $\mathcal{L}_\delta(t_\text{g})$ is the superoperator realized by the Lindblad master equation. Note that in the absence of decoherence, we have $\mathcal{L}_\delta(t_\text{g})=\hat{U}_\delta(t_\text{g})\otimes \hat{U}_\delta^*(t_\text{g})$.

\subsubsection{Optimization procedure}
\begin{figure*}
    \centering
    \includegraphics[width=\textwidth]{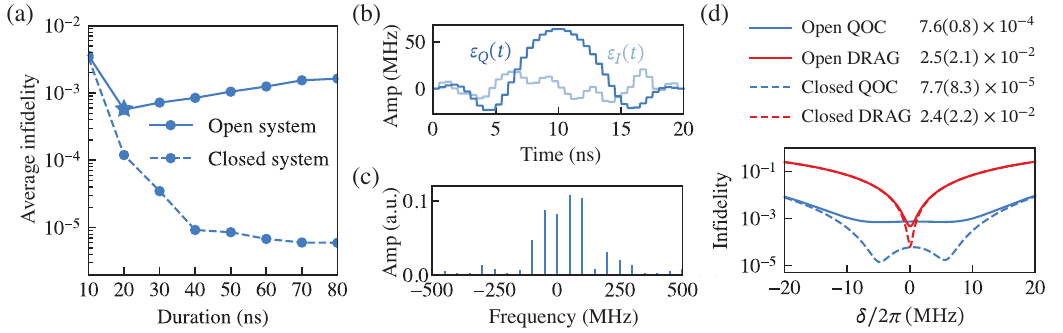}
    \caption{
    Optimization results for the $\hat{X}_\pi$ gate. 
(a) Average infidelities in Eq.~\eqref{eq:avg_inf} versus pulse durations for both closed (dashed line) and open (solid line) systems. The optimal duration selected from the range considered for this optimization is 20 ns (indicated by a star). 
(b) $I$ and $Q$ quadratures of the pulse envelopes for the selected 20-ns pulse from (a).
(c) The frequency components of the pulse envelopes, corresponding to the 20-ns pulse. 
(d) Comparative analysis of the 20-ns pulses, from both QOC (blue) and DRAG (red) pulses, in terms of their infidelities as a function of detuning.
The average and standard deviation of the infidelities are tabulated correspondingly by considering a uniform distribution of $\rho(\delta)$ for simplicity. The result shows that the optimized pulse is more robust than the commonly used DRAG pulse.
[Parameters: $\delta_i/2\pi\in\{\pm 1,\pm 3,\pm 4,\pm 7,\pm 10\}$ MHz, $K_\text{q}/2\pi=-200$ MHz, $T_1=50$ $\mu$s, $T_\phi=50$ $\mu$s.]\label{fig:fig_qoc} }
\end{figure*}

Quantum optimal control has been extensively used to develop robust control schemes by minimizing cost functions that encode robustness metrics. One approach is to penalize the derivative of the closed- or open-system fidelity with respect to detuning $\delta$~\cite{Watanabe,hai2023universalcompare,qctrl}. However, a more intuitive and pragmatic method is to penalize the average infidelity~\cite{reinhold2019controllingthesisrobustmetric,allen2020robustmetric,kosut2013robustmetric,khaneja2005optimalrobustmetric,rembold2020introductionrobustmetric}
\begin{equation}\label{eq:avg_inf}
C_{\text{o},\text{c}} = \frac{1}{N}\sum_{i=1}^N\mathcal{I}_{\text{o},\text{c}}(\delta_i).
\end{equation}
Here, $\delta_i$ denotes sample values of detunings within the specified range of $\mathbb{D}$. We choose a range $\mathbb{D}/2\pi=[-10, 10]$ MHz, since the frequency shifts of qubit may reach tens of MHz due to both large cavity photon numbers and dispersive shifts in our study. We find that minimization of the above cost function can suppress both mean infidelity and its standard deviation, as evidenced by subsequent numerical results.

To minimize the cost function, we optimize the pulses, including both the envelopes $\varepsilon_{I,Q}(t)$ and the pulse duration $t_\text{g}$. 
An optimal duration is determined by balancing two competing factors: a shorter duration leads to unwanted leakage, whereas a longer duration results in decoherence errors. 
A direct approach to optimize pulses involves using an open-system optimizer for various durations, and selecting the one that minimizes the cost function $C_\text{o}$. 
However, such optimization requires time-intensive open-system simulations. 
Instead, we propose a method, while not guaranteed to be equivalent, yields promising results. 
The method consists of two steps: (1) employing closed-system optimization to minimize $C_\text{c}$ for each potential duration, and (2) with given decoherence rates, evaluating $C_\text{o}$ for the optimized pulses from first step, then selecting the pulse with the lowest $C_\text{o}$. 
Our approach is summarized in Fig.~\ref{fig:optflow}.

For the closed-system optimization, we use Gradient Ascent Pulse Engineering (GRAPE), a method widely employed for optimal control in various contexts~\cite{khaneja2005optimalrobustmetric,lu2023optimalgrape,koch2022quantumqocreview}. We calculate the gradients required in GRAPE by using automatic differentiation, thereby eliminating the need for manual coding the analytical gradients of the cost function~\cite{adscalingbaydin2014automatic,PhysRevA.95.042318ad,PhysRevA.99.052327ad}. We use the initial pulse ansatz
\begin{align*}
    \varepsilon^\text{initial}_I(t) &=\frac{1}{t_\text{g}}\big[ (a-\theta/2)\cos(2\pi t /t_\text{g}) + (b-a)\cos(4\pi t /t_\text{g})\\
     &+(c-b)\cos(6\pi t / t_\text{g}) - c\cos(8\pi t / t_\text{g}) + \theta/2\big],\\
    \varepsilon_Q^\text{initial}(t) & = -\dot{\varepsilon}_I^\text{initial}(t) / K_\text{q},
\end{align*}
where $\theta$ is the rotation angle for the target $\hat{X}_\theta$ gate ($\theta=\pi,\pi/2$ in our study). This ansatz for $I$-quadrature control has been used to obtain robust control for two-level systems~\cite{pasini2009optimizedinitialpulse}. The initial guess for the $Q$-quadrature is inspired by the DRAG pulse scheme~\cite{Motzoi2009} to suppress leakage. 
Then, we optimize the pulse starting from various initial guesses for parameters $a,b,c\in\{-10,-8,-6,\cdots,10\}$ and select the pulse with the lowest cost value $C_\text{c}$. We accelerate this process by parallelizing with different initial values.

Throughout the closed-system optimization process, we ensure pulse smoothness by filtering out frequency components that exceed maximum allowed bandwidths of $5/t_\text{g}$ and $10/t_\text{g}$ for $\varepsilon_I(t)$ and $\varepsilon_Q(t)$, respectively. These empirical values effectively balance the pulse smoothness and the level of robustness required for our study.

\subsubsection{Benchmarking for an isolated transmon ancilla}
\label{optimization result}
We benchmark our optimization procedure for a transmon with typical parameters: $K_\text{q}/2\pi=-200$ MHz, $T_1=50$ $\mu$s, and $T_\phi=50$ $\mu$s. The optimization results for the $\hat{X}_\pi$ gate are shown in Fig.\ \ref{fig:fig_qoc}. 
Further details for the $\hat{X}_{\pi/2}$ gate are provided in Appendix~\ref{app:robust_pulse}.
Figure~\ref{fig:fig_qoc}(a) shows the average infidelities obtained for closed- and open-system simulations with varying pulse durations. 
Without decoherence, extending the pulse duration generally leads to a reduction in the cost value. In contrast, in the presence of decoherence, utilizing shorter pulses is more effective in reducing the infidelity. 
Given the particular transmon parameters under consideration, a pulse duration of 20 ns achieves the lowest cost.
The temporal profiles and frequency components of the optimized 20-ns pulse are shown in Fig.\ \ref{fig:fig_qoc}(b) and (c), respectively.
In Fig.~\ref{fig:fig_qoc}(d), we study the infidelity as a function of detuning, comparing the performance between the optimized and the DRAG pulses. 
In both closed- and open-system simulations, the optimized pulse consistently demonstrates substantially reduced infidelities across almost all detunings. 
Consequently, the optimized pulse exhibits enhanced robustness against frequency detuning in the transmon ancilla. 
The open-system infidelities obtained from these two pulses are limited by distinct factors: the infidelity derived from the QOC pulse is dominated by decoherence errors, while the one generated from the DRAG pulse is limited by coherent crosstalk. 
Moreover, we demonstrate that the optimized pulse we propose exhibits enhanced robustness compared to the one featured in Ref.~\onlinecite{hai2023universalcompare}, as detailed in Appendix~\ref{app:robust_pulse}.

\subsection{Robust control of a composite cavity-ancilla system}\label{sec:robust_composite}
For an isolated ancilla, the propagator at the end of a robust pulse may include a phase that depends on the detuning, \(\hat{U}_\delta(t_\text{g}) \approx \hat{U}_\text{target}e^{i\theta(\delta)}\), for $\delta\in \mathbb{D}$. 
This is merely a global phase and can be disregarded. 
Generally, the situation differs when a cavity mode is coupled to the ancilla. 
In instances where the cavity is prepared in a Fock state $|n\rangle$, the detuning of the ancilla arises from the Stark shift $\delta=n\chi$. Provided that $n\chi\in \mathbb{D}$, the evolution can be described by 
\begin{equation}\label{eq:definite_state}
    \hat{U}_\delta'(t_\text{g}) |q\rangle|n\rangle 
    \approx \hat{U}_\text{target}|q\rangle   e^{i\theta(n\chi)}|n\rangle.
\end{equation}
Here, $\hat{U}_\delta' = \hat{U}_\text{target}\otimes\sum_n e^{i \theta(n\chi)}|n\rangle \langle n|$ is the propagator of the composite system. 
When the cavity is in a superposition state $\sum_n c_n |n\rangle$, it follows that
\begin{equation}\label{eq:relative_phase}
    \hat{U}_\delta'(t_\text{g}) |q\rangle \sum_n c_n |n\rangle 
    \approx \hat{U}_\text{target}|q\rangle \sum_n c_n  e^{i\theta(n\chi)}|n\rangle.
\end{equation}
This implies that $\theta(n\chi)$ leads to relative phases among various Fock states, lowering the fidelity of unconditional ancilla rotations.
Nevertheless, as we demonstrate subsequently, the magnitude of the relative phase is minor for robust pulses. Furthermore, we illustrate that errors stemming from this minor relative phase can be mostly corrected, thus validating the usage of previously-developed optimal pulses.

We first illustrate that the phase $\theta
(\delta)$ is approximately a constant, with in the range $\delta\in \mathbb{D}$ for a robust ancilla pulse.
Consider the Hamiltonian of a two-level qubit in the frame co-rotating with the drive frequency,
\begin{equation}
    \hat{H}_\text{q} = \delta\hat{\sigma}_z/2 + \varepsilon_I(t)\hat{\sigma}_x +  \varepsilon_Q(t)\hat{\sigma}_y.
\end{equation}
Here, \(\delta\) represents the detuning of the drive from the bare qubit frequency, and \(\varepsilon_{I,Q}(t)\) denote the envelopes of the two quadratures of the qubit drive. 
Without detuning (i.e., $\delta=0$), the propagator is 
\begin{equation*}
    \hat{U}_0 = \mathcal{T}\exp\left[-i\int_0^t\mathrm{d}\tau\,\varepsilon_I(\tau)\,\hat{\sigma}_x -i\int_0^t\mathrm{d}\tau\,\varepsilon_Q(\tau)\,\hat{\sigma}_y\right],
\end{equation*}
and correspondingly the interaction Hamiltonian is \(\hat{H}_\text{int} = \hat{U}_0^\dag \hat{H}_\text{q} \hat{U}_0 - i\hat{U}_0^\dag \dot{\hat{U}}_0\). 
The propagator in the interaction picture is then given by
\begin{equation}\label{eq:propagator_int}
    \hat{U}_\text{int}(t) = \mathcal{T}\exp\left[-i\int_0^t \hat{H}_\text{int}(\tau)\,\mathrm{d}\tau\right].
\end{equation}
Equivalently, we can express it in the form of
\begin{equation}
    \hat{U}_\text{int}(t) = \exp[i A(t)\hat{\sigma}_x + i B(t)\hat{\sigma}_y + i C(t) \hat{\sigma}_z)],
\end{equation}
where $A(t)$, $B(t)$, $C(t)$ are real-valued functions.
Note that the parenthesis does not include a term proportional to $\hat{\mathbb{I}}$, which can be confirmed by conducting a Magnus expansion of the time-ordered exponential in Eq.~\eqref{eq:propagator_int}.
Denoting $\lambda = \sqrt{A^2+B^2+C^2}$, we have
\begin{equation}
    \hat{U}_\text{int}(t) = \cos(\lambda) \hat{\mathbb{I}} + i \text{sinc}(\lambda)[A\hat{\sigma}_x + B\hat{\sigma}_y + C \hat{\sigma}_z)].
\end{equation}
The propagator in the Schr\"{o}dinger picture at the end of the pulse is thus 
\begin{equation}\label{eq:target_approx}
    \hat{U}_\delta(t_\text{g}) = \hat{U}_0(t_\text{g})\hat{U}_\text{int}(t_\text{g}) = \hat{U}_\text{target}e^{i\theta_0} \hat{U}_\text{int}(t_\text{g}),
\end{equation}
where the second equality reflects the fact that the evolution without detuning realizes the target gate up to a constant phase $\theta_0$.
Therefore, we can express the infidelity of the target operation on a detuned ancilla as 
\begin{equation}\label{eq:ancilla_fidelity}
    \mathcal{I}_\text{c}(\delta) = 1-
\frac{1}{4}\left|\Tr\left[\hat{U}_\text{target}^\dagger \hat{U}_\delta(t_\text{g})\right]\right|^2 = 1 - |\cos(\lambda)|^2. 
\end{equation}
The realization of high-fidelity robust control on an isolated ancilla (i.e., $\mathcal{I}_\text{c}(\delta)\approx 0$) thus indicates that \(\cos(\lambda) \approx \pm 1\) and $ \hat{U}_\delta(t_\text{g}) \approx \pm \hat{U}_\text{target}e^{i\theta_0}$.
The above result shows that, approximately, the phase $\theta(\delta)$ is restricted to two discrete values: $\theta_0$ or $\theta_0+\pi$, for $\delta\in\mathbb{D}$. 
In the absence of detuning, $\delta=0$, the phase takes $\theta(0)=\theta_0$ by definition. 
Since the mapping between the detuning and the phase is continuous~\footnote{We have conducted numerical simulations to verify this point.}, the phase does not oscillate between the above two values, and remains close to $\theta(\delta)=\theta_0$. 
In such a scenario, we can perform a Taylor series expansion of the phase around \(\delta = 0\),
\begin{equation}
    \theta(\delta) = \theta_0 + \partial_\delta \theta |_{\delta = 0} \delta + \mathcal{O}(\delta^2).
\end{equation}
The analysis above is performed for two-level system. For a transmon driven by the robust pulse, we assume such analysis still holds. 
In the cavity-ancilla system with \(\delta = n\chi\), the leading-order contribution can be corrected via a virtual cavity phase rotation
\begin{equation}\label{eq:first_order}
    \hat{U}_\text{cavity} = \exp(- i \chi \partial_\delta \theta |_{\delta = 0} \hat{a}^\dag \hat{a} ).
\end{equation}
If required, residual higher-order contributions could be further suppressed by incorporating a cost function in the pulse optimization process to penalize the nonlinear component.

\begin{figure}
    \centering
    \includegraphics[width=\columnwidth]{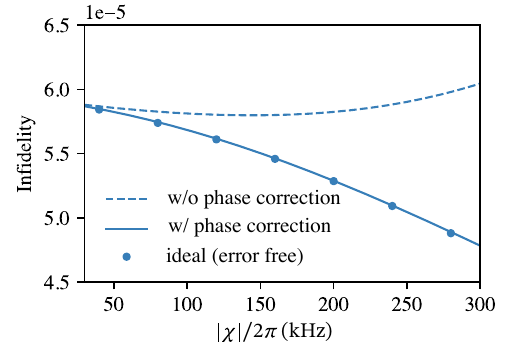}
    \caption{
    Infidelities of $\hat{X}_\pi$ gate pulses acting on an ancilla dispersively coupled to a cavity mode. 
    The horizontal axis denotes the dispersive shift $\chi$.
    Solid and dashed lines show the infidelities achieved with and without the cavity phase corrections in Eq.~\eqref{eq:first_order}, respectively. 
    For reference, infidelities obtained without relative cavity phase are displayed using discrete markers. 
    [Parameters: $K_\text{q}/2\pi=-200\,\text{MHz}$, $d_\text{c}=10$, with other nonlinearities derived from Eq.~\eqref{eq:dispersive_relation}.]}
    \label{fig:fig_coupled_pulse}
\end{figure}

To validate the effectiveness of our correction scheme, we execute optimized 20-ns \(\hat{X}_{\pi}\) pulses [as shown in Fig.~\ref{fig:fig_qoc}(b)] on a coupled cavity-ancilla system and calculate the resulting gate infidelity. 
The infidelity for the coupled system generalizes Eq.~\eqref{eq:fid_single},
\begin{equation}
\mathcal{I}'_\text{c}(\delta) = 1-\frac{1}{4 d_\text{c}^2}\left|\Tr\left[\hat{P}'_{0,1}\hat{U}_\text{target}^{'\dagger} \hat{U}'_\delta(t_\text{g})\right] \right|^2,
\end{equation}
where $\hat{P}'_{0,1}=\hat{P}_{0,1}\otimes\hat{\mathbb{I}}$ represents the projector onto the ancilla computational subspace, $\hat{U}_\text{target}'=\hat{U}_\text{target}\otimes \hat{\mathbb{I}}$ describes the target operation of the full system, and $d_\text{c}$ denotes the cavity cutoff dimension. 
In Fig.~\ref{fig:fig_coupled_pulse}, we present the ideal scenario where there is no relative phase associated with the cavity mode, i.e., \(\theta(\delta)\) remains constant. 
This case is used as a reference and is depicted by circular dots. 
In practice, the infidelity, displayed with the dashed curve, shows a minor deviation from the ideal case. 
Remarkably, this deviation can be almost entirely eliminated using the correction method outlined in Eq.~\eqref{eq:first_order}, with the outcomes displayed as the solid line.
Hence, the numerical simulation corroborates the preceding analytical proof, affirming the applicability of a robust pulse optimized for a single ancilla to coupled systems, with minor modifications.
It is crucial to emphasize that the virtual phase correction is only effective when the detuning remains within the range of robustness, making it primarily compatible with pulses obtained through optimal control techniques. In Appendix~\ref{app:virtual}, we elaborate on the limitations of this method, supported by examples involving Gaussian and DRAG pulses. 

\section{Benchmarking the robustness to crosstalk in multimode control}\label{sec:main}
We illustrate the reduction of crosstalk errors through robust ancilla control in two distinct multimode control scenarios. 
First, we focus on a typical quantum state transfer in one of a pair of bosonic modes, with the second mode idling.
Here, the population in the idling (spectator) mode can alter the ancilla frequency due to dispersive interactions. This issue becomes more pronounced as the number of modes linked to the same ancilla increases, presenting a significant challenge in scaling up multimode systems.
The second scenario addresses the creation of two-mode entangled states, a critical process in quantum computation. During the gate sequence, the photon population in both modes varies, leading to shifts in the ancilla frequency and resulting in crosstalk error. 

\subsection{Quantum state transfer with an idling mode}

\begin{figure}
    \centering
    \includegraphics[width=\columnwidth]{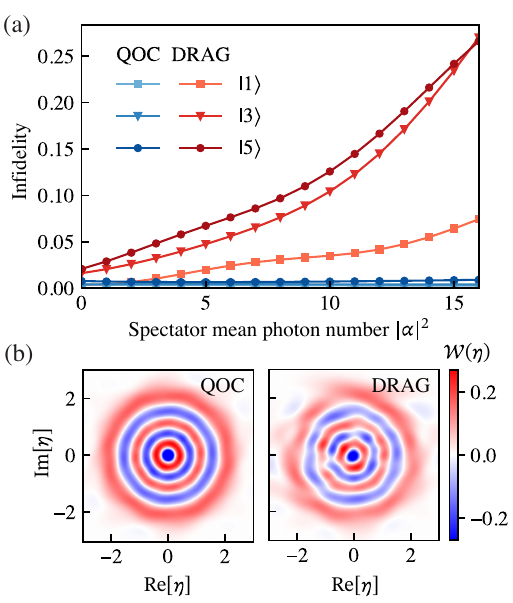}
    \caption{
    Benchmarking of QOC and DRAG pulses in Fock state generation, $|0\rangle |\alpha \rangle |g\rangle\to |n\rangle |\alpha \rangle |g\rangle$. 
    (a) Infidelity as a function of mean photon number $|\alpha|^2$ in the idling mode, for the preparation of target Fock states with $n=\{1,3,5\}$.
    DRAG pulse infidelity increases with $|\alpha|^2$, while QOC pulse infidelity remains nearly independent of $|\alpha|^2$.
    (b) Wigner tomography $\mathcal{W}(\eta)$ visualizing the state of the target mode subsystem for $|\alpha|^2=16$. The fidelities of states obtained using QOC and DRAG are 99.1\% and 73.3\%, respectively. 
    [Parameters: $K_\text{q}/2\pi=-200$\,MHz, $\chi_{1}/2\pi=\chi_{2}/2\pi=-300$\,kHz, $\Delta_{1}/2\pi = \Delta_{2}/2\pi = 2\,\text{GHz}$, with other nonlinearities derived from Eq.~\eqref{eq:dispersive_relation}.]}
    \label{fig:fig_fock_spectator} 
\end{figure}

To benchmark the performance of bosonic control, a common approach is to examine the generation of a Fock state \(|n\rangle\) from the vacuum state in the target mode~\cite{Heeres2015b,Chakram2020,Eickbusch2021,Chakram2022,Trowbridge2023}. 
Additionally, we ensure the idling mode maintains a prepared state. 
To accommodate the potential acquisition of nontrivial relative phases in the idling mode, we initialize it in a coherent state \(|\alpha\rangle\), rather than a Fock state.
Overall, the state-transfer process is denoted as \(|0\rangle |\alpha \rangle |g\rangle\to |n\rangle |\alpha \rangle |g\rangle\), with $|g\rangle$ being the ancilla ground state.
We implement the steps detailed in Sec.~\ref{sec:ecd_intro} to construct cavity control pulses. 
Employing two distinct ancilla control schemes, QOC and DRAG, we perform pulse-level simulations with the Hamiltonian in Eq.~\eqref{eq:displaced_frame} to evaluate the infidelities of the operation. 
This enables a direct comparison of the two control strategies in the presence of crosstalk.
Specifically, we vary the average photon number in the idling mode to examine the spurious effects of crosstalk.

We compute the state-transfer infidelity of the composite system as $1-|\langle\psi_\text{f}|\psi_\text{target}\rangle|^2$, for the final state $|\psi_\text{f}\rangle$ and the target state $|\psi_\text{target}\rangle=|n\rangle |\alpha\rangle | g\rangle$, and plot its dependence on $|\alpha|^2$ in Fig.~\ref{fig:fig_fock_spectator}(a).
When the idling mode is in the vacuum state (i.e., \(\alpha=0\)), the corresponding infidelities act as reference values, excluding crosstalk errors from the idling mode. 
As the average photon number in the idling mode increases, the infidelities associated with the DRAG scheme exhibit a marked escalation. 
Conversely, in the case of QOC, the infidelities barely deviate from the reference one, despite large variations in the average photon number of the idling mode. 
For instance, we consider the case of creating the Fock state \(n=5\), when the idling mode has a mean photon number of \(|\alpha|^2=16\). The infidelity of the final state generated by DRAG pulses surpasses \(0.25\), whereas it remains below \(0.01\) for QOC pulses. 
This marked difference illustrates the limited performance of conventional DRAG pulses in the presence of crosstalk, and underscores the superior robustness of the developed QOC pulses.

To visualize the differences in infidelities, we calculate the Wigner function \(\mathcal{W}(\eta)\) for the target mode subsystem, where the variable \(\eta\) denotes the coordinates in the complex phase space. The Wigner function, for the target Fock state \(n=5\) with an average photon number of \(|\alpha|^2=16\) in the idling mode, is calculated by tracing out the idling and ancilla degrees of freedom. These results are displayed in Fig.~\ref{fig:fig_fock_spectator}(b). 
The Wigner function of the final state derived from DRAG pulses manifests significant deviations from the expected ring pattern, indicative of diminished fidelity ($<0.75$).
In contrast, the results from QOC pulses mirror the ideal quantum state distribution, denoting high fidelity ($>0.99$). 
This example highlights the efficacy of the robust ancilla control protocol in operating the target mode, particularly in scenarios involving photons in spurious idling modes.

\subsection{Two-mode entanglement operations}
\begin{figure*}
    \centering
    \includegraphics[width=\textwidth]{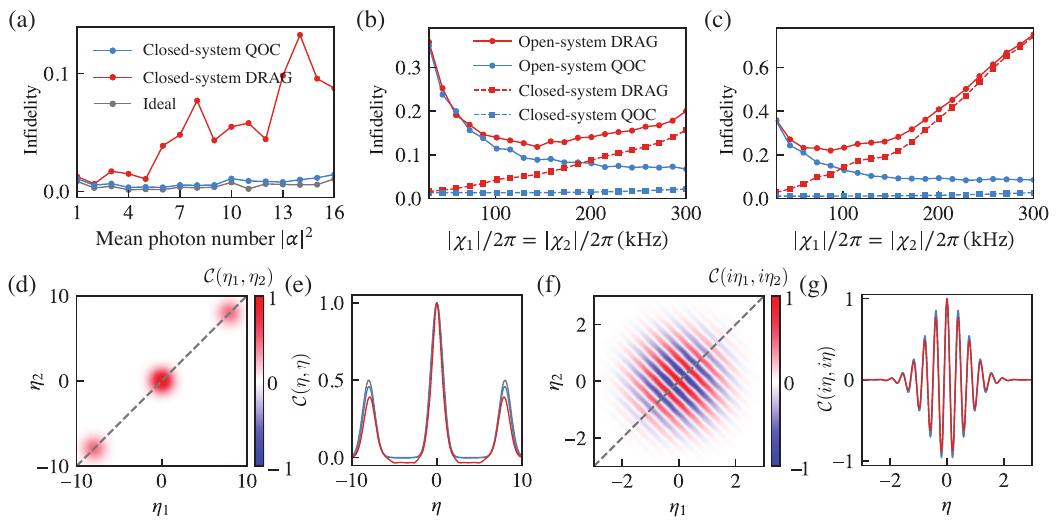}
    \caption{
    Benchmarking the robustness of QOC and DRAG pulses in the generation of entangled Bell-cat states, $|0\rangle |0 \rangle |g\rangle \to\mathcal{N}(|\alpha\rangle|\alpha\rangle + |{-}\alpha\rangle|{-}\alpha\rangle) |g\rangle$, with $\mathcal{N}\approx 1/\sqrt{2}$. 
    (a) Infidelity as a function of the mean photon number $|\alpha|^2$, with fixed dispersive shifts $|\chi_{1}|/2\pi=|\chi_{2}|/2\pi=200$\,kHz. Both pulses have a duration of 20 ns. 
    (b) and (c) Infidelity versus the dispersive shifts, with a constant $|\alpha|^2=16$, for ancilla pulse durations of 20 and 50 ns, respectively. Solid and dashed lines are obtained from simulations with and without decoherence errors.
    In all scenarios considered here, the infidelities derived from QOC pulses are significantly lower compared with the ones obtained from DRAG pulses. 
    (d) and (f) Joint characteristic function of the Bell-cat state, evaluated at real $(\eta_1,\eta_2)$ and imaginary arguments $(i\eta_1,i\eta_2)$, respectively. The simulation is performed for an open system with 20-ns QOC pulses. 
    (e) and (g) 1D line cuts obtained from (d) and (f). 
    While the results derived from QOC pulses (blue) closely align with the ideal case (gray), those obtained from DRAG pulses (red) exhibit clear deviations. 
    [Parameters: $K_\text{q}/2\pi=-200$ MHz, $\Delta_{1}/2\pi = \Delta_{2}/2\pi = 2\,\text{GHz}$, $T_1=T_\phi=50$\,$\mu$s, $T_{\phi 1,2}=150$ ms, with other nonlinearities and decoherence parameters derived from Eq.~\eqref{eq:dispersive_relation} and Eq.~\eqref{eq:disp_decay}, respectively.]}
    \label{fig:fig_bell_cat} 
\end{figure*}

In the second scenario, we consider the creation of two-mode Bell-cat states~\cite{cat_two,Diringer2023}, $\mathcal{N}(|\alpha\rangle|\alpha\rangle + |{-}\alpha\rangle|{-}\alpha\rangle)$, where $\mathcal{N}\approx 1/\sqrt{2}$ is the normalization factor. 
The ability to produce such entangled cat states is crucial for various applications, including quantum error correction~\cite{cat1,cat2,Touzard2018} and quantum teleportation~\cite{Lee2011}. 
Unlike the previous scenario, there is no idling mode in the entanglement operations. 
In the following, we focus on characterizing the detrimental effects of crosstalk. This is achieved by examining the dependence of infidelity on both the cavity-ancilla coupling strength and the average photon population in the cavity modes. 
These are the two key factors in the dispersive term $\chi_{i}\hat{a}^\dag_{i} \hat{a}_{i} \hat{q}^\dag \hat{q}$ ($i=1,2$).

In Fig.~\ref{fig:fig_bell_cat}(a), we set the dispersive shifts of both cavity modes to $\chi_{1}/2\pi=\chi_{2}/2\pi=-200$ kHz, and vary the cat size from $\alpha=0$ to $\alpha=4$. For each value of $|\alpha|^2$, we first carry out a circuit-level optimization to determine the parameters $\{\beta_{1,i}, \beta_{2,i}, \varphi_{1,i}, \varphi_{2,i}, \theta_{1,i}, \theta_{2,i}\}$ for two-mode ECD gates and ancilla rotations.
The resulting infidelity (which is exclusively due to gate decomposition errors), is shown as the gray line, and serves as a reference for subsequent pulse-level simulations. 
The abrupt jump in the infidelity as a function of photon number arises because the optimizations for circuit parameters are conducted independently for each value of \(|\alpha|^2\). Consequently, the local minimum achieved in every optimization may correspond to circuit parameters that are not continuous with respect to \(|\alpha|^2\).
We then perform pulse-level simulations by realizing the ancilla rotations with DRAG and QOC pulses, respectively. 
The infidelity obtained with DRAG pulses (red lines) significantly increases with larger cat size. By contrast, the infidelity obtained from QOC pulses (blue lines), closely aligns with the reference values. 
These results again highlight the impressive effectiveness of QOC pulses in achieving robust control against crosstalk. 

In addition to the photon number, we next explore the dependence of infidelities on the dispersive shift, which is another key factor in characterizing the crosstalk. Specifically, we fix $|\alpha|^2=16$ and adjust the dispersive shifts in the range $|\chi_{1}|/2\pi=|\chi_{2}|/2\pi\in[30,300]$ kHz. 
We start with closed-system simulations, the results of which are shown as dashed lines in Fig.~\ref{fig:fig_bell_cat}(b). 
Consistent with the results when varying photon numbers, we find distinct behaviors of the infidelities obtained from DRAG and QOC pulses. 
While the infidelity associated with DRAG pulses notably escalates for a larger dispersive shift, the one derived from QOC pulses remains negligible across the selected range of cavity-ancilla coupling.
This resilience of QOC pulses underscores their superior ability to achieve crosstalk-robust quantum control in complex quantum systems. 

The contrast becomes even more pronounced when we conduct open-system simulations~\footnote{To efficiently simulate an open system with a large Hilbert space, we perform Monte Carlo simulations with 2000 quantum trajectories.}.
At $|\chi_{1}|/2\pi=|\chi_{2}|/2\pi<100$ kHz, infidelities obtained from both DRAG and QOC decrease. This is attributed to the shorter gate duration, which leads to less decoherence error.
However, with a further increase in dispersive shifts, the infidelities exhibit distinct behaviors. 
For DRAG pulses, the infidelity reaches a minimum and then rises again. 
This pattern exhibits the competition between decreasing decoherence error and increasing crosstalk-induced coherent error as dispersive shifts increase.
In contrast, the infidelity derived from QOC pulses decreases monotonically, reducing the decoherence error while simultaneously maintaining negligible coherent error. 
Within the selected range of $|\chi_{1}|/2\pi=|\chi_{2}|/2\pi\in[30,300]$ kHz, the QOC protocol results in a suppression of infidelity by more than threefold. 
As a comparison for different pulse durations, we extend our simulation to include 50-ns pulses, as depicted in Fig.~\ref{fig:fig_bell_cat}(c). 
The use of longer DRAG pulses results in a notable increase in infidelity, particularly at larger values of dispersive shifts. In contrast, the behavior of QOC pulses remains consistently negligible for both short and long pulse durations.
At the maximum value of $|\chi_{1}|/2\pi=|\chi_{2}|/2\pi=300$ kHz considered here, switching from DRAG to QOC pulses results in an infidelity reduction exceeding sevenfold. This dramatic decrease clearly illustrates the effectiveness of the robust control protocol we have developed with QOC pulses.

To provide a visual representation of the fidelities of the prepared entangled states and establish a reference for future experimental investigations, we compute the joint characteristic function $\textrm{Tr}[\mathcal{C}(\eta_1,\eta_2)\equiv\hat{D}_1(\eta_1)\hat{D}_2(\eta_2)\hat{\rho}]$~\cite{characteristic_function,Diringer2023} from an open-system simulation with $|\chi_{1}|/2\pi=|\chi_{2}|/2\pi=300$ kHz and $|\alpha|^2=16$.
Given that each cavity displacement can be a complex number, the arguments of the two-mode characteristic function span a four-dimensional parameter space. For simplicity, we opt to plot the two-dimensional cuts by setting either $\Im[\eta_{1,2}]=0$ or $\Re[\eta_{1,2}]=0$ in Fig.~\ref{fig:fig_bell_cat}(d) and (f), respectively.
When the displacements are real-valued for both cavity modes, the characteristic function of a Bell-cat state $\mathcal{N}(|\alpha\rangle|\alpha\rangle + |{-}\alpha\rangle|{-}\alpha\rangle)$ exhibits three blobs centered at $(0,0)$, $(2\alpha,2\alpha)$, and $({-}2\alpha,{-}2\alpha)$. For $\alpha \gg 1$, these blobs have peak values $1$, $0.5$, and $0.5$, respectively. This pattern is accurately mirrored in the result obtained with QOC pulses, as shown in Fig.~\ref{fig:fig_bell_cat}(d). For a quantitative comparison, we further take a one-dimensional line cut along the diagonal line. The results (blue) are shown in Fig.~\ref{fig:fig_bell_cat}(e), along with the one obtained with DRAG pulses (red). The characteristic function for an ideal entangled cat state (shown in gray) obeys
\begin{equation}
    \mathcal{C}(\eta,\eta) = e^{-\eta^2} + \tfrac{1}{2}e^{-(\eta-2\alpha)^2} + \tfrac{1}{2}e^{-(\eta+2\alpha)^2}.
\end{equation}
While the result from QOC closely aligns with the ideal case, the one obtained from DRAG shows clear deviations in the two lower peaks, and even exhibits negative values between the peaks.
When the cavity displacements are purely imaginary, the characteristic functions, shown in Fig.~\ref{fig:fig_bell_cat}(f), display fringes that signify the entangled nature of the states. We take another line cut with $i\eta_{1}=i\eta_{2}$, see Fig.~\ref{fig:fig_bell_cat}(g). In this scenario, the QOC result again outperforms the one obtained from DRAG, and follows the ideal expression
\begin{equation}
    \mathcal{C}(i\eta,i\eta) = e^{-4\alpha^2-\eta^2}[1+e^{4\alpha^2}\cos{(4\alpha\eta)}]. 
\end{equation}
Therefore, this example clearly demonstrates the effectiveness of the robust control protocol in reducing crosstalk during two-mode entanglement operations.

\section{Discussion and outlook}\label{sec:discussion}
The robust control protocol for two-mode ECD gates significantly suppresses crosstalk errors in ``all-to-one" multimode cavity-based quantum processor architectures.
However, neither the ``all-to-one" architecture nor the ECD gate is the only viable option for controlling multimode oscillators. 
In this section, we first examine the advantages and challenges of the chosen architecture and compare it with other existing architectures in the literature. Subsequently, we discuss how our robust control technique can be applied to other multimode gate protocols, extending its benefits beyond the specific implementations discussed in this paper.

\subsection{Comparison with other multimode architectures}
In the ``all-to-one'' multimode architecture discussed in this work, a single ancilla is coupled to every cavity mode to execute single-mode gate operations. For multimode operations, the same ancilla entangles with multiple cavity modes to mediate the entanglement between modes that are otherwise not directly coupled. In addition to the two-mode ECD protocol presented here, this architecture supports various other multimode protocols that create multimode entanglement deterministically~\cite{Rosenblum2018,Chakram2020,Diringer2023}. 
In general, this architecture features several advantages: First, the usage of a single transmon ancilla greatly reduces the effort in device fabrication and operation. Additionally, the architecture minimizes the number of lossy ancillas, which typically Purcell-limits the cavity coherence. Moreover, weakly coupled ancilla can be used to further suppress Purcell loss while still performing fast gates through various acceleration strategies~\cite{Eickbusch2021,Diringer2023}. However, this architecture also presents challenges. 
The ancilla becomes entangled with the cavity modes during multimode operations, making ancilla decoherence the dominant source of error.

In an alternative architecture, each cavity mode is controlled by its own transmon ancilla, with a coupler facilitating the parametric beamsplitter interaction between cavity modes. 
For example, this can be realized with three-wave or four-wave mixing process in a Superconducting Nonlinear Asymmetric Inductive eLement (SNAIL)~\cite{Zhou2023,Chapman2022}, a DC Superconducting QUantum Interference Device (SQUID)~\cite{Lu2023}, and a transmon coupler~\cite{gao2018,Gao2019}.
When the coupler is undriven, different cavity modes are decoupled, eliminating crosstalk errors. 
This then allows the usage of selective single-mode control such as the SNAP gate. 
Additionally, the coupler is virtually populated during the beamsplitter interaction~\cite{gao2018,yaxing}, so the fidelity of multimode operations is less affected by coupler decoherence. 
However, this architecture demands a more complex experimental setup, requiring a transmon ancilla for each cavity mode in addition to the coupler. 
All the individual ancillas and the coupler introduce additional Purcell loss to the cavity modes. Moreover, strong coupling between the cavity modes and the coupler is typically required to enable fast inter-cavity operations, which further limits the cavity coherence.

The optimal architecture for multimode control remains an open question. A potential hybrid solution is the ``manipulator-storage'' architecture~\cite{manipulator,Pietikainen2024,jordan}, which combines features of both strategies. It consists of a manipulator unit with a few cavity modes directly coupled to an ancilla for operations, and a storage unit with a larger number of modes for information storage. A coupler then enables SWAP-type operations between the two units, facilitating state transfer between storage and manipulator modes. In this way, the architecture potentially allows for hardware-efficient multimode control while minimizing crosstalk. 

\subsection{Generalization to other unselective cavity control protocols}
Within the ``all-to-one'' multimode architecture, crosstalk error is ubiquitous across various cavity control protocols, though the degree of susceptibility varies. For protocols involving selective ancilla operations, such as the SNAP gate, crosstalk error is unavoidable due to the inherent selective nature of the protocol. This selectivity to photons in one cavity mode naturally makes it vulnerable to photons in other cavity modes. 
Conversely, there are protocols that rely on unselective ancilla operations, where ideally the ancilla operation does not depend on the photon number. 

Generally, our technique to suppress crosstalk errors can be applied to multimode control protocols that involve unselective ancilla operations. In addition to the two-mode ECD protocol discussed in the main text, other protocols include, for example, the conditional not displacement gate proposed in Ref.~\onlinecite{Diringer2023}. Furthermore, developing robust pulses for a coupled ancilla-cavity system could be extended to protocols with sideband drives~\cite{sideband,Rosenblum2018,Chakram2020}. By adapting our robust pulse development approach to these contexts, it is possible to mitigate crosstalk errors and enhance the fidelity of a broader range of quantum control protocols. 

\section{Conclusions}\label{sec:conclusions}
In summary, we propose a protocol for controlling multiple bosonic modes that substantially suppresses crosstalk error. This crosstalk is a prevalent challenge in current superconducting circuit architectures, where a single transmon ancilla is often coupled to multiple cavity modes, resulting in spurious transmon-mediated inter-cavity interactions.
To address this challenge, we leverage quantum optimal control techniques to generate ancilla control pulses. These pulses are robust to variations in ancilla frequency, which are induced by Stark shifts originating from the interaction with coupled bosonic modes.
We demonstrate the efficiency of our protocol by integrating these robust ancilla pulses into a promising multimode cavity control scheme based on the echoed conditional displacement (ECD) gate. We perform comprehensive numerical simulations for two scenarios prone to crosstalk error: the creation of single-mode Fock states in the presence of photons in idling (spectator) modes, and the generation of two-mode Bell-cat states.
Our analysis spans a wide range of photon occupations and cavity-transmon couplings, revealing that our protocol consistently yields high-fidelity operations. Notably, the performance markedly exceeds that achieved with conventional Gaussian or DRAG pulses.
Crucially, the advantages of our protocol extend beyond the ECD gate. They can be readily carried over to other multimode schemes with non-selective ancilla operations. These findings not only validate the feasibility of current architectures but also provide essential guidelines for accomplishing high-fidelity operations in multimode bosonic systems.

\begin{acknowledgments}
We thank E. Gupta and S. Joshi for illuminating discussions.
This material is based upon work supported by the U.S. Department of Energy, Office of Science, National Quantum Information Science Research Centers, Superconducting Quantum Materials and Systems Center (SQMS) under contract number DE-AC02-07CH11359.

X. Y. and Y. L. contributed equally to this work.
\end{acknowledgments}

\appendix
\section{Multimode universal control with ECD protocol}\label{app:beamsplitter}
The universal control of multiple bosonic modes encompasses the capability to enact arbitrary unitary transformations, generated by linear combinations of Hamiltonians of the form $\hat{x}_{j}^m \hat{p}_{k}^n \hat{\sigma}_i$ ($m,n\in \mathbb{Z}^+$).
Here, $\hat{x}_{j}=(1/\sqrt{2})(\hat{a}_{j}^\dag+\hat{a}_{j})$ and $\hat{p}_{k}=(i/\sqrt{2})(\hat{a}_{k}^\dag-\hat{a}_{k})$ denote the dimensionless position and momentum operators of modes $j$ and $k$, respectively, and $\hat{\sigma}_i\in\{\hat{\mathbb{I}}, \hat{\sigma}_x, \hat{\sigma}_y, \hat{\sigma}_z\}$ refers to an ancilla Pauli operator.

Given that universal control of a single mode using the ECD protocol has been established~\cite{Eickbusch2021}, it suffices to construct the unitary of the beamsplitter interaction $\hat{B}_{jk} = \hat{x}_j \hat{p}_k - \hat{p}_j \hat{x}_k$, in order to demonstrate multimode universality~\cite{Braunstein2005}. 
The generators of the ECD gate and the qubit rotation $\hat{R}_\varphi (\theta)$ are $\{\hat{x}_{j,k}\hat{\sigma}_z, \hat{p}_{j,k}\hat{\sigma}_z, \hat{\sigma}_x, \hat{\sigma}_y\}$. 
In the limit $\delta t\to 0$, geometric analysis reveals that the ability to evolve under generators $\hat{A}$ and $\hat{B}$ also enables evolving under generators $-i[\hat{A},\hat{B}]$ and $\hat{A}+\hat{B}$,
\begin{align*}
    \begin{split}
        e^{-i\hat{A}\delta t}e^{-i\hat{B}\delta t}e^{i\hat{A}\delta t}e^{i\hat{B}\delta t} = e^{[\hat{A},\hat{B}]\delta t^2} + \mathcal{O}(\delta t^3), \\
         e^{i\hat{A}\delta t/2}e^{i\hat{B}\delta t/2}e^{i\hat{B}\delta t/2}e^{i\hat{A}\delta t/2} = e^{i(\hat{A}+\hat{B})\delta t} + \mathcal{O}(\delta t^3) .
    \end{split}
\end{align*}
Utilizing these identities, we first extend the generator set to $\{\hat{x}_{j,k}\hat{\sigma}_i, \hat{p}_{j,k}\hat{\sigma}_i, \hat{\sigma}_i\}$. 
We then construct the entangling operation between two bosonic modes, facilitated by the commutator $[\hat{x}_{j,k}\hat{\sigma}_y, \hat{p}_{k,j}\hat{\sigma}_z]\propto \hat{x}_{j,k} \hat{p}_{k,j} \hat{\sigma}_x$. 
Next, we disentangle the ancilla by taking $[\hat{x}_{j,k} \hat{p}_{k,j} \hat{\sigma}_x, \hat{x}_{j,k}\hat{\sigma}_y] \propto \hat{x}_{j,k}^2 \hat{p}_{k,j} \hat{\sigma}_z$, and subsequently $[\hat{x}_{j,k}^2 \hat{p}_{k,j} \hat{\sigma}_z, \hat{p}_{j,k}\hat{\sigma}_z]\propto \hat{x}_{j,k}\hat{p}_{k,j}$, where we have used the commutation relation $[\hat{x}_{j,k}, \hat{p}_{j,k}]=i$. Finally, through the application of the second identity above, we successfully approximate the unitary generated by the beamsplitter interaction $\hat{x}_j \hat{p}_k - \hat{p}_j \hat{x}_k$, thereby establishing the framework for multimode universal control.

\section{Detailed derivation of numerical simulation for a noisy driven coupled cavity-ancilla system}\label{app:theory}
In this section, we provide a detailed step-by-step derivation of the procedure for numerically simulating a driven cavity-ancilla system in the presence of noise. 
\subsection{Dispersive Hamiltonian for the undriven system}
In the absence of external drives, the Hamiltonian governing the coupled cavity-ancilla system in the lab frame is
\begin{equation*}
    \hat{H}_0 = \omega_\text{q} \hat{q}^\dag \hat{q} + \frac{K_\text{q}}{2}\hat{q}^{\dag 2}\hat{q}^{2} + \sum_{i=1,2} \left[ \omega_{i}\hat{a}^\dag_i \hat{a}_i + g_i (\hat{a}^\dag_i \hat{q}+\hat{a}_i \hat{q}^\dag) \right].
\end{equation*}
Here, $\omega_\text{q}$ and $K_\text{q}$ parameterize the frequency and anharmonicity of the transmon ancilla. Additionally, $\omega_i$ and $g_i$ denote the frequencies of the two cavity modes and their coupling strengths to the ancilla. 
In the dispersive regime, characterized by large detunings $\Delta_i=\omega_i - \omega_\text{q}$ relative to the coupling strengths, $|\Delta_i| \gg g_i$, we diagonalize the Hamiltonian through a dispersive transformation, yielding
\begin{align*}
\hat{H}_\text{disp} &= \sum_{i=1,2} \left[\omega_i\hat{a}_i^\dag \hat{a}_i
+ (\chi_i \hat{a}_i^\dag \hat{a}_i + \frac{\chi_i'}{2} \hat{a}_i^{\dag 2} \hat{a}_i^2) \hat{q}^\dag \hat{q} 
+ \frac{K_{i}}{2} \hat{a}_i^{\dag 2} \hat{a}_i^2 \right] \nonumber\\ 
& + K_{12} \hat{a}_1^\dag \hat{a}_1 \hat{a}_2^\dag \hat{a}_2  + \omega_\text{q} \hat{q}^\dag \hat{q} + \frac{K_\text{q}}{2}\hat{q}^{\dag 2}\hat{q}^{2}.
\end{align*}
The nonlinearities manifest in several terms. 
The self- and cross-Kerr interaction strengths for the two cavity modes are described by \(K_{i}\) and  \(K_{12}\), and the first- and second-order dispersive shifts between the cavity modes and the ancilla are symbolized by $\chi_i$ and $\chi_{i}'$, respectively.
The above nonlinearities are inherited from the transmon anharmonicity and are interconnected through the relations~\cite{Wang2021},
\begin{align}\label{eq:dispersive_relation}
    \begin{split}
        \chi_{i} &\approx  2(g_{i}/\Delta_{i})^2 K_\text{q} - 8 ( g_{i}/\Delta_{i})^4 K_\text{q},  \\
        \chi_{i}' &\approx 9 ( g_{i}/\Delta_{i}) ^4 K_\text{q}^2/\Delta_{i}, \\
        K_{i} &\approx ({g_{i}/\Delta_{i}})^4 K_\text{q}, \\
        K_{12} & \approx 2(g_{1}g_2/\Delta_{1}\Delta_{2})^2 K_\text{q}. 
    \end{split}
\end{align}
The above approximations are valid in the regime where $|\Delta_i| \gg |K_\text{q}|$, consistent with the parameter regime considered throughout the paper. 
Specifically, we fix the parameter set: $K_\text{q}/2\pi = -200\,\text{MHz}, \Delta_{1}/2\pi = \Delta_{2}/2\pi = 2\,\text{GHz}$, and vary the coupling strengths $g_{i}$, taken to be identical for both cavity modes. Given that the dispersive shift $\chi_{i}$ is a directly measurable experimental quantity, we employ it to serve as the proxy for variations in the coupling strengths.

\subsection{Rotating-frame Hamiltonian for the driven system}
During the execution of ECD gates, microwave drives are applied on both the ancilla and the two cavity modes. Within the rotating-wave approximation, the Hamiltonian describing the drives takes the form of 
\begin{equation*}
    \hat{H}_\text{drive}(t) = \sum_{i=1,2}  \left[ \Omega_i(t)\hat{a}_i^\dag e^{-i\omega_{\text{d},i} t} + \varepsilon(t) \hat{q}^\dag e^{-i\omega_{\text{q}}t} + \textrm{h.c.} \right].
\end{equation*}
Here, the ancilla is driven resonantly with the temporal profile $\varepsilon(t)$, while the two cavity modes are driven off-resonantly with detunings $\omega_i - \omega_{\text{d},i} = -\chi_{i}/2$ and temporal profiles $\Omega_i(t)$. 
Moving into a co-rotating frame aligned with both the cavity and ancilla drives, we obtain the rotating-frame Hamiltonian
\begin{align}\label{eq:rot_ham}
\begin{split}
\hat{H}(t) &= \sum_{i=1,2} \bigg[ -\frac{\chi_i}{2}\hat{a}_i^\dag \hat{a}_i
+ (\chi_i \hat{a}_i^\dag \hat{a}_i + \frac{\chi_i'}{2} \hat{a}_i^{\dag 2} \hat{a}_i^2) \hat{q}^\dag \hat{q}  \\
& + \frac{K_{i}}{2} \hat{a}_i^{\dag 2} \hat{a}_i^2 \bigg]  + K_{12} \hat{a}_1^\dag \hat{a}_1 \hat{a}_2^\dag \hat{a}_2 + \frac{K_\text{q}}{2}\hat{q}^{\dag 2}\hat{q}^{2} \\ 
& +   \bigg( \sum_{i=1,2}  \Omega_i(t)\hat{a}_i^\dag + \varepsilon(t)\hat{q}^\dag + \textrm{h.c.}\bigg) .
\end{split}
\end{align}

\subsection{Displaced-frame Hamiltonian for the driven system}\label{app:displaced_frame}
When dealing with weak cavity drives, the rotating-frame Hamiltonian can be used to conveniently simulate the system dynamics. However, for large drive amplitudes where the cavity is highly populated, this approach demands a large truncation level of the cavity Hilbert space to ensure numerical convergence, thereby increasing computational cost.
A more computationally efficient strategy involves transforming the Hamiltonian to a drive-dependent displaced frame, in which the cavity displacement is eliminated (i.e., terms proportional to $\hat{a}_i$ or $\hat{a}_i^\dag$ are removed, see details in Ref.~\onlinecite{blais2007}).
As a result, simulations in the displaced frame require a considerably smaller dimension for the truncated Hilbert space.

To facilitate this transformation, we use the unitary $\hat{U}_\alpha(t) = \Pi_{i=1,2}\hat{D}_i^\dag[\alpha_i(t)]$, where $\alpha_i(t)$ obeys the following differential equation,
\begin{equation}\label{eq:ode}
    \partial_t \alpha_i = i\chi_i \alpha_i/2 - iK_i |\alpha_i|^2\alpha_i  - iK_{12}\alpha_i|\alpha_{\bar{i}}|^2 - i\Omega_i(t),
\end{equation}
subject to the initial condition $\alpha_i(0) = 0$.
Here, the subscript $\bar{i}$ denotes the index different than $i$, e.g., $\bar{1}=2$. 
Note that we have dropped the explicit time dependence of $\alpha_i(t)$ for simplicity.  
Applying this unitary transformation $\hat{H}'[\alpha_i(t), t] = \hat{U}_\alpha(t)\hat{H}(t)\hat{U}_\alpha^\dag(t) + i\partial_t\hat{U}_\alpha(t)\hat{U}_\alpha^\dag(t)$, we arrive at the displaced-frame Hamiltonian,
\begin{equation}\label{eq:displaced_frame}
    \hat{H}'[\alpha_i(t), t] = \hat{H}_\text{static} + \hat{H}_\text{diag}(t) + \hat{H}_\text{off-diag}(t),
\end{equation}
where we have emphasized its functional dependence on the specific $\alpha_i(t)$. 
The first term $\hat{H}_\text{static}$ consists of time-independent components of $\hat{H}(t)$ in Eq.~\eqref{eq:rot_ham}. The second term $\hat{H}_\text{diag}(t)$ includes time-dependent terms that are diagonal in the number basis of the transmon and cavity modes,
\begin{align}
     \hat{H}_\text{diag}(t) &= \sum_{i=1,2} 
     \bigg[ (2K_i|\alpha_i|^2 + K_{12}|\alpha_{\bar{i}}|^2 )\hat{a}_i^\dag \hat{a}_i  \nonumber\\ 
     & + 2\chi_i'|\alpha_i|^2 \hat{a}_i^\dag \hat{a}_i \hat{q}^\dag \hat{q} 
     + (\chi_i|\alpha_i|^2 - \frac{\chi'_i}{2}|\alpha_i|^4)\hat{q}^\dag \hat{q}\bigg]. \nonumber
\end{align}
The remaining term of $\hat{H}_\text{off-diag}(t)$ incorporates other off-diagonal, time-dependent terms,
\begin{align}
       \hat{H}_\text{off-diag}(t) &=  \sum_{i=1,2}  \bigg[\frac{K_i}{2}(2\alpha_i\hat{a}_i^{\dag 2} \hat{a}_i + \alpha_i^2 a_i^{\dag 2}) + \chi_i\alpha_i^* \hat{a}_i \hat{q}^\dag \hat{q} \nonumber \\
      &  + \frac{\chi'_i}{2}(2\alpha_i \hat{a}_i^{\dag 2} \hat{a}_i + \alpha_i^2 \hat{a}_i^{\dag 2}  + 2|\alpha_i|^2\alpha_i^* \hat{a}_i )\hat{q}^\dag \hat{q} \bigg] \nonumber\\
      & + K_{12} (  \alpha_2 \hat{a}_1^\dag \hat{a}_1 \hat{a}_2^\dag  + \alpha_1 \hat{a}_1^\dag  \hat{a}_2^\dag \hat{a}_2 + \alpha_1\alpha_2 \hat{a}_1^\dag \hat{a}_2^\dag  \nonumber\\
      & + \alpha_1\alpha_2^* \hat{a}_1^\dag \hat{a}_2  ) + \varepsilon(t) \hat{q}^\dag + \text{h.c.}. \label{eq:off_diag}
\end{align}

\subsection{Displaced-frame master equation for the driven open system}
In the presence of Markovian noise, the evolution of the coupled-system density matrix $\hat{\rho}$, within the same co-rotating frame as defined in Eq.~\eqref{eq:rot_ham}, is governed by the Lindblad master equation 
\begin{align}
    \begin{split}
        \partial_t \hat{\rho} &= -i[\hat{H}(t),\hat{\rho}]  + \gamma \mathcal{D}[\hat{q}]\hat{\rho} + 2\gamma_\phi \mathcal{D}[\hat{q}^\dag \hat{q}]\hat{\rho} \\
        & + \sum_{i=1,2} \left(\kappa_i \mathcal{D}[\hat{a}_i]\hat{\rho}  + 2\kappa_{\phi i}\mathcal{D}[\hat{a}_i^\dag \hat{a}_i]\hat{\rho} \right),
    \end{split}
\end{align}
with the dissipator $\mathcal{D}[\hat{\mathcal{O}}] \hat{\rho}= \hat{\mathcal{O}}\hat{\rho}\hat{\mathcal{O}}^\dag - \{ \hat{\mathcal{O}}^\dag \hat{\mathcal{O}}, \hat{\rho} \}/2$. 
Here, $\kappa_{i}$ and $\gamma$ denote the decay rates of the cavity modes and the ancilla, and $\kappa_{\phi i}$ and $\gamma_\phi$ represent the corresponding pure dephasing rates.
In the dispersive regime, the various decoherence rates follow the relations~\cite{Boissonneault2009a}
\begin{align}\label{eq:disp_decay}
    \begin{split}
        \kappa_{i} &\approx \tilde{\kappa}_{i} + (g_{i}/\Delta_{i})^2 \tilde{\gamma}, \\ 
        \gamma &\approx \tilde{\gamma} + \sum_{i=1,2} (g_{i}/\Delta_{i})^2 \tilde{\kappa}_i,  \\ 
        \kappa_{\phi i} &\approx \tilde{\kappa}_{\phi i} + (g_{i}/\Delta_{i})^4 \tilde{\gamma}_{\phi i},\\ 
        \gamma_{\phi} &\approx \tilde{\gamma}_{\phi} + \sum_{i=1,2} (g_{i}/\Delta_{i})^4 \tilde{\kappa}_{\phi i},
    \end{split}
\end{align}
with the quantities denoted by tildes representing the intrinsic decoherence rates for cavity and ancilla modes that are decoupled with each other, i.e., $g_i=0$.
For high-Q cavities, those rates are dominated by transmon decoherence.
Note that the above relations for dephasing terms are derived for a nonsingular (e.g., white) noise spectrum near zero frequency. For $1/f$ type noise, a Gaussian decay occurs instead of an exponential one, necessitating appropriate modifications to the above relationship~\cite{Ithier2005}. 

Similar to the case of a closed system, it is also efficient to simulate the open system dynamics in a displaced frame described by the unitary $\hat{U}_\eta(t) = \Pi_{i=1,2}\hat{D}^\dag[\eta_i(t)]$. 
Compared with Eq.~\eqref{eq:ode}, $\eta_i(t)$ satisfies the modified differential equation with an additional decay term,
\begin{equation*}
    \partial_t \eta_i = i\chi_i \eta_i/2 - iK_i |\eta_i|^2\eta_i  -iK_{12}\eta_i|\eta_{\bar{i}}|^2 - i\Omega_i(t) - \kappa_i\eta_i/2,
\end{equation*}
and is subject to the initial condition $\eta_i(0) = 0$.
Thus, the transformed density matrix $\hat{\rho}'(t)=\hat{U}_\eta(t)\hat{\rho}(t) \hat{U}_\eta^\dag(t)$ obeys the following master equation
\begin{align*}
    \begin{split}
        \partial_t \hat{\rho}' &= -i [\hat{H}'[\eta_i(t), t], \hat{\rho}']  + \gamma \mathcal{D}[\hat{q}]\hat{\rho}'+ 2\gamma_\phi \mathcal{D}[\hat{q}^\dag \hat{q}]\hat{\rho}'  \\
        & + \sum_{i=1,2} \left\{ \kappa_i \mathcal{D}[\hat{a}_i] \hat{\rho}' + 2 \kappa_{\phi i} \mathcal{D}[(\hat{a}_i^\dag + \eta_i^*)(\hat{a}_i+\eta_i)]\hat{\rho}' \right\}.
    \end{split}
\end{align*}
Here, $\hat{H}'[\eta_i(t), t]$ follows the same functional form as in Eq.~\eqref{eq:displaced_frame}, but is now dependent on  $\eta_i(t)$. 

\section{Effects from sixth-order transmon nonlinearity}\label{app:six_order}
Here, we study how the sixth-order nonlinearity of the transmon affects the ECD protocol. Specifically, we consider the following transmon Hamiltonian:
\begin{equation}
    \hat{H}_\text{trans} = \omega_\text{q}\hat{q}^\dag\hat{q}
    + \frac{K_\text{q}}{2}\hat{q}^{\dag2}\hat{q}^2
    + \frac{G_\text{q}}{3}\hat{q}^{\dag3}\hat{q}^3,
\end{equation}
with the nonlinearities $K_\text{q}\approx -E_\text{C}$ and $G_\text{q}\approx \frac{E_\text{C}}{6}\sqrt{\frac{2E_\text{C}}{E_\text{J}}}$. 
In the typical transmon regime with $E_\text{C}\ll E_\text{J}$, we have $G_\text{q}\ll K_\text{q}$. 
When the transmon is dispersively coupled to a cavity, the sixth-order nonlinearity contributes in two ways: First, it adds minor corrections to the nonlinearities in Eq.~\eqref{eq:dispersive_relation}. 
Second, it generates high-order cavity nonlinear terms such as $G_i\hat{a}_i^{\dag3}\hat{a}^3_i/3$, with $G_i\approx(g_i/\Delta_i)^6 G_\text{q}$. 
When the cavity is strongly displaced in the ECD protocol, the above term might be non-negligible compared with the contribution from the cavity self-Kerr. 
This can be analyzed by generalizing Eq.~\eqref{eq:ode}, which determines the classical trajectory,
\begin{equation}
\begin{split}
    \partial_t \alpha_i =& i\chi_i \alpha_i/2 - iK_i |\alpha_i|^2\alpha_i - iG_i |\alpha_i|^4\alpha_i \\
    &- iK_{12}\alpha_i|\alpha_{\bar{i}}|^2 - i\Omega_i(t).
\end{split}
\end{equation}
Therefore, the effect from the sixth-order nonlinearity is negligible when the following condition is satisfied:
\begin{equation}
    |\alpha|\ll\sqrt{\frac{K_i}{G_i}} \approx \sqrt{6}\left(\frac{g_i}{\Delta_i}\right)^{-1}\left( \frac{E_\text{J}}{2E_\text{C}}\right)^{1/4}.
\end{equation}
Within the transmon regime ($E_\text{J}\gg E_\text{C}$), the above bound is much larger than the one imposed by the critical photon number~\cite{Eickbusch2021}, with a ratio of $6(2E_\text{J}/E_\text{C})^{1/4}$. 
This suggests that the contribution from sixth-order transmon nonlinearity is negligible as long as the average photon in the cavity is below the critical photon number. 

\section{Ancilla gate decomposition for efficient experimental calibration}\label{app:decomposition}
In the multimode cavity control protocol outlined in Sec.~\ref{sec:ecd_intro}, the ancilla operations involve two key elements: the echo pulse \(\hat{X}_\pi\) used in ECD gates, and the qubit rotations represented by \(\hat{R}_\varphi(\theta)\).
To reduce the experimental overhead for calibrating different pulses for various rotation parameters, we decompose the arbitrary qubit rotation in terms of qubit $\hat{X}_{\pi/2}$ gates and virtual $\hat{Z}$ rotations~\cite{McKay2017},
\begin{equation}
    \hat{R}_\varphi(\theta) = \hat{Z}_{\varphi-\pi/2} \hat{X}_{\pi/2} \hat{Z}_{\pi-\theta} \hat{X}_{\pi/2} \hat{Z}_{-\varphi- \pi/2}.
\end{equation}
Consequently, it is sufficient to optimize robust ancilla pulses only for $\hat{X}_{\pi/2}$ and $\hat{X}_{\pi}$ gates.

\section{Supplement information on frequency-robust control of a transmon}\label{app:robust_pulse}

\subsection{Optimized 20-ns pulse for $\hat{X}_{\pi/2}$ gate}
\begin{figure*}
    \centering
    \includegraphics[width=\textwidth]{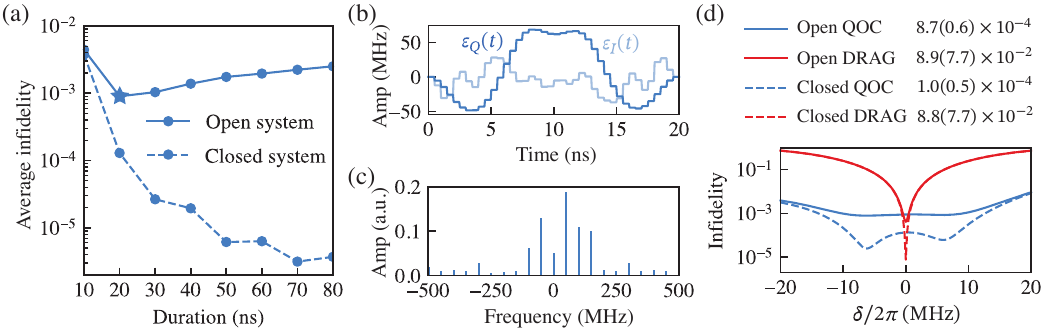}
    \caption{
    Optimization results for the $\hat{X}_{\pi/2}$ gate. 
(a) Average infidelities in Eq.~\eqref{eq:avg_inf} versus pulse durations for both closed (dashed line) and open (solid line) systems. The optimal duration selected from the range considered for this optimization is 20 ns (indicated by a star). 
(b) $I$ and $Q$ quadratures of the pulse envelopes for the selected 20-ns pulse from (a).
(c) The frequency components of the pulse envelopes, corresponding to the 20-ns pulse. 
(d) Comparative analysis of the 20-ns pulses, from both QOC (blue) and DRAG (red) pulses, in terms of their infidelities as a function of detuning.
The average and standard deviation of the infidelities are tabulated correspondingly by considering a uniform distribution of $\rho(\delta)$ for simplicity. The result shows that the optimized pulse is more robust than the commonly used DRAG pulse.
[Parameters: $\delta_i/2\pi\in\{\pm 1,\pm 3,\pm 4,\pm 7,\pm 10\}$ MHz, $K_\text{q}/2\pi=-200$ MHz, $T_1=50$ $\mu$s, $T_\phi=50$ $\mu$s.]\label{fig:fig_qoc_app} }
\end{figure*}

In Fig.~\ref{fig:fig_qoc}, we characterize the optimized 20-ns QOC pulse for the $\hat{X}_{\pi}$ gate. For completeness, we also provide the details of the pulse optimized for the $\hat{X}_{\pi/2}$ gate in Fig.~\ref{fig:fig_qoc_app}.

\subsection{Comparison of pulse robustness with recent literature}
\begin{figure*}
    \centering
    \includegraphics[width=\textwidth]{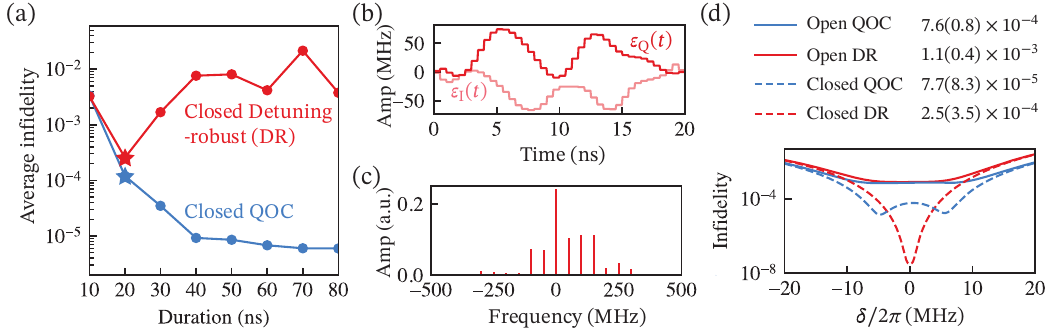}
    \caption{
Comparison of optimization results between Detuning-robust (DR) and QOC for the $\hat{X}{\pi}$ gate.
(a) Average infidelities of a closed system [Eq.~\eqref{eq:avg_inf}] versus pulse durations derived from DR (red) and QOC (blue). The optimal duration selected for DR from the range considered is 20 ns (indicated by a red star).
(b) $I$ and $Q$ quadratures of the DR pulse envelopes for the selected 20-ns pulse from (a).
(c) Frequency components of the DR pulse envelopes corresponding to the 20-ns pulse.
(d) Comparative analysis of the 20-ns pulses from both QOC (blue) and DR (red) in terms of their infidelities as a function of detuning.
The average and standard deviation of the infidelities are tabulated considering a uniform distribution of $\rho(\delta)$ for simplicity.
[Parameters: $\delta_i/2\pi \in {\pm 1, \pm 3, \pm 4, \pm 7, \pm 10}$ MHz, $K\text{q}/2\pi = -200$ MHz, $T_1 = 50$ $\mu$s, $T_\phi = 50$ $\mu$s.]\label{fig:fig_qoc_qctrl}
}
\end{figure*}

Here, we compare the robustness of the pulses derived from our optimization procedure with selected recent literature. First, we obtain robust pulses using the \texttt{Q-CTRL} package~\cite{ball2021softwarepulseengi,qctrl2022boulder}. 
To ensure a fair comparison with our optimization protocol, which aims to generate ancilla pulses robust to frequency changes caused by dispersive shifts from coupled resonators, we adopt the same error Hamiltonian as in Sec.~\ref{sec:qoc}.A.1. 
Specifically, the noise operator in \texttt{Q-CTRL} is implemented as a detuning term $\delta \hat{q}^\dag\hat{q}$ with $\delta/2\pi = 10$ MHz, matching the maximum value considered in our protocol.  
Additionally, we apply the same bandwidth constraints as in our optimization, such as a 500 MHz bandwidth for a 20 ns pulse. 
To differentiate these results from those produced by our optimization procedure, we use the label ``Detuning-robust" for the former and ``QOC" for the latter.
The averaged infidelities [Eq.~\eqref{eq:avg_inf}] of the closed system as a function of pulse duration are shown in Fig.~\ref{fig:fig_qoc_qctrl}(a). While the average infidelities for Detuning-robust (red) and QOC (blue) are similar for shorter durations, QOC demonstrates orders of magnitude lower infidelity for longer durations. 
To illustrate the robustness details of the Detuning-robust results, we examine the 20-ns case, which exhibits the lowest infidelity. 
The $I$ and $Q$ quadratures of the pulse envelopes and their corresponding Fourier components are shown in Fig.~\ref{fig:fig_qoc_qctrl}(b) and (c). 
Note that although the bandwidth constraint is the same for both Detuning-robust and QOC optimizations, the pulse optimized by Detuning-robust exhibits a smaller bandwidth compared to the one obtained from QOC.
The infidelity as a function of detuning is presented in Fig.~\ref{fig:fig_qoc_qctrl}(d).
For a closed system, the Detuning-robust results (red dashed line) show lower infidelity near resonance but exhibit higher infidelity off-resonance ($\delta/2\pi > 5$ MHz) compared to the QOC results (blue dashed line). 
Notably, in the presence of ancilla decoherence ($T_1 = 50$ $\mu$s, $T_\phi = 50$ $\mu$s), the infidelity at small detuning is dominated by incoherent error (solid lines). This results in QOC having an overall lower infidelity within the considered range of detuning for robust pulses.

\begin{figure}
    \centering
    \includegraphics[width=\columnwidth]{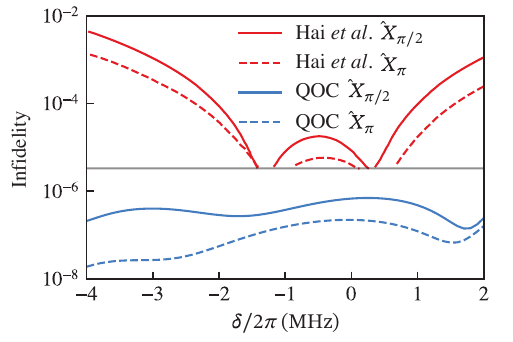}
    \caption{Infidelity of $\hat{X}_\pi$ and $\hat{X}_{\pi/2}$ as a function of detuning. The blue curves plot results obtained from the developed QOC pulses, while the red ones represent infidelities reported in Ref.~\onlinecite{hai2023universalcompare}. Note that the data below the gray line are not available in the reference. [Parameters: $K_\text{q}/2\pi=-260\,\text{MHz}$, durations for $\hat{X}_\pi$ and $\hat{X}_{\pi/2}$ gates of 70 ns and 80 ns, respectively.]}
    \label{fig:fig_compare} 
\end{figure}

Furthermore, we evaluate our pulse robustness against those reported in Ref.~\onlinecite{hai2023universalcompare}. For a fair comparison, we adopt the same parameters: transmon anharmonicity $K_\text{q}/2\pi = -260$ MHz, detuning range $\mathbb{D}/2\pi = [-4, 2]$ MHz, durations for $\hat{X}_\pi$ and $\hat{X}_{\pi/2}$ gates of 70 ns and 80 ns, respectively, and no decoherence error. The comparison is illustrated in Fig.~\ref{fig:fig_compare}, where the blue and red curves represent our results and those from Ref.~\onlinecite{hai2023universalcompare}, respectively. Within the chosen range of detuning, our results consistently exhibit significantly lower infidelity.

\section{Limits of the virtual phase correction in Eq.~\eqref{eq:first_order}}\label{app:virtual}

\begin{figure}
    \centering
    \includegraphics[width=\columnwidth]{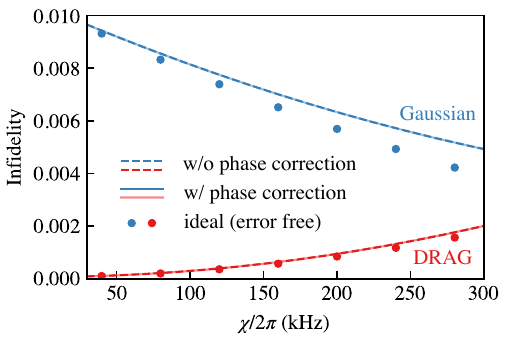}
    \caption{
    Infidelities of $\hat{X}_\pi$ gate acting on an ancilla dispersively coupled to a cavity mode. 
    Blue and red curves are realized with Gaussian and DRAG pulses, respectively. 
    The horizontal axis denotes the dispersive shift $\chi$.
    Solid and dashed lines show the infidelities achieved with and without the cavity phase corrections in Eq.~\eqref{eq:first_order}, respectively. 
    For reference, infidelities obtained without relative cavity phase are displayed using discrete markers. 
    [Parameters: $K_\text{q}/2\pi=-200\,\text{MHz}$, $d_\text{c}=10$, with other nonlinearities derived from Eq.~\eqref{eq:dispersive_relation}.]}
    \label{fig:fig_coupled_pulse_app}
\end{figure}

In Sec.~\ref{sec:robust_composite}, we demonstrate that the direct application of the robust pulse developed for an isolated ancilla to a composite cavity-ancilla system results in a small reduction in fidelity, which can be almost entirely recovered by applying the virtual phase correction outlined in Eq.~\eqref{eq:first_order}. It is important to emphasize that this correction is effective only under the condition that the detunings are within the range of robust control, i.e., $\mathcal{I}_\text{c}(\delta) \approx 0$ in Eq.~\eqref{eq:ancilla_fidelity}. 
To illustrate the limitations of this correction, we present a scenario where the above condition is not satisfied. Specifically, we calculate the infidelity of the $\hat{X}_\pi$ gate acting on a cavity-ancilla system, realized with both Gaussian and DRAG pulses. In Fig.~\ref{fig:fig_coupled_pulse_app}, the ideal cases with no relative phase associated with the cavity mode are shown as circular dots. In practice, when the relative phase is accounted for, the infidelities display deviations from the ideal cases, as shown by the dashed curves. Unlike the situation with optimized pulses shown in Fig.~\ref{fig:fig_coupled_pulse}, the infidelities here barely change when the phase corrections are applied (solid curves). 
This behavior is expected due to the weak robustness of the Gaussian and DRAG pulses, as indicated in Fig.~\ref{fig:fig_qoc}. Therefore, the virtual phase correction is only effective when the detuning is within the range of robustness and is primarily compatible with pulses obtained from optimal control.

\bibliography{main.bib}

\begin{thebibliography}{100}%
\makeatletter
\providecommand \@ifxundefined [1]{%
 \@ifx{#1\undefined}
}%
\providecommand \@ifnum [1]{%
 \ifnum #1\expandafter \@firstoftwo
 \else \expandafter \@secondoftwo
 \fi
}%
\providecommand \@ifx [1]{%
 \ifx #1\expandafter \@firstoftwo
 \else \expandafter \@secondoftwo
 \fi
}%
\providecommand \natexlab [1]{#1}%
\providecommand \enquote  [1]{``#1''}%
\providecommand \bibnamefont  [1]{#1}%
\providecommand \bibfnamefont [1]{#1}%
\providecommand \citenamefont [1]{#1}%
\providecommand \href@noop [0]{\@secondoftwo}%
\providecommand \href [0]{\begingroup \@sanitize@url \@href}%
\providecommand \@href[1]{\@@startlink{#1}\@@href}%
\providecommand \@@href[1]{\endgroup#1\@@endlink}%
\providecommand \@sanitize@url [0]{\catcode `\\12\catcode `\$12\catcode `\&12\catcode `\#12\catcode `\^12\catcode `\_12\catcode `\%12\relax}%
\providecommand \@@startlink[1]{}%
\providecommand \@@endlink[0]{}%
\providecommand \url  [0]{\begingroup\@sanitize@url \@url }%
\providecommand \@url [1]{\endgroup\@href {#1}{\urlprefix }}%
\providecommand \urlprefix  [0]{URL }%
\providecommand \Eprint [0]{\href }%
\providecommand \doibase [0]{https://doi.org/}%
\providecommand \selectlanguage [0]{\@gobble}%
\providecommand \bibinfo  [0]{\@secondoftwo}%
\providecommand \bibfield  [0]{\@secondoftwo}%
\providecommand \translation [1]{[#1]}%
\providecommand \BibitemOpen [0]{}%
\providecommand \bibitemStop [0]{}%
\providecommand \bibitemNoStop [0]{.\EOS\space}%
\providecommand \EOS [0]{\spacefactor3000\relax}%
\providecommand \BibitemShut  [1]{\csname bibitem#1\endcsname}%
\let\auto@bib@innerbib\@empty
\bibitem [{\citenamefont {Preskill}(2018)}]{NISQ}%
  \BibitemOpen
  \bibfield  {author} {\bibinfo {author} {\bibfnamefont {J.}~\bibnamefont {Preskill}},\ }\bibfield  {title} {\bibinfo {title} {Quantum {C}omputing in the {NISQ} era and beyond},\ }\href {https://doi.org/10.22331/q-2018-08-06-79} {\bibfield  {journal} {\bibinfo  {journal} {{Quantum}}\ }\textbf {\bibinfo {volume} {2}},\ \bibinfo {pages} {79} (\bibinfo {year} {2018})}\BibitemShut {NoStop}%
\bibitem [{\citenamefont {Shor}(1995)}]{QEC1}%
  \BibitemOpen
  \bibfield  {author} {\bibinfo {author} {\bibfnamefont {P.~W.}\ \bibnamefont {Shor}},\ }\bibfield  {title} {\bibinfo {title} {Scheme for reducing decoherence in quantum computer memory},\ }\href {https://doi.org/10.1103/PhysRevA.52.R2493} {\bibfield  {journal} {\bibinfo  {journal} {Phys. Rev. A}\ }\textbf {\bibinfo {volume} {52}},\ \bibinfo {pages} {R2493} (\bibinfo {year} {1995})}\BibitemShut {NoStop}%
\bibitem [{\citenamefont {Knill}\ and\ \citenamefont {Laflamme}(1997)}]{QEC2}%
  \BibitemOpen
  \bibfield  {author} {\bibinfo {author} {\bibfnamefont {E.}~\bibnamefont {Knill}}\ and\ \bibinfo {author} {\bibfnamefont {R.}~\bibnamefont {Laflamme}},\ }\bibfield  {title} {\bibinfo {title} {Theory of quantum error-correcting codes},\ }\href {https://doi.org/10.1103/PhysRevA.55.900} {\bibfield  {journal} {\bibinfo  {journal} {Phys. Rev. A}\ }\textbf {\bibinfo {volume} {55}},\ \bibinfo {pages} {900} (\bibinfo {year} {1997})}\BibitemShut {NoStop}%
\bibitem [{\citenamefont {Gambetta}(2023)}]{ibm}%
  \BibitemOpen
  \bibfield  {author} {\bibinfo {author} {\bibfnamefont {J.}~\bibnamefont {Gambetta}},\ }\href {https://research.ibm.com/blog/quantum-roadmap-2033} {\bibinfo {title} {{The hardware and software for the era of quantum utility is here.}}} (\bibinfo {year} {2023})\BibitemShut {NoStop}%
\bibitem [{\citenamefont {Rosenberg}\ \emph {et~al.}(2017)\citenamefont {Rosenberg}, \citenamefont {Kim}, \citenamefont {Das}, \citenamefont {Yost}, \citenamefont {Gustavsson}, \citenamefont {Hover}, \citenamefont {Krantz}, \citenamefont {Melville}, \citenamefont {Racz}, \citenamefont {Samach} \emph {et~al.}}]{Rosenberg2017}%
  \BibitemOpen
  \bibfield  {author} {\bibinfo {author} {\bibfnamefont {D.}~\bibnamefont {Rosenberg}}, \bibinfo {author} {\bibfnamefont {D.}~\bibnamefont {Kim}}, \bibinfo {author} {\bibfnamefont {R.}~\bibnamefont {Das}}, \bibinfo {author} {\bibfnamefont {D.}~\bibnamefont {Yost}}, \bibinfo {author} {\bibfnamefont {S.}~\bibnamefont {Gustavsson}}, \bibinfo {author} {\bibfnamefont {D.}~\bibnamefont {Hover}}, \bibinfo {author} {\bibfnamefont {P.}~\bibnamefont {Krantz}}, \bibinfo {author} {\bibfnamefont {A.}~\bibnamefont {Melville}}, \bibinfo {author} {\bibfnamefont {L.}~\bibnamefont {Racz}}, \bibinfo {author} {\bibfnamefont {G.~O.}\ \bibnamefont {Samach}}, \emph {et~al.},\ }\bibfield  {title} {\bibinfo {title} {{3D integrated superconducting qubits}},\ }\href {https://doi.org/10.1038/s41534-017-0044-0} {\bibfield  {journal} {\bibinfo  {journal} {npj Quantum Inf.}\ }\textbf {\bibinfo {volume} {3}},\ \bibinfo {pages} {42} (\bibinfo {year} {2017})}\BibitemShut {NoStop}%
\bibitem [{\citenamefont {McEwen}\ \emph {et~al.}(2022)\citenamefont {McEwen}, \citenamefont {Faoro}, \citenamefont {Arya}, \citenamefont {Dunsworth}, \citenamefont {Huang}, \citenamefont {Kim}, \citenamefont {Burkett}, \citenamefont {Fowler}, \citenamefont {Arute}, \citenamefont {Bardin} \emph {et~al.}}]{McEwen2022}%
  \BibitemOpen
  \bibfield  {author} {\bibinfo {author} {\bibfnamefont {M.}~\bibnamefont {McEwen}}, \bibinfo {author} {\bibfnamefont {L.}~\bibnamefont {Faoro}}, \bibinfo {author} {\bibfnamefont {K.}~\bibnamefont {Arya}}, \bibinfo {author} {\bibfnamefont {A.}~\bibnamefont {Dunsworth}}, \bibinfo {author} {\bibfnamefont {T.}~\bibnamefont {Huang}}, \bibinfo {author} {\bibfnamefont {S.}~\bibnamefont {Kim}}, \bibinfo {author} {\bibfnamefont {B.}~\bibnamefont {Burkett}}, \bibinfo {author} {\bibfnamefont {A.}~\bibnamefont {Fowler}}, \bibinfo {author} {\bibfnamefont {F.}~\bibnamefont {Arute}}, \bibinfo {author} {\bibfnamefont {J.~C.}\ \bibnamefont {Bardin}}, \emph {et~al.},\ }\bibfield  {title} {\bibinfo {title} {Resolving catastrophic error bursts from cosmic rays in large arrays of superconducting qubits},\ }\href@noop {} {\bibfield  {journal} {\bibinfo  {journal} {Nat. Phys.}\ }\textbf {\bibinfo {volume} {18}},\ \bibinfo {pages} {107} (\bibinfo {year} {2022})}\BibitemShut {NoStop}%
\bibitem [{\citenamefont {Romanenko}\ \emph {et~al.}(2020)\citenamefont {Romanenko}, \citenamefont {Pilipenko}, \citenamefont {Zorzetti}, \citenamefont {Frolov}, \citenamefont {Awida}, \citenamefont {Belomestnykh}, \citenamefont {Posen},\ and\ \citenamefont {Grassellino}}]{Romanenko2020}%
  \BibitemOpen
  \bibfield  {author} {\bibinfo {author} {\bibfnamefont {A.}~\bibnamefont {Romanenko}}, \bibinfo {author} {\bibfnamefont {R.}~\bibnamefont {Pilipenko}}, \bibinfo {author} {\bibfnamefont {S.}~\bibnamefont {Zorzetti}}, \bibinfo {author} {\bibfnamefont {D.}~\bibnamefont {Frolov}}, \bibinfo {author} {\bibfnamefont {M.}~\bibnamefont {Awida}}, \bibinfo {author} {\bibfnamefont {S.}~\bibnamefont {Belomestnykh}}, \bibinfo {author} {\bibfnamefont {S.}~\bibnamefont {Posen}},\ and\ \bibinfo {author} {\bibfnamefont {A.}~\bibnamefont {Grassellino}},\ }\bibfield  {title} {\bibinfo {title} {{Three-Dimensional Superconducting Resonators at T$<$20 mK with Photon Lifetimes up to $\tau$=2 s}},\ }\href {https://doi.org/10.1103/PhysRevApplied.13.034032} {\bibfield  {journal} {\bibinfo  {journal} {Phys. Rev. Appl.}\ }\textbf {\bibinfo {volume} {13}},\ \bibinfo {pages} {034032} (\bibinfo {year} {2020})}\BibitemShut {NoStop}%
\bibitem [{\citenamefont {Milul}\ \emph {et~al.}(2023)\citenamefont {Milul}, \citenamefont {Guttel}, \citenamefont {Goldblatt}, \citenamefont {Hazanov}, \citenamefont {Joshi}, \citenamefont {Chausovsky}, \citenamefont {Kahn}, \citenamefont {\ifmmode~\mbox{\c{C}}\else \c{C}\fi{}ifty\"urek}, \citenamefont {Lafont},\ and\ \citenamefont {Rosenblum}}]{Milul2023}%
  \BibitemOpen
  \bibfield  {author} {\bibinfo {author} {\bibfnamefont {O.}~\bibnamefont {Milul}}, \bibinfo {author} {\bibfnamefont {B.}~\bibnamefont {Guttel}}, \bibinfo {author} {\bibfnamefont {U.}~\bibnamefont {Goldblatt}}, \bibinfo {author} {\bibfnamefont {S.}~\bibnamefont {Hazanov}}, \bibinfo {author} {\bibfnamefont {L.~M.}\ \bibnamefont {Joshi}}, \bibinfo {author} {\bibfnamefont {D.}~\bibnamefont {Chausovsky}}, \bibinfo {author} {\bibfnamefont {N.}~\bibnamefont {Kahn}}, \bibinfo {author} {\bibfnamefont {E.}~\bibnamefont {\ifmmode~\mbox{\c{C}}\else \c{C}\fi{}ifty\"urek}}, \bibinfo {author} {\bibfnamefont {F.}~\bibnamefont {Lafont}},\ and\ \bibinfo {author} {\bibfnamefont {S.}~\bibnamefont {Rosenblum}},\ }\bibfield  {title} {\bibinfo {title} {Superconducting cavity qubit with tens of milliseconds single-photon coherence time},\ }\href {https://doi.org/10.1103/PRXQuantum.4.030336} {\bibfield  {journal} {\bibinfo  {journal} {PRX Quantum}\ }\textbf {\bibinfo {volume} {4}},\ \bibinfo {pages} {030336} (\bibinfo {year}
  {2023})}\BibitemShut {NoStop}%
\bibitem [{\citenamefont {Wang}\ \emph {et~al.}(2020)\citenamefont {Wang}, \citenamefont {Hu}, \citenamefont {Sanders},\ and\ \citenamefont {Kais}}]{Wang2020}%
  \BibitemOpen
  \bibfield  {author} {\bibinfo {author} {\bibfnamefont {Y.}~\bibnamefont {Wang}}, \bibinfo {author} {\bibfnamefont {Z.}~\bibnamefont {Hu}}, \bibinfo {author} {\bibfnamefont {B.~C.}\ \bibnamefont {Sanders}},\ and\ \bibinfo {author} {\bibfnamefont {S.}~\bibnamefont {Kais}},\ }\bibfield  {title} {\bibinfo {title} {{Qudits and High-Dimensional Quantum Computing}},\ }\href {https://doi.org/10.3389/fphy.2020.589504} {\bibfield  {journal} {\bibinfo  {journal} {Front. Phys.}\ }\textbf {\bibinfo {volume} {8}},\ \bibinfo {pages} {1} (\bibinfo {year} {2020})}\BibitemShut {NoStop}%
\bibitem [{\citenamefont {Wu}\ \emph {et~al.}(2020)\citenamefont {Wu}, \citenamefont {Tomarken}, \citenamefont {Petersson}, \citenamefont {Martinez}, \citenamefont {Rosen},\ and\ \citenamefont {DuBois}}]{Wu2020}%
  \BibitemOpen
  \bibfield  {author} {\bibinfo {author} {\bibfnamefont {X.}~\bibnamefont {Wu}}, \bibinfo {author} {\bibfnamefont {S.~L.}\ \bibnamefont {Tomarken}}, \bibinfo {author} {\bibfnamefont {N.~A.}\ \bibnamefont {Petersson}}, \bibinfo {author} {\bibfnamefont {L.~A.}\ \bibnamefont {Martinez}}, \bibinfo {author} {\bibfnamefont {Y.~J.}\ \bibnamefont {Rosen}},\ and\ \bibinfo {author} {\bibfnamefont {J.~L.}\ \bibnamefont {DuBois}},\ }\bibfield  {title} {\bibinfo {title} {{High-Fidelity Software-Defined Quantum Logic on a Superconducting Qudit}},\ }\href {https://doi.org/10.1103/PhysRevLett.125.170502} {\bibfield  {journal} {\bibinfo  {journal} {Phys. Rev. Lett.}\ }\textbf {\bibinfo {volume} {125}},\ \bibinfo {pages} {170502} (\bibinfo {year} {2020})}\BibitemShut {NoStop}%
\bibitem [{\citenamefont {Ringbauer}\ \emph {et~al.}(2022)\citenamefont {Ringbauer}, \citenamefont {Meth}, \citenamefont {Postler}, \citenamefont {Stricker}, \citenamefont {Blatt}, \citenamefont {Schindler},\ and\ \citenamefont {Monz}}]{Ringbauer2022}%
  \BibitemOpen
  \bibfield  {author} {\bibinfo {author} {\bibfnamefont {M.}~\bibnamefont {Ringbauer}}, \bibinfo {author} {\bibfnamefont {M.}~\bibnamefont {Meth}}, \bibinfo {author} {\bibfnamefont {L.}~\bibnamefont {Postler}}, \bibinfo {author} {\bibfnamefont {R.}~\bibnamefont {Stricker}}, \bibinfo {author} {\bibfnamefont {R.}~\bibnamefont {Blatt}}, \bibinfo {author} {\bibfnamefont {P.}~\bibnamefont {Schindler}},\ and\ \bibinfo {author} {\bibfnamefont {T.}~\bibnamefont {Monz}},\ }\bibfield  {title} {\bibinfo {title} {{A universal qudit quantum processor with trapped ions}},\ }\href {https://doi.org/10.1038/s41567-022-01658-0} {\bibfield  {journal} {\bibinfo  {journal} {Nat. Phys.}\ }\textbf {\bibinfo {volume} {18}},\ \bibinfo {pages} {1053} (\bibinfo {year} {2022})}\BibitemShut {NoStop}%
\bibitem [{\citenamefont {Chi}\ \emph {et~al.}(2022)\citenamefont {Chi}, \citenamefont {Huang}, \citenamefont {Zhang}, \citenamefont {Mao}, \citenamefont {Zhou}, \citenamefont {Chen}, \citenamefont {Zhai}, \citenamefont {Bao}, \citenamefont {Dai}, \citenamefont {Yuan} \emph {et~al.}}]{Chi2022}%
  \BibitemOpen
  \bibfield  {author} {\bibinfo {author} {\bibfnamefont {Y.}~\bibnamefont {Chi}}, \bibinfo {author} {\bibfnamefont {J.}~\bibnamefont {Huang}}, \bibinfo {author} {\bibfnamefont {Z.}~\bibnamefont {Zhang}}, \bibinfo {author} {\bibfnamefont {J.}~\bibnamefont {Mao}}, \bibinfo {author} {\bibfnamefont {Z.}~\bibnamefont {Zhou}}, \bibinfo {author} {\bibfnamefont {X.}~\bibnamefont {Chen}}, \bibinfo {author} {\bibfnamefont {C.}~\bibnamefont {Zhai}}, \bibinfo {author} {\bibfnamefont {J.}~\bibnamefont {Bao}}, \bibinfo {author} {\bibfnamefont {T.}~\bibnamefont {Dai}}, \bibinfo {author} {\bibfnamefont {H.}~\bibnamefont {Yuan}}, \emph {et~al.},\ }\bibfield  {title} {\bibinfo {title} {{A programmable qudit-based quantum processor}},\ }\href {https://doi.org/10.1038/s41467-022-28767-x} {\bibfield  {journal} {\bibinfo  {journal} {Nat. Commun.}\ }\textbf {\bibinfo {volume} {13}},\ \bibinfo {pages} {1166} (\bibinfo {year} {2022})}\BibitemShut {NoStop}%
\bibitem [{\citenamefont {MacDonell}\ \emph {et~al.}(2021)\citenamefont {MacDonell}, \citenamefont {Dickerson}, \citenamefont {Birch}, \citenamefont {Kumar}, \citenamefont {Edmunds}, \citenamefont {Biercuk}, \citenamefont {Hempel},\ and\ \citenamefont {Kassal}}]{MacDonell2021}%
  \BibitemOpen
  \bibfield  {author} {\bibinfo {author} {\bibfnamefont {R.~J.}\ \bibnamefont {MacDonell}}, \bibinfo {author} {\bibfnamefont {C.~E.}\ \bibnamefont {Dickerson}}, \bibinfo {author} {\bibfnamefont {C.~J.}\ \bibnamefont {Birch}}, \bibinfo {author} {\bibfnamefont {A.}~\bibnamefont {Kumar}}, \bibinfo {author} {\bibfnamefont {C.~L.}\ \bibnamefont {Edmunds}}, \bibinfo {author} {\bibfnamefont {M.~J.}\ \bibnamefont {Biercuk}}, \bibinfo {author} {\bibfnamefont {C.}~\bibnamefont {Hempel}},\ and\ \bibinfo {author} {\bibfnamefont {I.}~\bibnamefont {Kassal}},\ }\bibfield  {title} {\bibinfo {title} {{Analog quantum simulation of chemical dynamics}},\ }\href {https://doi.org/10.1039/d1sc02142g} {\bibfield  {journal} {\bibinfo  {journal} {Chem. Sci.}\ }\textbf {\bibinfo {volume} {12}},\ \bibinfo {pages} {9794} (\bibinfo {year} {2021})}\BibitemShut {NoStop}%
\bibitem [{\citenamefont {Rico}\ \emph {et~al.}(2018)\citenamefont {Rico}, \citenamefont {Dalmonte}, \citenamefont {Zoller}, \citenamefont {Banerjee}, \citenamefont {Bögli}, \citenamefont {Stebler},\ and\ \citenamefont {Wiese}}]{Rico2018}%
  \BibitemOpen
  \bibfield  {author} {\bibinfo {author} {\bibfnamefont {E.}~\bibnamefont {Rico}}, \bibinfo {author} {\bibfnamefont {M.}~\bibnamefont {Dalmonte}}, \bibinfo {author} {\bibfnamefont {P.}~\bibnamefont {Zoller}}, \bibinfo {author} {\bibfnamefont {D.}~\bibnamefont {Banerjee}}, \bibinfo {author} {\bibfnamefont {M.}~\bibnamefont {Bögli}}, \bibinfo {author} {\bibfnamefont {P.}~\bibnamefont {Stebler}},\ and\ \bibinfo {author} {\bibfnamefont {U.-J.}\ \bibnamefont {Wiese}},\ }\bibfield  {title} {\bibinfo {title} {{SO(3) “nuclear physics” with ultracold gases}},\ }\href {https://doi.org/https://doi.org/10.1016/j.aop.2018.03.020} {\bibfield  {journal} {\bibinfo  {journal} {Ann. Phys.}\ }\textbf {\bibinfo {volume} {393}},\ \bibinfo {pages} {466} (\bibinfo {year} {2018})}\BibitemShut {NoStop}%
\bibitem [{\citenamefont {Ma}\ \emph {et~al.}(2021)\citenamefont {Ma}, \citenamefont {Puri}, \citenamefont {Schoelkopf}, \citenamefont {Devoret}, \citenamefont {Girvin},\ and\ \citenamefont {Jiang}}]{QEC_review}%
  \BibitemOpen
  \bibfield  {author} {\bibinfo {author} {\bibfnamefont {W.-L.}\ \bibnamefont {Ma}}, \bibinfo {author} {\bibfnamefont {S.}~\bibnamefont {Puri}}, \bibinfo {author} {\bibfnamefont {R.~J.}\ \bibnamefont {Schoelkopf}}, \bibinfo {author} {\bibfnamefont {M.~H.}\ \bibnamefont {Devoret}}, \bibinfo {author} {\bibfnamefont {S.}~\bibnamefont {Girvin}},\ and\ \bibinfo {author} {\bibfnamefont {L.}~\bibnamefont {Jiang}},\ }\bibfield  {title} {\bibinfo {title} {Quantum control of bosonic modes with superconducting circuits},\ }\href {https://doi.org/https://doi.org/10.1016/j.scib.2021.05.024} {\bibfield  {journal} {\bibinfo  {journal} {Sci. Bull.}\ }\textbf {\bibinfo {volume} {66}},\ \bibinfo {pages} {1789} (\bibinfo {year} {2021})}\BibitemShut {NoStop}%
\bibitem [{\citenamefont {Leghtas}\ \emph {et~al.}(2013)\citenamefont {Leghtas}, \citenamefont {Kirchmair}, \citenamefont {Vlastakis}, \citenamefont {Schoelkopf}, \citenamefont {Devoret},\ and\ \citenamefont {Mirrahimi}}]{cat1}%
  \BibitemOpen
  \bibfield  {author} {\bibinfo {author} {\bibfnamefont {Z.}~\bibnamefont {Leghtas}}, \bibinfo {author} {\bibfnamefont {G.}~\bibnamefont {Kirchmair}}, \bibinfo {author} {\bibfnamefont {B.}~\bibnamefont {Vlastakis}}, \bibinfo {author} {\bibfnamefont {R.~J.}\ \bibnamefont {Schoelkopf}}, \bibinfo {author} {\bibfnamefont {M.~H.}\ \bibnamefont {Devoret}},\ and\ \bibinfo {author} {\bibfnamefont {M.}~\bibnamefont {Mirrahimi}},\ }\bibfield  {title} {\bibinfo {title} {Hardware-efficient autonomous quantum memory protection},\ }\href {https://doi.org/10.1103/PhysRevLett.111.120501} {\bibfield  {journal} {\bibinfo  {journal} {Phys. Rev. Lett.}\ }\textbf {\bibinfo {volume} {111}},\ \bibinfo {pages} {120501} (\bibinfo {year} {2013})}\BibitemShut {NoStop}%
\bibitem [{\citenamefont {Ofek}\ \emph {et~al.}(2016)\citenamefont {Ofek}, \citenamefont {Petrenko}, \citenamefont {Heeres}, \citenamefont {Reinhold}, \citenamefont {Leghtas}, \citenamefont {Vlastakis}, \citenamefont {Liu}, \citenamefont {Frunzio}, \citenamefont {Girvin}, \citenamefont {Jiang} \emph {et~al.}}]{cat2}%
  \BibitemOpen
  \bibfield  {author} {\bibinfo {author} {\bibfnamefont {N.}~\bibnamefont {Ofek}}, \bibinfo {author} {\bibfnamefont {A.}~\bibnamefont {Petrenko}}, \bibinfo {author} {\bibfnamefont {R.}~\bibnamefont {Heeres}}, \bibinfo {author} {\bibfnamefont {P.}~\bibnamefont {Reinhold}}, \bibinfo {author} {\bibfnamefont {Z.}~\bibnamefont {Leghtas}}, \bibinfo {author} {\bibfnamefont {B.}~\bibnamefont {Vlastakis}}, \bibinfo {author} {\bibfnamefont {Y.}~\bibnamefont {Liu}}, \bibinfo {author} {\bibfnamefont {L.}~\bibnamefont {Frunzio}}, \bibinfo {author} {\bibfnamefont {S.~M.}\ \bibnamefont {Girvin}}, \bibinfo {author} {\bibfnamefont {L.}~\bibnamefont {Jiang}}, \emph {et~al.},\ }\bibfield  {title} {\bibinfo {title} {{Extending the lifetime of a quantum bit with error correction in superconducting circuits}},\ }\href {https://doi.org/10.1038/nature18949} {\bibfield  {journal} {\bibinfo  {journal} {Nature}\ }\textbf {\bibinfo {volume} {536}},\ \bibinfo {pages} {441} (\bibinfo {year} {2016})}\BibitemShut {NoStop}%
\bibitem [{\citenamefont {Touzard}\ \emph {et~al.}(2018)\citenamefont {Touzard}, \citenamefont {Grimm}, \citenamefont {Leghtas}, \citenamefont {Mundhada}, \citenamefont {Reinhold}, \citenamefont {Axline}, \citenamefont {Reagor}, \citenamefont {Chou}, \citenamefont {Blumoff}, \citenamefont {Sliwa} \emph {et~al.}}]{Touzard2018}%
  \BibitemOpen
  \bibfield  {author} {\bibinfo {author} {\bibfnamefont {S.}~\bibnamefont {Touzard}}, \bibinfo {author} {\bibfnamefont {A.}~\bibnamefont {Grimm}}, \bibinfo {author} {\bibfnamefont {Z.}~\bibnamefont {Leghtas}}, \bibinfo {author} {\bibfnamefont {S.~O.}\ \bibnamefont {Mundhada}}, \bibinfo {author} {\bibfnamefont {P.}~\bibnamefont {Reinhold}}, \bibinfo {author} {\bibfnamefont {C.}~\bibnamefont {Axline}}, \bibinfo {author} {\bibfnamefont {M.}~\bibnamefont {Reagor}}, \bibinfo {author} {\bibfnamefont {K.}~\bibnamefont {Chou}}, \bibinfo {author} {\bibfnamefont {J.}~\bibnamefont {Blumoff}}, \bibinfo {author} {\bibfnamefont {K.~M.}\ \bibnamefont {Sliwa}}, \emph {et~al.},\ }\bibfield  {title} {\bibinfo {title} {Coherent oscillations inside a quantum manifold stabilized by dissipation},\ }\href {https://doi.org/10.1103/PhysRevX.8.021005} {\bibfield  {journal} {\bibinfo  {journal} {Phys. Rev. X}\ }\textbf {\bibinfo {volume} {8}},\ \bibinfo {pages} {021005} (\bibinfo {year} {2018})}\BibitemShut {NoStop}%
\bibitem [{\citenamefont {Michael}\ \emph {et~al.}(2016)\citenamefont {Michael}, \citenamefont {Silveri}, \citenamefont {Brierley}, \citenamefont {Albert}, \citenamefont {Salmilehto}, \citenamefont {Jiang},\ and\ \citenamefont {Girvin}}]{binomial}%
  \BibitemOpen
  \bibfield  {author} {\bibinfo {author} {\bibfnamefont {M.~H.}\ \bibnamefont {Michael}}, \bibinfo {author} {\bibfnamefont {M.}~\bibnamefont {Silveri}}, \bibinfo {author} {\bibfnamefont {R.~T.}\ \bibnamefont {Brierley}}, \bibinfo {author} {\bibfnamefont {V.~V.}\ \bibnamefont {Albert}}, \bibinfo {author} {\bibfnamefont {J.}~\bibnamefont {Salmilehto}}, \bibinfo {author} {\bibfnamefont {L.}~\bibnamefont {Jiang}},\ and\ \bibinfo {author} {\bibfnamefont {S.~M.}\ \bibnamefont {Girvin}},\ }\bibfield  {title} {\bibinfo {title} {New class of quantum error-correcting codes for a bosonic mode},\ }\href {https://doi.org/10.1103/PhysRevX.6.031006} {\bibfield  {journal} {\bibinfo  {journal} {Phys. Rev. X}\ }\textbf {\bibinfo {volume} {6}},\ \bibinfo {pages} {031006} (\bibinfo {year} {2016})}\BibitemShut {NoStop}%
\bibitem [{\citenamefont {Hu}\ \emph {et~al.}(2019)\citenamefont {Hu}, \citenamefont {Ma}, \citenamefont {Cai}, \citenamefont {Mu}, \citenamefont {Xu}, \citenamefont {Wang}, \citenamefont {Wu}, \citenamefont {Wang}, \citenamefont {Song}, \citenamefont {Zou} \emph {et~al.}}]{Hu2019}%
  \BibitemOpen
  \bibfield  {author} {\bibinfo {author} {\bibfnamefont {L.}~\bibnamefont {Hu}}, \bibinfo {author} {\bibfnamefont {Y.}~\bibnamefont {Ma}}, \bibinfo {author} {\bibfnamefont {W.}~\bibnamefont {Cai}}, \bibinfo {author} {\bibfnamefont {X.}~\bibnamefont {Mu}}, \bibinfo {author} {\bibfnamefont {Y.}~\bibnamefont {Xu}}, \bibinfo {author} {\bibfnamefont {W.}~\bibnamefont {Wang}}, \bibinfo {author} {\bibfnamefont {Y.}~\bibnamefont {Wu}}, \bibinfo {author} {\bibfnamefont {H.}~\bibnamefont {Wang}}, \bibinfo {author} {\bibfnamefont {Y.~P.}\ \bibnamefont {Song}}, \bibinfo {author} {\bibfnamefont {C.~L.}\ \bibnamefont {Zou}}, \emph {et~al.},\ }\bibfield  {title} {\bibinfo {title} {{Quantum error correction and universal gate set operation on a binomial bosonic logical qubit}},\ }\href {https://doi.org/10.1038/s41567-018-0414-3} {\bibfield  {journal} {\bibinfo  {journal} {Nat. Phys.}\ }\textbf {\bibinfo {volume} {15}},\ \bibinfo {pages} {503} (\bibinfo {year} {2019})}\BibitemShut {NoStop}%
\bibitem [{\citenamefont {Ni}\ \emph {et~al.}(2023)\citenamefont {Ni}, \citenamefont {Li}, \citenamefont {Deng}, \citenamefont {Cai}, \citenamefont {Zhang}, \citenamefont {Wang}, \citenamefont {Yang}, \citenamefont {Yu}, \citenamefont {Yan}, \citenamefont {Liu} \emph {et~al.}}]{Ni2023}%
  \BibitemOpen
  \bibfield  {author} {\bibinfo {author} {\bibfnamefont {Z.}~\bibnamefont {Ni}}, \bibinfo {author} {\bibfnamefont {S.}~\bibnamefont {Li}}, \bibinfo {author} {\bibfnamefont {X.}~\bibnamefont {Deng}}, \bibinfo {author} {\bibfnamefont {Y.}~\bibnamefont {Cai}}, \bibinfo {author} {\bibfnamefont {L.}~\bibnamefont {Zhang}}, \bibinfo {author} {\bibfnamefont {W.}~\bibnamefont {Wang}}, \bibinfo {author} {\bibfnamefont {Z.-B.}\ \bibnamefont {Yang}}, \bibinfo {author} {\bibfnamefont {H.}~\bibnamefont {Yu}}, \bibinfo {author} {\bibfnamefont {F.}~\bibnamefont {Yan}}, \bibinfo {author} {\bibfnamefont {S.}~\bibnamefont {Liu}}, \emph {et~al.},\ }\bibfield  {title} {\bibinfo {title} {{Beating the break-even point with a discrete-variable-encoded logical qubit}},\ }\href {https://doi.org/10.1038/s41586-023-05784-4} {\bibfield  {journal} {\bibinfo  {journal} {Nature}\ }\textbf {\bibinfo {volume} {616}},\ \bibinfo {pages} {56} (\bibinfo {year} {2023})}\BibitemShut {NoStop}%
\bibitem [{\citenamefont {Gottesman}\ \emph {et~al.}(2001)\citenamefont {Gottesman}, \citenamefont {Kitaev},\ and\ \citenamefont {Preskill}}]{GKP}%
  \BibitemOpen
  \bibfield  {author} {\bibinfo {author} {\bibfnamefont {D.}~\bibnamefont {Gottesman}}, \bibinfo {author} {\bibfnamefont {A.}~\bibnamefont {Kitaev}},\ and\ \bibinfo {author} {\bibfnamefont {J.}~\bibnamefont {Preskill}},\ }\bibfield  {title} {\bibinfo {title} {Encoding a qubit in an oscillator},\ }\href {https://doi.org/10.1103/PhysRevA.64.012310} {\bibfield  {journal} {\bibinfo  {journal} {Phys. Rev. A}\ }\textbf {\bibinfo {volume} {64}},\ \bibinfo {pages} {012310} (\bibinfo {year} {2001})}\BibitemShut {NoStop}%
\bibitem [{\citenamefont {Campagne-Ibarcq}\ \emph {et~al.}(2020)\citenamefont {Campagne-Ibarcq}, \citenamefont {Eickbusch}, \citenamefont {Touzard}, \citenamefont {Zalys-Geller}, \citenamefont {Frattini}, \citenamefont {Sivak}, \citenamefont {Reinhold}, \citenamefont {Puri}, \citenamefont {Shankar}, \citenamefont {Schoelkopf} \emph {et~al.}}]{Campagne-Ibarcq2019}%
  \BibitemOpen
  \bibfield  {author} {\bibinfo {author} {\bibfnamefont {P.}~\bibnamefont {Campagne-Ibarcq}}, \bibinfo {author} {\bibfnamefont {A.}~\bibnamefont {Eickbusch}}, \bibinfo {author} {\bibfnamefont {S.}~\bibnamefont {Touzard}}, \bibinfo {author} {\bibfnamefont {E.}~\bibnamefont {Zalys-Geller}}, \bibinfo {author} {\bibfnamefont {N.~E.}\ \bibnamefont {Frattini}}, \bibinfo {author} {\bibfnamefont {V.~V.}\ \bibnamefont {Sivak}}, \bibinfo {author} {\bibfnamefont {P.}~\bibnamefont {Reinhold}}, \bibinfo {author} {\bibfnamefont {S.}~\bibnamefont {Puri}}, \bibinfo {author} {\bibfnamefont {S.}~\bibnamefont {Shankar}}, \bibinfo {author} {\bibfnamefont {R.~J.}\ \bibnamefont {Schoelkopf}}, \emph {et~al.},\ }\bibfield  {title} {\bibinfo {title} {{Quantum error correction of a qubit encoded in grid states of an oscillator}},\ }\href {https://doi.org/10.1038/s41586-020-2603-3} {\bibfield  {journal} {\bibinfo  {journal} {Nature}\ }\textbf {\bibinfo {volume} {584}},\ \bibinfo {pages} {368} (\bibinfo {year} {2020})}\BibitemShut
  {NoStop}%
\bibitem [{\citenamefont {Grimsmo}\ and\ \citenamefont {Puri}(2021)}]{gkp_review}%
  \BibitemOpen
  \bibfield  {author} {\bibinfo {author} {\bibfnamefont {A.~L.}\ \bibnamefont {Grimsmo}}\ and\ \bibinfo {author} {\bibfnamefont {S.}~\bibnamefont {Puri}},\ }\bibfield  {title} {\bibinfo {title} {Quantum error correction with the gottesman-kitaev-preskill code},\ }\href {https://doi.org/10.1103/PRXQuantum.2.020101} {\bibfield  {journal} {\bibinfo  {journal} {PRX Quantum}\ }\textbf {\bibinfo {volume} {2}},\ \bibinfo {pages} {020101} (\bibinfo {year} {2021})}\BibitemShut {NoStop}%
\bibitem [{\citenamefont {Sivak}\ \emph {et~al.}(2023)\citenamefont {Sivak}, \citenamefont {Eickbusch}, \citenamefont {Royer}, \citenamefont {Singh}, \citenamefont {Tsioutsios}, \citenamefont {Ganjam}, \citenamefont {Miano}, \citenamefont {Brock}, \citenamefont {Ding}, \citenamefont {Frunzio} \emph {et~al.}}]{Sivak2022}%
  \BibitemOpen
  \bibfield  {author} {\bibinfo {author} {\bibfnamefont {V.~V.}\ \bibnamefont {Sivak}}, \bibinfo {author} {\bibfnamefont {A.}~\bibnamefont {Eickbusch}}, \bibinfo {author} {\bibfnamefont {B.}~\bibnamefont {Royer}}, \bibinfo {author} {\bibfnamefont {S.}~\bibnamefont {Singh}}, \bibinfo {author} {\bibfnamefont {I.}~\bibnamefont {Tsioutsios}}, \bibinfo {author} {\bibfnamefont {S.}~\bibnamefont {Ganjam}}, \bibinfo {author} {\bibfnamefont {A.}~\bibnamefont {Miano}}, \bibinfo {author} {\bibfnamefont {B.~L.}\ \bibnamefont {Brock}}, \bibinfo {author} {\bibfnamefont {A.~Z.}\ \bibnamefont {Ding}}, \bibinfo {author} {\bibfnamefont {L.}~\bibnamefont {Frunzio}}, \emph {et~al.},\ }\bibfield  {title} {\bibinfo {title} {{Real-time quantum error correction beyond break-even}},\ }\href {https://doi.org/10.1038/s41586-023-05782-6} {\bibfield  {journal} {\bibinfo  {journal} {Nature}\ }\textbf {\bibinfo {volume} {616}},\ \bibinfo {pages} {50} (\bibinfo {year} {2023})}\BibitemShut {NoStop}%
\bibitem [{\citenamefont {Naik}\ \emph {et~al.}(2017)\citenamefont {Naik}, \citenamefont {Leung}, \citenamefont {Chakram}, \citenamefont {Groszkowski}, \citenamefont {Lu}, \citenamefont {Earnest}, \citenamefont {McKay}, \citenamefont {Koch},\ and\ \citenamefont {Schuster}}]{Naik2017a}%
  \BibitemOpen
  \bibfield  {author} {\bibinfo {author} {\bibfnamefont {R.~K.}\ \bibnamefont {Naik}}, \bibinfo {author} {\bibfnamefont {N.}~\bibnamefont {Leung}}, \bibinfo {author} {\bibfnamefont {S.}~\bibnamefont {Chakram}}, \bibinfo {author} {\bibfnamefont {P.}~\bibnamefont {Groszkowski}}, \bibinfo {author} {\bibfnamefont {Y.}~\bibnamefont {Lu}}, \bibinfo {author} {\bibfnamefont {N.}~\bibnamefont {Earnest}}, \bibinfo {author} {\bibfnamefont {D.~C.}\ \bibnamefont {McKay}}, \bibinfo {author} {\bibfnamefont {J.}~\bibnamefont {Koch}},\ and\ \bibinfo {author} {\bibfnamefont {D.~I.}\ \bibnamefont {Schuster}},\ }\bibfield  {title} {\bibinfo {title} {{Random access quantum information processors using multimode circuit quantum electrodynamics}},\ }\href {https://doi.org/10.1038/s41467-017-02046-6} {\bibfield  {journal} {\bibinfo  {journal} {Nat. Commun.}\ }\textbf {\bibinfo {volume} {8}},\ \bibinfo {pages} {1904} (\bibinfo {year} {2017})}\BibitemShut {NoStop}%
\bibitem [{\citenamefont {Chakram}\ \emph {et~al.}(2021)\citenamefont {Chakram}, \citenamefont {Oriani}, \citenamefont {Naik}, \citenamefont {Dixit}, \citenamefont {He}, \citenamefont {Agrawal}, \citenamefont {Kwon},\ and\ \citenamefont {Schuster}}]{Chakram2020}%
  \BibitemOpen
  \bibfield  {author} {\bibinfo {author} {\bibfnamefont {S.}~\bibnamefont {Chakram}}, \bibinfo {author} {\bibfnamefont {A.~E.}\ \bibnamefont {Oriani}}, \bibinfo {author} {\bibfnamefont {R.~K.}\ \bibnamefont {Naik}}, \bibinfo {author} {\bibfnamefont {A.~V.}\ \bibnamefont {Dixit}}, \bibinfo {author} {\bibfnamefont {K.}~\bibnamefont {He}}, \bibinfo {author} {\bibfnamefont {A.}~\bibnamefont {Agrawal}}, \bibinfo {author} {\bibfnamefont {H.}~\bibnamefont {Kwon}},\ and\ \bibinfo {author} {\bibfnamefont {D.~I.}\ \bibnamefont {Schuster}},\ }\bibfield  {title} {\bibinfo {title} {{Seamless High-Q Microwave Cavities for Multimode Circuit Quantum Electrodynamics}},\ }\href {https://doi.org/10.1103/PhysRevLett.127.107701} {\bibfield  {journal} {\bibinfo  {journal} {Phys. Rev. Lett.}\ }\textbf {\bibinfo {volume} {127}},\ \bibinfo {pages} {107701} (\bibinfo {year} {2021})}\BibitemShut {NoStop}%
\bibitem [{\citenamefont {Chakram}\ \emph {et~al.}(2022)\citenamefont {Chakram}, \citenamefont {He}, \citenamefont {Dixit}, \citenamefont {Oriani}, \citenamefont {Naik}, \citenamefont {Leung}, \citenamefont {Kwon}, \citenamefont {Ma}, \citenamefont {Jiang},\ and\ \citenamefont {Schuster}}]{Chakram2022}%
  \BibitemOpen
  \bibfield  {author} {\bibinfo {author} {\bibfnamefont {S.}~\bibnamefont {Chakram}}, \bibinfo {author} {\bibfnamefont {K.}~\bibnamefont {He}}, \bibinfo {author} {\bibfnamefont {A.~V.}\ \bibnamefont {Dixit}}, \bibinfo {author} {\bibfnamefont {A.~E.}\ \bibnamefont {Oriani}}, \bibinfo {author} {\bibfnamefont {R.~K.}\ \bibnamefont {Naik}}, \bibinfo {author} {\bibfnamefont {N.}~\bibnamefont {Leung}}, \bibinfo {author} {\bibfnamefont {H.}~\bibnamefont {Kwon}}, \bibinfo {author} {\bibfnamefont {W.-L.}\ \bibnamefont {Ma}}, \bibinfo {author} {\bibfnamefont {L.}~\bibnamefont {Jiang}},\ and\ \bibinfo {author} {\bibfnamefont {D.~I.}\ \bibnamefont {Schuster}},\ }\bibfield  {title} {\bibinfo {title} {{Multimode photon blockade}},\ }\href {https://doi.org/10.1038/s41567-022-01630-y} {\bibfield  {journal} {\bibinfo  {journal} {Nat. Phys.}\ }\textbf {\bibinfo {volume} {18}},\ \bibinfo {pages} {879} (\bibinfo {year} {2022})}\BibitemShut {NoStop}%
\bibitem [{\citenamefont {Reineri}\ \emph {et~al.}(2023)\citenamefont {Reineri}, \citenamefont {Zorzetti}, \citenamefont {Roy},\ and\ \citenamefont {You}}]{Alessandro}%
  \BibitemOpen
  \bibfield  {author} {\bibinfo {author} {\bibfnamefont {A.}~\bibnamefont {Reineri}}, \bibinfo {author} {\bibfnamefont {S.}~\bibnamefont {Zorzetti}}, \bibinfo {author} {\bibfnamefont {T.}~\bibnamefont {Roy}},\ and\ \bibinfo {author} {\bibfnamefont {X.}~\bibnamefont {You}},\ }\bibfield  {title} {\bibinfo {title} {Exploration of superconducting multi-mode cavity architectures for quantum computing},\ }in\ \href {https://doi.org/10.1109/QCE57702.2023.00152} {\emph {\bibinfo {booktitle} {2023 IEEE International Conference on Quantum Computing and Engineering (QCE)}}},\ Vol.~\bibinfo {volume} {01}\ (\bibinfo {year} {2023})\ pp.\ \bibinfo {pages} {1342--1348}\BibitemShut {NoStop}%
\bibitem [{\citenamefont {Blais}\ \emph {et~al.}(2021)\citenamefont {Blais}, \citenamefont {Grimsmo}, \citenamefont {Girvin},\ and\ \citenamefont {Wallraff}}]{blais2021}%
  \BibitemOpen
  \bibfield  {author} {\bibinfo {author} {\bibfnamefont {A.}~\bibnamefont {Blais}}, \bibinfo {author} {\bibfnamefont {A.~L.}\ \bibnamefont {Grimsmo}}, \bibinfo {author} {\bibfnamefont {S.~M.}\ \bibnamefont {Girvin}},\ and\ \bibinfo {author} {\bibfnamefont {A.}~\bibnamefont {Wallraff}},\ }\bibfield  {title} {\bibinfo {title} {{Circuit quantum electrodynamics}},\ }\href {https://doi.org/10.1103/RevModPhys.93.025005} {\bibfield  {journal} {\bibinfo  {journal} {Rev. Mod. Phys.}\ }\textbf {\bibinfo {volume} {93}},\ \bibinfo {pages} {025005} (\bibinfo {year} {2021})}\BibitemShut {NoStop}%
\bibitem [{Note1()}]{Note1}%
  \BibitemOpen
  \bibinfo {note} {It's crucial to differentiate this form of crosstalk from that caused by the cross-Kerr effect, which is generally negligible in comparison to the Stark shift-induced crosstalk.}\BibitemShut {Stop}%
\bibitem [{\citenamefont {Heeres}\ \emph {et~al.}(2015)\citenamefont {Heeres}, \citenamefont {Vlastakis}, \citenamefont {Holland}, \citenamefont {Krastanov}, \citenamefont {Albert}, \citenamefont {Frunzio}, \citenamefont {Jiang},\ and\ \citenamefont {Schoelkopf}}]{Heeres2015b}%
  \BibitemOpen
  \bibfield  {author} {\bibinfo {author} {\bibfnamefont {R.~W.}\ \bibnamefont {Heeres}}, \bibinfo {author} {\bibfnamefont {B.}~\bibnamefont {Vlastakis}}, \bibinfo {author} {\bibfnamefont {E.}~\bibnamefont {Holland}}, \bibinfo {author} {\bibfnamefont {S.}~\bibnamefont {Krastanov}}, \bibinfo {author} {\bibfnamefont {V.~V.}\ \bibnamefont {Albert}}, \bibinfo {author} {\bibfnamefont {L.}~\bibnamefont {Frunzio}}, \bibinfo {author} {\bibfnamefont {L.}~\bibnamefont {Jiang}},\ and\ \bibinfo {author} {\bibfnamefont {R.~J.}\ \bibnamefont {Schoelkopf}},\ }\bibfield  {title} {\bibinfo {title} {{Cavity State Manipulation Using Photon-Number Selective Phase Gates}},\ }\href {https://doi.org/10.1103/PhysRevLett.115.137002} {\bibfield  {journal} {\bibinfo  {journal} {Phys. Rev. Lett.}\ }\textbf {\bibinfo {volume} {115}},\ \bibinfo {pages} {137002} (\bibinfo {year} {2015})}\BibitemShut {NoStop}%
\bibitem [{\citenamefont {Krastanov}\ \emph {et~al.}(2015)\citenamefont {Krastanov}, \citenamefont {Albert}, \citenamefont {Shen}, \citenamefont {Zou}, \citenamefont {Heeres}, \citenamefont {Vlastakis}, \citenamefont {Schoelkopf},\ and\ \citenamefont {Jiang}}]{Krastanov2015b}%
  \BibitemOpen
  \bibfield  {author} {\bibinfo {author} {\bibfnamefont {S.}~\bibnamefont {Krastanov}}, \bibinfo {author} {\bibfnamefont {V.~V.}\ \bibnamefont {Albert}}, \bibinfo {author} {\bibfnamefont {C.}~\bibnamefont {Shen}}, \bibinfo {author} {\bibfnamefont {C.-L.}\ \bibnamefont {Zou}}, \bibinfo {author} {\bibfnamefont {R.~W.}\ \bibnamefont {Heeres}}, \bibinfo {author} {\bibfnamefont {B.}~\bibnamefont {Vlastakis}}, \bibinfo {author} {\bibfnamefont {R.~J.}\ \bibnamefont {Schoelkopf}},\ and\ \bibinfo {author} {\bibfnamefont {L.}~\bibnamefont {Jiang}},\ }\bibfield  {title} {\bibinfo {title} {{Universal control of an oscillator with dispersive coupling to a qubit}},\ }\href {https://doi.org/10.1103/PhysRevA.92.040303} {\bibfield  {journal} {\bibinfo  {journal} {Phys. Rev. A}\ }\textbf {\bibinfo {volume} {92}},\ \bibinfo {pages} {040303(R)} (\bibinfo {year} {2015})}\BibitemShut {NoStop}%
\bibitem [{\citenamefont {Wang}\ \emph {et~al.}(2021)\citenamefont {Wang}, \citenamefont {Noh}, \citenamefont {Lebreuilly}, \citenamefont {Girvin},\ and\ \citenamefont {Jiang}}]{Wang2021}%
  \BibitemOpen
  \bibfield  {author} {\bibinfo {author} {\bibfnamefont {C.-H.}\ \bibnamefont {Wang}}, \bibinfo {author} {\bibfnamefont {K.}~\bibnamefont {Noh}}, \bibinfo {author} {\bibfnamefont {J.}~\bibnamefont {Lebreuilly}}, \bibinfo {author} {\bibfnamefont {S.~M.}\ \bibnamefont {Girvin}},\ and\ \bibinfo {author} {\bibfnamefont {L.}~\bibnamefont {Jiang}},\ }\bibfield  {title} {\bibinfo {title} {{Photon-Number-Dependent Hamiltonian Engineering for Cavities}},\ }\href {https://doi.org/10.1103/PhysRevApplied.15.044026} {\bibfield  {journal} {\bibinfo  {journal} {Phys. Rev. Appl.}\ }\textbf {\bibinfo {volume} {15}},\ \bibinfo {pages} {044026} (\bibinfo {year} {2021})}\BibitemShut {NoStop}%
\bibitem [{\citenamefont {Kudra}\ \emph {et~al.}(2022)\citenamefont {Kudra}, \citenamefont {Kervinen}, \citenamefont {Strandberg}, \citenamefont {Ahmed}, \citenamefont {Scigliuzzo}, \citenamefont {Osman}, \citenamefont {Lozano}, \citenamefont {Thol\'en}, \citenamefont {Borgani}, \citenamefont {Haviland} \emph {et~al.}}]{Kudra2022}%
  \BibitemOpen
  \bibfield  {author} {\bibinfo {author} {\bibfnamefont {M.}~\bibnamefont {Kudra}}, \bibinfo {author} {\bibfnamefont {M.}~\bibnamefont {Kervinen}}, \bibinfo {author} {\bibfnamefont {I.}~\bibnamefont {Strandberg}}, \bibinfo {author} {\bibfnamefont {S.}~\bibnamefont {Ahmed}}, \bibinfo {author} {\bibfnamefont {M.}~\bibnamefont {Scigliuzzo}}, \bibinfo {author} {\bibfnamefont {A.}~\bibnamefont {Osman}}, \bibinfo {author} {\bibfnamefont {D.~P.}\ \bibnamefont {Lozano}}, \bibinfo {author} {\bibfnamefont {M.~O.}\ \bibnamefont {Thol\'en}}, \bibinfo {author} {\bibfnamefont {R.}~\bibnamefont {Borgani}}, \bibinfo {author} {\bibfnamefont {D.~B.}\ \bibnamefont {Haviland}}, \emph {et~al.},\ }\bibfield  {title} {\bibinfo {title} {Robust preparation of wigner-negative states with optimized snap-displacement sequences},\ }\href {https://doi.org/10.1103/PRXQuantum.3.030301} {\bibfield  {journal} {\bibinfo  {journal} {PRX Quantum}\ }\textbf {\bibinfo {volume} {3}},\ \bibinfo {pages} {030301} (\bibinfo {year} {2022})}\BibitemShut
  {NoStop}%
\bibitem [{\citenamefont {Eickbusch}\ \emph {et~al.}(2022)\citenamefont {Eickbusch}, \citenamefont {Sivak}, \citenamefont {Ding}, \citenamefont {Elder}, \citenamefont {Jha}, \citenamefont {Venkatraman}, \citenamefont {Royer}, \citenamefont {Girvin}, \citenamefont {Schoelkopf},\ and\ \citenamefont {Devoret}}]{Eickbusch2021}%
  \BibitemOpen
  \bibfield  {author} {\bibinfo {author} {\bibfnamefont {A.}~\bibnamefont {Eickbusch}}, \bibinfo {author} {\bibfnamefont {V.}~\bibnamefont {Sivak}}, \bibinfo {author} {\bibfnamefont {A.~Z.}\ \bibnamefont {Ding}}, \bibinfo {author} {\bibfnamefont {S.~S.}\ \bibnamefont {Elder}}, \bibinfo {author} {\bibfnamefont {S.~R.}\ \bibnamefont {Jha}}, \bibinfo {author} {\bibfnamefont {J.}~\bibnamefont {Venkatraman}}, \bibinfo {author} {\bibfnamefont {B.}~\bibnamefont {Royer}}, \bibinfo {author} {\bibfnamefont {S.~M.}\ \bibnamefont {Girvin}}, \bibinfo {author} {\bibfnamefont {R.~J.}\ \bibnamefont {Schoelkopf}},\ and\ \bibinfo {author} {\bibfnamefont {M.~H.}\ \bibnamefont {Devoret}},\ }\bibfield  {title} {\bibinfo {title} {{Fast universal control of an oscillator with weak dispersive coupling to a qubit}},\ }\href {https://doi.org/10.1038/s41567-022-01776-9} {\bibfield  {journal} {\bibinfo  {journal} {Nat. Phys.}\ }\textbf {\bibinfo {volume} {18}},\ \bibinfo {pages} {1464} (\bibinfo {year} {2022})}\BibitemShut {NoStop}%
\bibitem [{\citenamefont {Diringer}\ \emph {et~al.}(2024)\citenamefont {Diringer}, \citenamefont {Blumenthal}, \citenamefont {Grinberg}, \citenamefont {Jiang},\ and\ \citenamefont {Hacohen-Gourgy}}]{Diringer2023}%
  \BibitemOpen
  \bibfield  {author} {\bibinfo {author} {\bibfnamefont {A.~A.}\ \bibnamefont {Diringer}}, \bibinfo {author} {\bibfnamefont {E.}~\bibnamefont {Blumenthal}}, \bibinfo {author} {\bibfnamefont {A.}~\bibnamefont {Grinberg}}, \bibinfo {author} {\bibfnamefont {L.}~\bibnamefont {Jiang}},\ and\ \bibinfo {author} {\bibfnamefont {S.}~\bibnamefont {Hacohen-Gourgy}},\ }\bibfield  {title} {\bibinfo {title} {Conditional-not displacement: Fast multioscillator control with a single qubit},\ }\href {https://doi.org/10.1103/PhysRevX.14.011055} {\bibfield  {journal} {\bibinfo  {journal} {Phys. Rev. X}\ }\textbf {\bibinfo {volume} {14}},\ \bibinfo {pages} {011055} (\bibinfo {year} {2024})}\BibitemShut {NoStop}%
\bibitem [{\citenamefont {Kelly}\ \emph {et~al.}(2014)\citenamefont {Kelly}, \citenamefont {Barends}, \citenamefont {Campbell}, \citenamefont {Chen}, \citenamefont {Chen}, \citenamefont {Chiaro}, \citenamefont {Dunsworth}, \citenamefont {Fowler}, \citenamefont {Hoi}, \citenamefont {Jeffrey} \emph {et~al.}}]{Kelly2014}%
  \BibitemOpen
  \bibfield  {author} {\bibinfo {author} {\bibfnamefont {J.}~\bibnamefont {Kelly}}, \bibinfo {author} {\bibfnamefont {R.}~\bibnamefont {Barends}}, \bibinfo {author} {\bibfnamefont {B.}~\bibnamefont {Campbell}}, \bibinfo {author} {\bibfnamefont {Y.}~\bibnamefont {Chen}}, \bibinfo {author} {\bibfnamefont {Z.}~\bibnamefont {Chen}}, \bibinfo {author} {\bibfnamefont {B.}~\bibnamefont {Chiaro}}, \bibinfo {author} {\bibfnamefont {A.}~\bibnamefont {Dunsworth}}, \bibinfo {author} {\bibfnamefont {A.~G.}\ \bibnamefont {Fowler}}, \bibinfo {author} {\bibfnamefont {I.-C.}\ \bibnamefont {Hoi}}, \bibinfo {author} {\bibfnamefont {E.}~\bibnamefont {Jeffrey}}, \emph {et~al.},\ }\bibfield  {title} {\bibinfo {title} {{Optimal Quantum Control Using Randomized Benchmarking}},\ }\href {https://doi.org/10.1103/PhysRevLett.112.240504} {\bibfield  {journal} {\bibinfo  {journal} {Phys. Rev. Lett.}\ }\textbf {\bibinfo {volume} {112}},\ \bibinfo {pages} {240504} (\bibinfo {year} {2014})}\BibitemShut {NoStop}%
\bibitem [{\citenamefont {Machnes}\ \emph {et~al.}(2018)\citenamefont {Machnes}, \citenamefont {Ass{\'{e}}mat}, \citenamefont {Tannor},\ and\ \citenamefont {Wilhelm}}]{Machnes2018}%
  \BibitemOpen
  \bibfield  {author} {\bibinfo {author} {\bibfnamefont {S.}~\bibnamefont {Machnes}}, \bibinfo {author} {\bibfnamefont {E.}~\bibnamefont {Ass{\'{e}}mat}}, \bibinfo {author} {\bibfnamefont {D.}~\bibnamefont {Tannor}},\ and\ \bibinfo {author} {\bibfnamefont {F.~K.}\ \bibnamefont {Wilhelm}},\ }\bibfield  {title} {\bibinfo {title} {{Tunable, Flexible, and Efficient Optimization of Control Pulses for Practical Qubits}},\ }\href {https://doi.org/10.1103/PhysRevLett.120.150401} {\bibfield  {journal} {\bibinfo  {journal} {Phys. Rev. Lett.}\ }\textbf {\bibinfo {volume} {120}},\ \bibinfo {pages} {150401} (\bibinfo {year} {2018})}\BibitemShut {NoStop}%
\bibitem [{\citenamefont {Werninghaus}\ \emph {et~al.}(2021)\citenamefont {Werninghaus}, \citenamefont {Egger}, \citenamefont {Roy}, \citenamefont {Machnes}, \citenamefont {Wilhelm},\ and\ \citenamefont {Filipp}}]{Werninghaus2021}%
  \BibitemOpen
  \bibfield  {author} {\bibinfo {author} {\bibfnamefont {M.}~\bibnamefont {Werninghaus}}, \bibinfo {author} {\bibfnamefont {D.~J.}\ \bibnamefont {Egger}}, \bibinfo {author} {\bibfnamefont {F.}~\bibnamefont {Roy}}, \bibinfo {author} {\bibfnamefont {S.}~\bibnamefont {Machnes}}, \bibinfo {author} {\bibfnamefont {F.~K.}\ \bibnamefont {Wilhelm}},\ and\ \bibinfo {author} {\bibfnamefont {S.}~\bibnamefont {Filipp}},\ }\bibfield  {title} {\bibinfo {title} {{Leakage reduction in fast superconducting qubit gates via optimal control}},\ }\href {https://doi.org/10.1038/s41534-020-00346-2} {\bibfield  {journal} {\bibinfo  {journal} {npj Quantum Inf.}\ }\textbf {\bibinfo {volume} {7}},\ \bibinfo {pages} {14} (\bibinfo {year} {2021})}\BibitemShut {NoStop}%
\bibitem [{\citenamefont {Goerz}\ \emph {et~al.}(2017)\citenamefont {Goerz}, \citenamefont {Motzoi}, \citenamefont {Whaley},\ and\ \citenamefont {Koch}}]{Goerz2017}%
  \BibitemOpen
  \bibfield  {author} {\bibinfo {author} {\bibfnamefont {M.~H.}\ \bibnamefont {Goerz}}, \bibinfo {author} {\bibfnamefont {F.}~\bibnamefont {Motzoi}}, \bibinfo {author} {\bibfnamefont {K.~B.}\ \bibnamefont {Whaley}},\ and\ \bibinfo {author} {\bibfnamefont {C.~P.}\ \bibnamefont {Koch}},\ }\bibfield  {title} {\bibinfo {title} {{Charting the circuit QED design landscape using optimal control theory}},\ }\href {https://doi.org/10.1038/s41534-017-0036-0} {\bibfield  {journal} {\bibinfo  {journal} {npj Quantum Inf.}\ }\textbf {\bibinfo {volume} {3}},\ \bibinfo {pages} {37} (\bibinfo {year} {2017})}\BibitemShut {NoStop}%
\bibitem [{\citenamefont {Hahn}(1950)}]{hahn1950spin}%
  \BibitemOpen
  \bibfield  {author} {\bibinfo {author} {\bibfnamefont {E.~L.}\ \bibnamefont {Hahn}},\ }\bibfield  {title} {\bibinfo {title} {Spin echoes},\ }\href {https://doi.org/10.1103/PhysRev.80.580} {\bibfield  {journal} {\bibinfo  {journal} {Phys. Rev.}\ }\textbf {\bibinfo {volume} {80}},\ \bibinfo {pages} {580} (\bibinfo {year} {1950})}\BibitemShut {NoStop}%
\bibitem [{\citenamefont {Dridi}\ \emph {et~al.}(2020{\natexlab{a}})\citenamefont {Dridi}, \citenamefont {Mejatty}, \citenamefont {Glaser},\ and\ \citenamefont {Sugny}}]{dridi2020robust}%
  \BibitemOpen
  \bibfield  {author} {\bibinfo {author} {\bibfnamefont {G.}~\bibnamefont {Dridi}}, \bibinfo {author} {\bibfnamefont {M.}~\bibnamefont {Mejatty}}, \bibinfo {author} {\bibfnamefont {S.~J.}\ \bibnamefont {Glaser}},\ and\ \bibinfo {author} {\bibfnamefont {D.}~\bibnamefont {Sugny}},\ }\bibfield  {title} {\bibinfo {title} {Robust control of a not gate by composite pulses},\ }\href {https://doi.org/10.1103/PhysRevA.101.012321} {\bibfield  {journal} {\bibinfo  {journal} {Phys. Rev. A}\ }\textbf {\bibinfo {volume} {101}},\ \bibinfo {pages} {012321} (\bibinfo {year} {2020}{\natexlab{a}})}\BibitemShut {NoStop}%
\bibitem [{\citenamefont {Motzoi}\ \emph {et~al.}(2009)\citenamefont {Motzoi}, \citenamefont {Gambetta}, \citenamefont {Rebentrost},\ and\ \citenamefont {Wilhelm}}]{Motzoi2009}%
  \BibitemOpen
  \bibfield  {author} {\bibinfo {author} {\bibfnamefont {F.}~\bibnamefont {Motzoi}}, \bibinfo {author} {\bibfnamefont {J.~M.}\ \bibnamefont {Gambetta}}, \bibinfo {author} {\bibfnamefont {P.}~\bibnamefont {Rebentrost}},\ and\ \bibinfo {author} {\bibfnamefont {F.~K.}\ \bibnamefont {Wilhelm}},\ }\bibfield  {title} {\bibinfo {title} {{Simple Pulses for Elimination of Leakage in Weakly Nonlinear Qubits}},\ }\href {https://doi.org/10.1103/PhysRevLett.103.110501} {\bibfield  {journal} {\bibinfo  {journal} {Phys. Rev. Lett.}\ }\textbf {\bibinfo {volume} {103}},\ \bibinfo {pages} {110501} (\bibinfo {year} {2009})}\BibitemShut {NoStop}%
\bibitem [{\citenamefont {Chow}\ \emph {et~al.}(2010)\citenamefont {Chow}, \citenamefont {DiCarlo}, \citenamefont {Gambetta}, \citenamefont {Motzoi}, \citenamefont {Frunzio}, \citenamefont {Girvin},\ and\ \citenamefont {Schoelkopf}}]{Chow2010}%
  \BibitemOpen
  \bibfield  {author} {\bibinfo {author} {\bibfnamefont {J.~M.}\ \bibnamefont {Chow}}, \bibinfo {author} {\bibfnamefont {L.}~\bibnamefont {DiCarlo}}, \bibinfo {author} {\bibfnamefont {J.~M.}\ \bibnamefont {Gambetta}}, \bibinfo {author} {\bibfnamefont {F.}~\bibnamefont {Motzoi}}, \bibinfo {author} {\bibfnamefont {L.}~\bibnamefont {Frunzio}}, \bibinfo {author} {\bibfnamefont {S.~M.}\ \bibnamefont {Girvin}},\ and\ \bibinfo {author} {\bibfnamefont {R.~J.}\ \bibnamefont {Schoelkopf}},\ }\bibfield  {title} {\bibinfo {title} {{Optimized driving of superconducting artificial atoms for improved single-qubit gates}},\ }\href {https://doi.org/10.1103/PhysRevA.82.040305} {\bibfield  {journal} {\bibinfo  {journal} {Phys. Rev. A}\ }\textbf {\bibinfo {volume} {82}},\ \bibinfo {pages} {040305(R)} (\bibinfo {year} {2010})}\BibitemShut {NoStop}%
\bibitem [{\citenamefont {Gambetta}\ \emph {et~al.}(2011)\citenamefont {Gambetta}, \citenamefont {Motzoi}, \citenamefont {Merkel},\ and\ \citenamefont {Wilhelm}}]{Gambetta2011}%
  \BibitemOpen
  \bibfield  {author} {\bibinfo {author} {\bibfnamefont {J.~M.}\ \bibnamefont {Gambetta}}, \bibinfo {author} {\bibfnamefont {F.}~\bibnamefont {Motzoi}}, \bibinfo {author} {\bibfnamefont {S.~T.}\ \bibnamefont {Merkel}},\ and\ \bibinfo {author} {\bibfnamefont {F.~K.}\ \bibnamefont {Wilhelm}},\ }\bibfield  {title} {\bibinfo {title} {{Analytic control methods for high-fidelity unitary operations in a weakly nonlinear oscillator}},\ }\href {https://doi.org/10.1103/PhysRevA.83.012308} {\bibfield  {journal} {\bibinfo  {journal} {Phys. Rev. A}\ }\textbf {\bibinfo {volume} {83}},\ \bibinfo {pages} {012308} (\bibinfo {year} {2011})}\BibitemShut {NoStop}%
\bibitem [{\citenamefont {Motzoi}\ and\ \citenamefont {Wilhelm}(2013)}]{Motzoi2013c}%
  \BibitemOpen
  \bibfield  {author} {\bibinfo {author} {\bibfnamefont {F.}~\bibnamefont {Motzoi}}\ and\ \bibinfo {author} {\bibfnamefont {F.~K.}\ \bibnamefont {Wilhelm}},\ }\bibfield  {title} {\bibinfo {title} {{Improving frequency selection of driven pulses using derivative-based transition suppression}},\ }\href {https://doi.org/10.1103/PhysRevA.88.062318} {\bibfield  {journal} {\bibinfo  {journal} {Phys. Rev. A}\ }\textbf {\bibinfo {volume} {88}},\ \bibinfo {pages} {062318} (\bibinfo {year} {2013})}\BibitemShut {NoStop}%
\bibitem [{\citenamefont {Chen}\ \emph {et~al.}(2016)\citenamefont {Chen}, \citenamefont {Kelly}, \citenamefont {Quintana}, \citenamefont {Barends}, \citenamefont {Campbell}, \citenamefont {Chen}, \citenamefont {Chiaro}, \citenamefont {Dunsworth}, \citenamefont {Fowler}, \citenamefont {Lucero} \emph {et~al.}}]{Chen2016a}%
  \BibitemOpen
  \bibfield  {author} {\bibinfo {author} {\bibfnamefont {Z.}~\bibnamefont {Chen}}, \bibinfo {author} {\bibfnamefont {J.}~\bibnamefont {Kelly}}, \bibinfo {author} {\bibfnamefont {C.}~\bibnamefont {Quintana}}, \bibinfo {author} {\bibfnamefont {R.}~\bibnamefont {Barends}}, \bibinfo {author} {\bibfnamefont {B.}~\bibnamefont {Campbell}}, \bibinfo {author} {\bibfnamefont {Y.}~\bibnamefont {Chen}}, \bibinfo {author} {\bibfnamefont {B.}~\bibnamefont {Chiaro}}, \bibinfo {author} {\bibfnamefont {A.}~\bibnamefont {Dunsworth}}, \bibinfo {author} {\bibfnamefont {A.~G.}\ \bibnamefont {Fowler}}, \bibinfo {author} {\bibfnamefont {E.}~\bibnamefont {Lucero}}, \emph {et~al.},\ }\bibfield  {title} {\bibinfo {title} {{Measuring and Suppressing Quantum State Leakage in a Superconducting Qubit}},\ }\href {https://doi.org/10.1103/PhysRevLett.116.020501} {\bibfield  {journal} {\bibinfo  {journal} {Phys. Rev. Lett.}\ }\textbf {\bibinfo {volume} {116}},\ \bibinfo {pages} {020501} (\bibinfo {year} {2016})}\BibitemShut {NoStop}%
\bibitem [{\citenamefont {Theis}\ \emph {et~al.}(2018)\citenamefont {Theis}, \citenamefont {Motzoi}, \citenamefont {Machnes},\ and\ \citenamefont {Wilhelm}}]{Theis2018a}%
  \BibitemOpen
  \bibfield  {author} {\bibinfo {author} {\bibfnamefont {L.~S.}\ \bibnamefont {Theis}}, \bibinfo {author} {\bibfnamefont {F.}~\bibnamefont {Motzoi}}, \bibinfo {author} {\bibfnamefont {S.}~\bibnamefont {Machnes}},\ and\ \bibinfo {author} {\bibfnamefont {F.~K.}\ \bibnamefont {Wilhelm}},\ }\bibfield  {title} {\bibinfo {title} {{Counteracting systems of diabaticities using DRAG controls: The status after 10 years}},\ }\href {https://doi.org/10.1209/0295-5075/123/60001} {\bibfield  {journal} {\bibinfo  {journal} {EPL}\ }\textbf {\bibinfo {volume} {123}},\ \bibinfo {pages} {60001} (\bibinfo {year} {2018})}\BibitemShut {NoStop}%
\bibitem [{\citenamefont {Gupta}\ \emph {et~al.}(2023)\citenamefont {Gupta}, \citenamefont {DiNapoli}, \citenamefont {Huang}, \citenamefont {Yuan}, \citenamefont {He}, \citenamefont {Jiang}, \citenamefont {Schuster},\ and\ \citenamefont {Chakram}}]{eesh}%
  \BibitemOpen
  \bibfield  {author} {\bibinfo {author} {\bibfnamefont {E.}~\bibnamefont {Gupta}}, \bibinfo {author} {\bibfnamefont {T.}~\bibnamefont {DiNapoli}}, \bibinfo {author} {\bibfnamefont {J.}~\bibnamefont {Huang}}, \bibinfo {author} {\bibfnamefont {M.}~\bibnamefont {Yuan}}, \bibinfo {author} {\bibfnamefont {K.}~\bibnamefont {He}}, \bibinfo {author} {\bibfnamefont {L.}~\bibnamefont {Jiang}}, \bibinfo {author} {\bibfnamefont {D.}~\bibnamefont {Schuster}},\ and\ \bibinfo {author} {\bibfnamefont {S.}~\bibnamefont {Chakram}},\ }\bibfield  {title} {\bibinfo {title} {{Fast Control of Multimode Cavities with Conditional Displacements}},\ }\href {https://meetings.aps.org/Meeting/MAR23/Session/W67.6} {\bibfield  {journal} {\bibinfo  {journal} {Bull. Am. Phys. Soc.}\ } (\bibinfo {year} {2023})}\BibitemShut {NoStop}%
\bibitem [{Note2()}]{Note2}%
  \BibitemOpen
  \bibinfo {note} {This is not strictly valid in an open system where the ancilla can decay. However, considering that the gate duration is much shorter than a typical transmon lifetime, this is still a good approximation.}\BibitemShut {Stop}%
\bibitem [{\citenamefont {McKay}\ \emph {et~al.}(2017)\citenamefont {McKay}, \citenamefont {Wood}, \citenamefont {Sheldon}, \citenamefont {Chow},\ and\ \citenamefont {Gambetta}}]{McKay2017}%
  \BibitemOpen
  \bibfield  {author} {\bibinfo {author} {\bibfnamefont {D.~C.}\ \bibnamefont {McKay}}, \bibinfo {author} {\bibfnamefont {C.~J.}\ \bibnamefont {Wood}}, \bibinfo {author} {\bibfnamefont {S.}~\bibnamefont {Sheldon}}, \bibinfo {author} {\bibfnamefont {J.~M.}\ \bibnamefont {Chow}},\ and\ \bibinfo {author} {\bibfnamefont {J.~M.}\ \bibnamefont {Gambetta}},\ }\bibfield  {title} {\bibinfo {title} {Efficient $z$ gates for quantum computing},\ }\href {https://doi.org/10.1103/PhysRevA.96.022330} {\bibfield  {journal} {\bibinfo  {journal} {Phys. Rev. A}\ }\textbf {\bibinfo {volume} {96}},\ \bibinfo {pages} {022330} (\bibinfo {year} {2017})}\BibitemShut {NoStop}%
\bibitem [{\citenamefont {Banerjee}\ \emph {et~al.}(2023)\citenamefont {Banerjee}, \citenamefont {Chakram}, \citenamefont {Oriani}, \citenamefont {Zhao}, \citenamefont {He}, \citenamefont {Agrawal},\ and\ \citenamefont {Schuster}}]{manipulator}%
  \BibitemOpen
  \bibfield  {author} {\bibinfo {author} {\bibfnamefont {R.}~\bibnamefont {Banerjee}}, \bibinfo {author} {\bibfnamefont {S.}~\bibnamefont {Chakram}}, \bibinfo {author} {\bibfnamefont {A.}~\bibnamefont {Oriani}}, \bibinfo {author} {\bibfnamefont {F.}~\bibnamefont {Zhao}}, \bibinfo {author} {\bibfnamefont {K.}~\bibnamefont {He}}, \bibinfo {author} {\bibfnamefont {A.}~\bibnamefont {Agrawal}},\ and\ \bibinfo {author} {\bibfnamefont {D.}~\bibnamefont {Schuster}},\ }\bibfield  {title} {\bibinfo {title} {{Fast flux control of a high-Q 3D multimode cavity}},\ }\href {https://meetings.aps.org/Meeting/MAR23/Session/K75.3} {\bibfield  {journal} {\bibinfo  {journal} {Bull. Am. Phys. Soc.}\ } (\bibinfo {year} {2023})}\BibitemShut {NoStop}%
\bibitem [{\citenamefont {Daems}\ \emph {et~al.}(2013)\citenamefont {Daems}, \citenamefont {Ruschhaupt}, \citenamefont {Sugny},\ and\ \citenamefont {Gu\'erin}}]{PhysRevLett.111.050404inverseengi}%
  \BibitemOpen
  \bibfield  {author} {\bibinfo {author} {\bibfnamefont {D.}~\bibnamefont {Daems}}, \bibinfo {author} {\bibfnamefont {A.}~\bibnamefont {Ruschhaupt}}, \bibinfo {author} {\bibfnamefont {D.}~\bibnamefont {Sugny}},\ and\ \bibinfo {author} {\bibfnamefont {S.}~\bibnamefont {Gu\'erin}},\ }\bibfield  {title} {\bibinfo {title} {Robust quantum control by a single-shot shaped pulse},\ }\href {https://doi.org/10.1103/PhysRevLett.111.050404} {\bibfield  {journal} {\bibinfo  {journal} {Phys. Rev. Lett.}\ }\textbf {\bibinfo {volume} {111}},\ \bibinfo {pages} {050404} (\bibinfo {year} {2013})}\BibitemShut {NoStop}%
\bibitem [{\citenamefont {Dridi}\ \emph {et~al.}(2020{\natexlab{b}})\citenamefont {Dridi}, \citenamefont {Liu},\ and\ \citenamefont {Gu\'erin}}]{PhysRevLett.125.250403inverseengi}%
  \BibitemOpen
  \bibfield  {author} {\bibinfo {author} {\bibfnamefont {G.}~\bibnamefont {Dridi}}, \bibinfo {author} {\bibfnamefont {K.}~\bibnamefont {Liu}},\ and\ \bibinfo {author} {\bibfnamefont {S.}~\bibnamefont {Gu\'erin}},\ }\bibfield  {title} {\bibinfo {title} {Optimal robust quantum control by inverse geometric optimization},\ }\href {https://doi.org/10.1103/PhysRevLett.125.250403} {\bibfield  {journal} {\bibinfo  {journal} {Phys. Rev. Lett.}\ }\textbf {\bibinfo {volume} {125}},\ \bibinfo {pages} {250403} (\bibinfo {year} {2020}{\natexlab{b}})}\BibitemShut {NoStop}%
\bibitem [{\citenamefont {Barnes}\ \emph {et~al.}(2022)\citenamefont {Barnes}, \citenamefont {Calderon-Vargas}, \citenamefont {Dong}, \citenamefont {Li}, \citenamefont {Zeng},\ and\ \citenamefont {Zhuang}}]{barnes2022dynamicallyspacecurve}%
  \BibitemOpen
  \bibfield  {author} {\bibinfo {author} {\bibfnamefont {E.}~\bibnamefont {Barnes}}, \bibinfo {author} {\bibfnamefont {F.~A.}\ \bibnamefont {Calderon-Vargas}}, \bibinfo {author} {\bibfnamefont {W.}~\bibnamefont {Dong}}, \bibinfo {author} {\bibfnamefont {B.}~\bibnamefont {Li}}, \bibinfo {author} {\bibfnamefont {J.}~\bibnamefont {Zeng}},\ and\ \bibinfo {author} {\bibfnamefont {F.}~\bibnamefont {Zhuang}},\ }\bibfield  {title} {\bibinfo {title} {Dynamically corrected gates from geometric space curves},\ }\href {https://doi.org/10.1088/2058-9565/ac4421} {\bibfield  {journal} {\bibinfo  {journal} {Quantum Sci. Technol.}\ }\textbf {\bibinfo {volume} {7}},\ \bibinfo {pages} {023001} (\bibinfo {year} {2022})}\BibitemShut {NoStop}%
\bibitem [{\citenamefont {Wang}\ \emph {et~al.}(2012)\citenamefont {Wang}, \citenamefont {Bishop}, \citenamefont {Kestner}, \citenamefont {Barnes}, \citenamefont {Sun},\ and\ \citenamefont {Das~Sarma}}]{wang2012compositepulse}%
  \BibitemOpen
  \bibfield  {author} {\bibinfo {author} {\bibfnamefont {X.}~\bibnamefont {Wang}}, \bibinfo {author} {\bibfnamefont {L.~S.}\ \bibnamefont {Bishop}}, \bibinfo {author} {\bibfnamefont {J.}~\bibnamefont {Kestner}}, \bibinfo {author} {\bibfnamefont {E.}~\bibnamefont {Barnes}}, \bibinfo {author} {\bibfnamefont {K.}~\bibnamefont {Sun}},\ and\ \bibinfo {author} {\bibfnamefont {S.}~\bibnamefont {Das~Sarma}},\ }\bibfield  {title} {\bibinfo {title} {Composite pulses for robust universal control of singlet--triplet qubits},\ }\href {https://www.nature.com/articles/ncomms2003} {\bibfield  {journal} {\bibinfo  {journal} {Nat. Commun.}\ }\textbf {\bibinfo {volume} {3}},\ \bibinfo {pages} {997} (\bibinfo {year} {2012})}\BibitemShut {NoStop}%
\bibitem [{\citenamefont {Torosov}\ and\ \citenamefont {Vitanov}(2011)}]{PhysRevA.83.053420pulse}%
  \BibitemOpen
  \bibfield  {author} {\bibinfo {author} {\bibfnamefont {B.~T.}\ \bibnamefont {Torosov}}\ and\ \bibinfo {author} {\bibfnamefont {N.~V.}\ \bibnamefont {Vitanov}},\ }\bibfield  {title} {\bibinfo {title} {Smooth composite pulses for high-fidelity quantum information processing},\ }\href {https://doi.org/10.1103/PhysRevA.83.053420} {\bibfield  {journal} {\bibinfo  {journal} {Phys. Rev. A}\ }\textbf {\bibinfo {volume} {83}},\ \bibinfo {pages} {053420} (\bibinfo {year} {2011})}\BibitemShut {NoStop}%
\bibitem [{\citenamefont {Owrutsky}\ and\ \citenamefont {Khaneja}(2012)}]{PhysRevA.86.022315pulse}%
  \BibitemOpen
  \bibfield  {author} {\bibinfo {author} {\bibfnamefont {P.}~\bibnamefont {Owrutsky}}\ and\ \bibinfo {author} {\bibfnamefont {N.}~\bibnamefont {Khaneja}},\ }\bibfield  {title} {\bibinfo {title} {Control of inhomogeneous ensembles on the bloch sphere},\ }\href {https://doi.org/10.1103/PhysRevA.86.022315} {\bibfield  {journal} {\bibinfo  {journal} {Phys. Rev. A}\ }\textbf {\bibinfo {volume} {86}},\ \bibinfo {pages} {022315} (\bibinfo {year} {2012})}\BibitemShut {NoStop}%
\bibitem [{\citenamefont {Tycko}(1983)}]{tycko1983broadbandpulse}%
  \BibitemOpen
  \bibfield  {author} {\bibinfo {author} {\bibfnamefont {R.}~\bibnamefont {Tycko}},\ }\bibfield  {title} {\bibinfo {title} {Broadband population inversion},\ }\href {https://doi.org/10.1103/PhysRevLett.51.775} {\bibfield  {journal} {\bibinfo  {journal} {Phys. Rev. Lett.}\ }\textbf {\bibinfo {volume} {51}},\ \bibinfo {pages} {775} (\bibinfo {year} {1983})}\BibitemShut {NoStop}%
\bibitem [{\citenamefont {Wu}\ \emph {et~al.}(2023)\citenamefont {Wu}, \citenamefont {Zhang}, \citenamefont {Song}, \citenamefont {Xia},\ and\ \citenamefont {Shi}}]{PhysRevA.107.023103pulse}%
  \BibitemOpen
  \bibfield  {author} {\bibinfo {author} {\bibfnamefont {H.-N.}\ \bibnamefont {Wu}}, \bibinfo {author} {\bibfnamefont {C.}~\bibnamefont {Zhang}}, \bibinfo {author} {\bibfnamefont {J.}~\bibnamefont {Song}}, \bibinfo {author} {\bibfnamefont {Y.}~\bibnamefont {Xia}},\ and\ \bibinfo {author} {\bibfnamefont {Z.-C.}\ \bibnamefont {Shi}},\ }\bibfield  {title} {\bibinfo {title} {Composite pulses for optimal robust control in two-level systems},\ }\href {https://doi.org/10.1103/PhysRevA.107.023103} {\bibfield  {journal} {\bibinfo  {journal} {Phys. Rev. A}\ }\textbf {\bibinfo {volume} {107}},\ \bibinfo {pages} {023103} (\bibinfo {year} {2023})}\BibitemShut {NoStop}%
\bibitem [{\citenamefont {Ball}\ \emph {et~al.}(2021)\citenamefont {Ball}, \citenamefont {Biercuk}, \citenamefont {Carvalho}, \citenamefont {Chen}, \citenamefont {Hush}, \citenamefont {De~Castro}, \citenamefont {Li}, \citenamefont {Liebermann}, \citenamefont {Slatyer}, \citenamefont {Edmunds} \emph {et~al.}}]{ball2021softwarepulseengi}%
  \BibitemOpen
  \bibfield  {author} {\bibinfo {author} {\bibfnamefont {H.}~\bibnamefont {Ball}}, \bibinfo {author} {\bibfnamefont {M.~J.}\ \bibnamefont {Biercuk}}, \bibinfo {author} {\bibfnamefont {A.~R.}\ \bibnamefont {Carvalho}}, \bibinfo {author} {\bibfnamefont {J.}~\bibnamefont {Chen}}, \bibinfo {author} {\bibfnamefont {M.}~\bibnamefont {Hush}}, \bibinfo {author} {\bibfnamefont {L.~A.}\ \bibnamefont {De~Castro}}, \bibinfo {author} {\bibfnamefont {L.}~\bibnamefont {Li}}, \bibinfo {author} {\bibfnamefont {P.~J.}\ \bibnamefont {Liebermann}}, \bibinfo {author} {\bibfnamefont {H.~J.}\ \bibnamefont {Slatyer}}, \bibinfo {author} {\bibfnamefont {C.}~\bibnamefont {Edmunds}}, \emph {et~al.},\ }\bibfield  {title} {\bibinfo {title} {Software tools for quantum control: Improving quantum computer performance through noise and error suppression},\ }\href {https://iopscience.iop.org/article/10.1088/2058-9565/abdca6/meta} {\bibfield  {journal} {\bibinfo  {journal} {Quantum Sci. Technol.}\ }\textbf {\bibinfo {volume} {6}},\ \bibinfo
  {pages} {044011} (\bibinfo {year} {2021})}\BibitemShut {NoStop}%
\bibitem [{\citenamefont {Wimperis}(1994)}]{wimperis1994broadbandpulse}%
  \BibitemOpen
  \bibfield  {author} {\bibinfo {author} {\bibfnamefont {S.}~\bibnamefont {Wimperis}},\ }\bibfield  {title} {\bibinfo {title} {Broadband, narrowband, and passband composite pulses for use in advanced nmr experiments},\ }\href {https://doi.org/https://doi.org/10.1006/jmra.1994.1159} {\bibfield  {journal} {\bibinfo  {journal} {J. Magn. Reson. A}\ }\textbf {\bibinfo {volume} {109}},\ \bibinfo {pages} {221} (\bibinfo {year} {1994})}\BibitemShut {NoStop}%
\bibitem [{\citenamefont {Pasini}\ \emph {et~al.}(2009)\citenamefont {Pasini}, \citenamefont {Karbach}, \citenamefont {Raas},\ and\ \citenamefont {Uhrig}}]{pasini2009optimizedinitialpulse}%
  \BibitemOpen
  \bibfield  {author} {\bibinfo {author} {\bibfnamefont {S.}~\bibnamefont {Pasini}}, \bibinfo {author} {\bibfnamefont {P.}~\bibnamefont {Karbach}}, \bibinfo {author} {\bibfnamefont {C.}~\bibnamefont {Raas}},\ and\ \bibinfo {author} {\bibfnamefont {G.~S.}\ \bibnamefont {Uhrig}},\ }\bibfield  {title} {\bibinfo {title} {Optimized pulses for the perturbative decoupling of a spin and a decoherence bath},\ }\href {https://doi.org/10.1103/PhysRevA.80.022328} {\bibfield  {journal} {\bibinfo  {journal} {Phys. Rev. A}\ }\textbf {\bibinfo {volume} {80}},\ \bibinfo {pages} {022328} (\bibinfo {year} {2009})}\BibitemShut {NoStop}%
\bibitem [{\citenamefont {Propson}\ \emph {et~al.}(2022)\citenamefont {Propson}, \citenamefont {Jackson}, \citenamefont {Koch}, \citenamefont {Manchester},\ and\ \citenamefont {Schuster}}]{propson2022robustcollocation}%
  \BibitemOpen
  \bibfield  {author} {\bibinfo {author} {\bibfnamefont {T.}~\bibnamefont {Propson}}, \bibinfo {author} {\bibfnamefont {B.~E.}\ \bibnamefont {Jackson}}, \bibinfo {author} {\bibfnamefont {J.}~\bibnamefont {Koch}}, \bibinfo {author} {\bibfnamefont {Z.}~\bibnamefont {Manchester}},\ and\ \bibinfo {author} {\bibfnamefont {D.~I.}\ \bibnamefont {Schuster}},\ }\bibfield  {title} {\bibinfo {title} {Robust quantum optimal control with trajectory optimization},\ }\href {https://doi.org/10.1103/PhysRevApplied.17.014036} {\bibfield  {journal} {\bibinfo  {journal} {Phys. Rev. Appl.}\ }\textbf {\bibinfo {volume} {17}},\ \bibinfo {pages} {014036} (\bibinfo {year} {2022})}\BibitemShut {NoStop}%
\bibitem [{\citenamefont {Trowbridge}\ \emph {et~al.}(2023)\citenamefont {Trowbridge}, \citenamefont {Bhardwaj}, \citenamefont {He}, \citenamefont {Schuster},\ and\ \citenamefont {Manchester}}]{Trowbridge2023}%
  \BibitemOpen
  \bibfield  {author} {\bibinfo {author} {\bibfnamefont {A.}~\bibnamefont {Trowbridge}}, \bibinfo {author} {\bibfnamefont {A.}~\bibnamefont {Bhardwaj}}, \bibinfo {author} {\bibfnamefont {K.}~\bibnamefont {He}}, \bibinfo {author} {\bibfnamefont {D.~I.}\ \bibnamefont {Schuster}},\ and\ \bibinfo {author} {\bibfnamefont {Z.}~\bibnamefont {Manchester}},\ }\bibfield  {title} {\bibinfo {title} {Direct collocation for quantum optimal control},\ }in\ \href {https://doi.org/10.1109/QCE57702.2023.00144} {\emph {\bibinfo {booktitle} {2023 IEEE International Conference on Quantum Computing and Engineering (QCE)}}},\ Vol.~\bibinfo {volume} {01}\ (\bibinfo {year} {2023})\ pp.\ \bibinfo {pages} {1278--1285}\BibitemShut {NoStop}%
\bibitem [{\citenamefont {Hai}\ \emph {et~al.}(2023)\citenamefont {Hai}, \citenamefont {Li}, \citenamefont {Zeng}, \citenamefont {Yu},\ and\ \citenamefont {Deng}}]{hai2023universalcompare}%
  \BibitemOpen
  \bibfield  {author} {\bibinfo {author} {\bibfnamefont {Y.-J.}\ \bibnamefont {Hai}}, \bibinfo {author} {\bibfnamefont {J.}~\bibnamefont {Li}}, \bibinfo {author} {\bibfnamefont {J.}~\bibnamefont {Zeng}}, \bibinfo {author} {\bibfnamefont {D.}~\bibnamefont {Yu}},\ and\ \bibinfo {author} {\bibfnamefont {X.-H.}\ \bibnamefont {Deng}},\ }\bibfield  {title} {\bibinfo {title} {Universal robust quantum gates by geometric correspondence of noisy quantum evolution},\ }\href {https://arxiv.org/abs/2210.14521} {\bibfield  {journal} {\bibinfo  {journal} {arXiv:2210.14521}\ } (\bibinfo {year} {2023})}\BibitemShut {NoStop}%
\bibitem [{\citenamefont {Carvalho}\ \emph {et~al.}(2021)\citenamefont {Carvalho}, \citenamefont {Ball}, \citenamefont {Biercuk}, \citenamefont {Hush},\ and\ \citenamefont {Thomsen}}]{qctrl}%
  \BibitemOpen
  \bibfield  {author} {\bibinfo {author} {\bibfnamefont {A.~R.~R.}\ \bibnamefont {Carvalho}}, \bibinfo {author} {\bibfnamefont {H.}~\bibnamefont {Ball}}, \bibinfo {author} {\bibfnamefont {M.~J.}\ \bibnamefont {Biercuk}}, \bibinfo {author} {\bibfnamefont {M.~R.}\ \bibnamefont {Hush}},\ and\ \bibinfo {author} {\bibfnamefont {F.}~\bibnamefont {Thomsen}},\ }\bibfield  {title} {\bibinfo {title} {Error-robust quantum logic optimization using a cloud quantum computer interface},\ }\href {https://doi.org/10.1103/PhysRevApplied.15.064054} {\bibfield  {journal} {\bibinfo  {journal} {Phys. Rev. Appl.}\ }\textbf {\bibinfo {volume} {15}},\ \bibinfo {pages} {064054} (\bibinfo {year} {2021})}\BibitemShut {NoStop}%
\bibitem [{\citenamefont {Abdelhafez}(2019)}]{abdelhafez2019quantumthesis}%
  \BibitemOpen
  \bibfield  {author} {\bibinfo {author} {\bibfnamefont {M.~R.}\ \bibnamefont {Abdelhafez}},\ }\emph {\bibinfo {title} {Quantum Optimal Control Using Automatic Differentiation}},\ \href@noop {} {Ph.D. thesis},\ \bibinfo  {school} {The University of Chicago} (\bibinfo {year} {2019})\BibitemShut {NoStop}%
\bibitem [{\citenamefont {Watanabe}\ \emph {et~al.}(2024)\citenamefont {Watanabe}, \citenamefont {Tabuchi}, \citenamefont {Heya}, \citenamefont {Tamate},\ and\ \citenamefont {Nakamura}}]{Watanabe}%
  \BibitemOpen
  \bibfield  {author} {\bibinfo {author} {\bibfnamefont {S.}~\bibnamefont {Watanabe}}, \bibinfo {author} {\bibfnamefont {Y.}~\bibnamefont {Tabuchi}}, \bibinfo {author} {\bibfnamefont {K.}~\bibnamefont {Heya}}, \bibinfo {author} {\bibfnamefont {S.}~\bibnamefont {Tamate}},\ and\ \bibinfo {author} {\bibfnamefont {Y.}~\bibnamefont {Nakamura}},\ }\bibfield  {title} {\bibinfo {title} {$zz$-interaction-free single-qubit-gate optimization in superconducting qubits},\ }\href {https://doi.org/10.1103/PhysRevA.109.012616} {\bibfield  {journal} {\bibinfo  {journal} {Phys. Rev. A}\ }\textbf {\bibinfo {volume} {109}},\ \bibinfo {pages} {012616} (\bibinfo {year} {2024})}\BibitemShut {NoStop}%
\bibitem [{\citenamefont {Reinhold}(2019)}]{reinhold2019controllingthesisrobustmetric}%
  \BibitemOpen
  \bibfield  {author} {\bibinfo {author} {\bibfnamefont {P.}~\bibnamefont {Reinhold}},\ }\emph {\bibinfo {title} {Controlling error-correctable bosonic qubits}},\ \href@noop {} {Ph.D. thesis},\ \bibinfo  {school} {Yale Univ.} (\bibinfo {year} {2019})\BibitemShut {NoStop}%
\bibitem [{\citenamefont {Allen}(2020)}]{allen2020robustmetric}%
  \BibitemOpen
  \bibfield  {author} {\bibinfo {author} {\bibfnamefont {J.}~\bibnamefont {Allen}},\ }\emph {\bibinfo {title} {Robust Optimal Control of the Cross-Resonance Gate in Superconducting Qubits}},\ \href@noop {} {Ph.D. thesis},\ \bibinfo  {school} {Univ. of Surrey} (\bibinfo {year} {2020})\BibitemShut {NoStop}%
\bibitem [{\citenamefont {Kosut}\ \emph {et~al.}(2013)\citenamefont {Kosut}, \citenamefont {Grace},\ and\ \citenamefont {Brif}}]{kosut2013robustmetric}%
  \BibitemOpen
  \bibfield  {author} {\bibinfo {author} {\bibfnamefont {R.~L.}\ \bibnamefont {Kosut}}, \bibinfo {author} {\bibfnamefont {M.~D.}\ \bibnamefont {Grace}},\ and\ \bibinfo {author} {\bibfnamefont {C.}~\bibnamefont {Brif}},\ }\bibfield  {title} {\bibinfo {title} {Robust control of quantum gates via sequential convex programming},\ }\href {https://doi.org/10.1103/PhysRevA.88.052326} {\bibfield  {journal} {\bibinfo  {journal} {Phys. Rev. A}\ }\textbf {\bibinfo {volume} {88}},\ \bibinfo {pages} {052326} (\bibinfo {year} {2013})}\BibitemShut {NoStop}%
\bibitem [{\citenamefont {Khaneja}\ \emph {et~al.}(2005)\citenamefont {Khaneja}, \citenamefont {Reiss}, \citenamefont {Kehlet}, \citenamefont {Schulte-Herbr{\"u}ggen},\ and\ \citenamefont {Glaser}}]{khaneja2005optimalrobustmetric}%
  \BibitemOpen
  \bibfield  {author} {\bibinfo {author} {\bibfnamefont {N.}~\bibnamefont {Khaneja}}, \bibinfo {author} {\bibfnamefont {T.}~\bibnamefont {Reiss}}, \bibinfo {author} {\bibfnamefont {C.}~\bibnamefont {Kehlet}}, \bibinfo {author} {\bibfnamefont {T.}~\bibnamefont {Schulte-Herbr{\"u}ggen}},\ and\ \bibinfo {author} {\bibfnamefont {S.~J.}\ \bibnamefont {Glaser}},\ }\bibfield  {title} {\bibinfo {title} {Optimal control of coupled spin dynamics: design of nmr pulse sequences by gradient ascent algorithms},\ }\href {https://www.sciencedirect.com/science/article/abs/pii/S1090780704003696} {\bibfield  {journal} {\bibinfo  {journal} {J. Magn. Reson.}\ }\textbf {\bibinfo {volume} {172}},\ \bibinfo {pages} {296} (\bibinfo {year} {2005})}\BibitemShut {NoStop}%
\bibitem [{\citenamefont {Rembold}\ \emph {et~al.}(2020)\citenamefont {Rembold}, \citenamefont {Oshnik}, \citenamefont {M{\"u}ller}, \citenamefont {Montangero}, \citenamefont {Calarco},\ and\ \citenamefont {Neu}}]{rembold2020introductionrobustmetric}%
  \BibitemOpen
  \bibfield  {author} {\bibinfo {author} {\bibfnamefont {P.}~\bibnamefont {Rembold}}, \bibinfo {author} {\bibfnamefont {N.}~\bibnamefont {Oshnik}}, \bibinfo {author} {\bibfnamefont {M.~M.}\ \bibnamefont {M{\"u}ller}}, \bibinfo {author} {\bibfnamefont {S.}~\bibnamefont {Montangero}}, \bibinfo {author} {\bibfnamefont {T.}~\bibnamefont {Calarco}},\ and\ \bibinfo {author} {\bibfnamefont {E.}~\bibnamefont {Neu}},\ }\bibfield  {title} {\bibinfo {title} {Introduction to quantum optimal control for quantum sensing with nitrogen-vacancy centers in diamond},\ }\href {https://doi.org/10.1116/5.0006785} {\bibfield  {journal} {\bibinfo  {journal} {AVS Quantum Sci.}\ }\textbf {\bibinfo {volume} {2}},\ \bibinfo {pages} {024701} (\bibinfo {year} {2020})}\BibitemShut {NoStop}%
\bibitem [{\citenamefont {Lu}\ \emph {et~al.}(2024)\citenamefont {Lu}, \citenamefont {Joshi}, \citenamefont {Dinh},\ and\ \citenamefont {Koch}}]{lu2023optimalgrape}%
  \BibitemOpen
  \bibfield  {author} {\bibinfo {author} {\bibfnamefont {Y.}~\bibnamefont {Lu}}, \bibinfo {author} {\bibfnamefont {S.}~\bibnamefont {Joshi}}, \bibinfo {author} {\bibfnamefont {V.~S.}\ \bibnamefont {Dinh}},\ and\ \bibinfo {author} {\bibfnamefont {J.}~\bibnamefont {Koch}},\ }\bibfield  {title} {\bibinfo {title} {Optimal control of large quantum systems: assessing memory and runtime performance of grape},\ }\href {https://doi.org/10.1088/2399-6528/ad22e5} {\bibfield  {journal} {\bibinfo  {journal} {J. Phys. Commun.}\ }\textbf {\bibinfo {volume} {8}},\ \bibinfo {pages} {025002} (\bibinfo {year} {2024})}\BibitemShut {NoStop}%
\bibitem [{\citenamefont {Koch}\ \emph {et~al.}(2022)\citenamefont {Koch}, \citenamefont {Boscain}, \citenamefont {Calarco}, \citenamefont {Dirr}, \citenamefont {Filipp}, \citenamefont {Glaser}, \citenamefont {Kosloff}, \citenamefont {Montangero}, \citenamefont {Schulte-Herbr{\"u}ggen}, \citenamefont {Sugny} \emph {et~al.}}]{koch2022quantumqocreview}%
  \BibitemOpen
  \bibfield  {author} {\bibinfo {author} {\bibfnamefont {C.~P.}\ \bibnamefont {Koch}}, \bibinfo {author} {\bibfnamefont {U.}~\bibnamefont {Boscain}}, \bibinfo {author} {\bibfnamefont {T.}~\bibnamefont {Calarco}}, \bibinfo {author} {\bibfnamefont {G.}~\bibnamefont {Dirr}}, \bibinfo {author} {\bibfnamefont {S.}~\bibnamefont {Filipp}}, \bibinfo {author} {\bibfnamefont {S.~J.}\ \bibnamefont {Glaser}}, \bibinfo {author} {\bibfnamefont {R.}~\bibnamefont {Kosloff}}, \bibinfo {author} {\bibfnamefont {S.}~\bibnamefont {Montangero}}, \bibinfo {author} {\bibfnamefont {T.}~\bibnamefont {Schulte-Herbr{\"u}ggen}}, \bibinfo {author} {\bibfnamefont {D.}~\bibnamefont {Sugny}}, \emph {et~al.},\ }\bibfield  {title} {\bibinfo {title} {Quantum optimal control in quantum technologies. strategic report on current status, visions and goals for research in europe},\ }\href {https://epjquantumtechnology.springeropen.com/articles/10.1140/epjqt/s40507-022-00138-x} {\bibfield  {journal} {\bibinfo  {journal} {EPJ Quantum Technol.}\
  }\textbf {\bibinfo {volume} {9}},\ \bibinfo {pages} {19} (\bibinfo {year} {2022})}\BibitemShut {NoStop}%
\bibitem [{\citenamefont {Baydin}\ and\ \citenamefont {Pearlmutter}(2014)}]{adscalingbaydin2014automatic}%
  \BibitemOpen
  \bibfield  {author} {\bibinfo {author} {\bibfnamefont {A.~G.}\ \bibnamefont {Baydin}}\ and\ \bibinfo {author} {\bibfnamefont {B.~A.}\ \bibnamefont {Pearlmutter}},\ }\bibfield  {title} {\bibinfo {title} {Automatic differentiation of algorithms for machine learning},\ }\href {http://arxiv.org/abs/1404.7456} {\bibfield  {journal} {\bibinfo  {journal} {arXiv:1404.7456}\ } (\bibinfo {year} {2014})}\BibitemShut {NoStop}%
\bibitem [{\citenamefont {Leung}\ \emph {et~al.}(2017)\citenamefont {Leung}, \citenamefont {Abdelhafez}, \citenamefont {Koch},\ and\ \citenamefont {Schuster}}]{PhysRevA.95.042318ad}%
  \BibitemOpen
  \bibfield  {author} {\bibinfo {author} {\bibfnamefont {N.}~\bibnamefont {Leung}}, \bibinfo {author} {\bibfnamefont {M.}~\bibnamefont {Abdelhafez}}, \bibinfo {author} {\bibfnamefont {J.}~\bibnamefont {Koch}},\ and\ \bibinfo {author} {\bibfnamefont {D.}~\bibnamefont {Schuster}},\ }\bibfield  {title} {\bibinfo {title} {Speedup for quantum optimal control from automatic differentiation based on graphics processing units},\ }\href {https://doi.org/10.1103/PhysRevA.95.042318} {\bibfield  {journal} {\bibinfo  {journal} {Phys. Rev. A}\ }\textbf {\bibinfo {volume} {95}},\ \bibinfo {pages} {042318} (\bibinfo {year} {2017})}\BibitemShut {NoStop}%
\bibitem [{\citenamefont {Abdelhafez}\ \emph {et~al.}(2019)\citenamefont {Abdelhafez}, \citenamefont {Schuster},\ and\ \citenamefont {Koch}}]{PhysRevA.99.052327ad}%
  \BibitemOpen
  \bibfield  {author} {\bibinfo {author} {\bibfnamefont {M.}~\bibnamefont {Abdelhafez}}, \bibinfo {author} {\bibfnamefont {D.~I.}\ \bibnamefont {Schuster}},\ and\ \bibinfo {author} {\bibfnamefont {J.}~\bibnamefont {Koch}},\ }\bibfield  {title} {\bibinfo {title} {Gradient-based optimal control of open quantum systems using quantum trajectories and automatic differentiation ad},\ }\href {https://doi.org/10.1103/PhysRevA.99.052327} {\bibfield  {journal} {\bibinfo  {journal} {Phys. Rev. A}\ }\textbf {\bibinfo {volume} {99}},\ \bibinfo {pages} {052327} (\bibinfo {year} {2019})}\BibitemShut {NoStop}%
\bibitem [{Note3()}]{Note3}%
  \BibitemOpen
  \bibinfo {note} {We have conducted numerical simulations to verify this point.}\BibitemShut {Stop}%
\bibitem [{\citenamefont {Wang}\ \emph {et~al.}(2016)\citenamefont {Wang}, \citenamefont {Gao}, \citenamefont {Reinhold}, \citenamefont {Heeres}, \citenamefont {Ofek}, \citenamefont {Chou}, \citenamefont {Axline}, \citenamefont {Reagor}, \citenamefont {Blumoff}, \citenamefont {Sliwa} \emph {et~al.}}]{cat_two}%
  \BibitemOpen
  \bibfield  {author} {\bibinfo {author} {\bibfnamefont {C.}~\bibnamefont {Wang}}, \bibinfo {author} {\bibfnamefont {Y.~Y.}\ \bibnamefont {Gao}}, \bibinfo {author} {\bibfnamefont {P.}~\bibnamefont {Reinhold}}, \bibinfo {author} {\bibfnamefont {R.~W.}\ \bibnamefont {Heeres}}, \bibinfo {author} {\bibfnamefont {N.}~\bibnamefont {Ofek}}, \bibinfo {author} {\bibfnamefont {K.}~\bibnamefont {Chou}}, \bibinfo {author} {\bibfnamefont {C.}~\bibnamefont {Axline}}, \bibinfo {author} {\bibfnamefont {M.}~\bibnamefont {Reagor}}, \bibinfo {author} {\bibfnamefont {J.}~\bibnamefont {Blumoff}}, \bibinfo {author} {\bibfnamefont {K.~M.}\ \bibnamefont {Sliwa}}, \emph {et~al.},\ }\bibfield  {title} {\bibinfo {title} {A schrödinger cat living in two boxes},\ }\href {https://doi.org/10.1126/science.aaf2941} {\bibfield  {journal} {\bibinfo  {journal} {Science}\ }\textbf {\bibinfo {volume} {352}},\ \bibinfo {pages} {1087} (\bibinfo {year} {2016})}\BibitemShut {NoStop}%
\bibitem [{\citenamefont {Lee}\ \emph {et~al.}(2011)\citenamefont {Lee}, \citenamefont {Benichi}, \citenamefont {Takeno}, \citenamefont {Takeda}, \citenamefont {Webb}, \citenamefont {Huntington},\ and\ \citenamefont {Furusawa}}]{Lee2011}%
  \BibitemOpen
  \bibfield  {author} {\bibinfo {author} {\bibfnamefont {N.}~\bibnamefont {Lee}}, \bibinfo {author} {\bibfnamefont {H.}~\bibnamefont {Benichi}}, \bibinfo {author} {\bibfnamefont {Y.}~\bibnamefont {Takeno}}, \bibinfo {author} {\bibfnamefont {S.}~\bibnamefont {Takeda}}, \bibinfo {author} {\bibfnamefont {J.}~\bibnamefont {Webb}}, \bibinfo {author} {\bibfnamefont {E.}~\bibnamefont {Huntington}},\ and\ \bibinfo {author} {\bibfnamefont {A.}~\bibnamefont {Furusawa}},\ }\bibfield  {title} {\bibinfo {title} {Teleportation of nonclassical wave packets of light},\ }\href {https://doi.org/10.1126/science.1201034} {\bibfield  {journal} {\bibinfo  {journal} {Science}\ }\textbf {\bibinfo {volume} {332}},\ \bibinfo {pages} {330} (\bibinfo {year} {2011})}\BibitemShut {NoStop}%
\bibitem [{Note4()}]{Note4}%
  \BibitemOpen
  \bibinfo {note} {To efficiently simulate an open system with a large Hilbert space, we perform Monte Carlo simulations with 2000 quantum trajectories.}\BibitemShut {Stop}%
\bibitem [{\citenamefont {Lai}\ and\ \citenamefont {Haus}(1989)}]{characteristic_function}%
  \BibitemOpen
  \bibfield  {author} {\bibinfo {author} {\bibfnamefont {Y.}~\bibnamefont {Lai}}\ and\ \bibinfo {author} {\bibfnamefont {H.~A.}\ \bibnamefont {Haus}},\ }\bibfield  {title} {\bibinfo {title} {Characteristic functions and quantum measurements of optical observables},\ }\href {https://doi.org/10.1088/0954-8998/1/2/003} {\bibfield  {journal} {\bibinfo  {journal} {Quantum Opt.}\ }\textbf {\bibinfo {volume} {1}},\ \bibinfo {pages} {99} (\bibinfo {year} {1989})}\BibitemShut {NoStop}%
\bibitem [{\citenamefont {Rosenblum}\ \emph {et~al.}(2018)\citenamefont {Rosenblum}, \citenamefont {Gao}, \citenamefont {Reinhold}, \citenamefont {Wang}, \citenamefont {Axline}, \citenamefont {Frunzio}, \citenamefont {Girvin}, \citenamefont {Jiang}, \citenamefont {Mirrahimi}, \citenamefont {Devoret},\ and\ \citenamefont {Schoelkopf}}]{Rosenblum2018}%
  \BibitemOpen
  \bibfield  {author} {\bibinfo {author} {\bibfnamefont {S.}~\bibnamefont {Rosenblum}}, \bibinfo {author} {\bibfnamefont {Y.~Y.}\ \bibnamefont {Gao}}, \bibinfo {author} {\bibfnamefont {P.}~\bibnamefont {Reinhold}}, \bibinfo {author} {\bibfnamefont {C.}~\bibnamefont {Wang}}, \bibinfo {author} {\bibfnamefont {C.~J.}\ \bibnamefont {Axline}}, \bibinfo {author} {\bibfnamefont {L.}~\bibnamefont {Frunzio}}, \bibinfo {author} {\bibfnamefont {S.~M.}\ \bibnamefont {Girvin}}, \bibinfo {author} {\bibfnamefont {L.}~\bibnamefont {Jiang}}, \bibinfo {author} {\bibfnamefont {M.}~\bibnamefont {Mirrahimi}}, \bibinfo {author} {\bibfnamefont {M.~H.}\ \bibnamefont {Devoret}},\ and\ \bibinfo {author} {\bibfnamefont {R.~J.}\ \bibnamefont {Schoelkopf}},\ }\bibfield  {title} {\bibinfo {title} {A {CNOT} gate between multiphoton qubits encoded in two cavities},\ }\href {https://doi.org/10.1038/s41467-018-03059-5} {\bibfield  {journal} {\bibinfo  {journal} {Nat. Comm.}\ }\textbf {\bibinfo {volume} {9}},\ \bibinfo {pages} {652} (\bibinfo
  {year} {2018})}\BibitemShut {NoStop}%
\bibitem [{\citenamefont {Zhou}\ \emph {et~al.}(2023)\citenamefont {Zhou}, \citenamefont {Lu}, \citenamefont {Praquin}, \citenamefont {Chien}, \citenamefont {Kaufman}, \citenamefont {Cao}, \citenamefont {Xia}, \citenamefont {Mong}, \citenamefont {Pfaff}, \citenamefont {Pekker} \emph {et~al.}}]{Zhou2023}%
  \BibitemOpen
  \bibfield  {author} {\bibinfo {author} {\bibfnamefont {C.}~\bibnamefont {Zhou}}, \bibinfo {author} {\bibfnamefont {P.}~\bibnamefont {Lu}}, \bibinfo {author} {\bibfnamefont {M.}~\bibnamefont {Praquin}}, \bibinfo {author} {\bibfnamefont {T.-C.}\ \bibnamefont {Chien}}, \bibinfo {author} {\bibfnamefont {R.}~\bibnamefont {Kaufman}}, \bibinfo {author} {\bibfnamefont {X.}~\bibnamefont {Cao}}, \bibinfo {author} {\bibfnamefont {M.}~\bibnamefont {Xia}}, \bibinfo {author} {\bibfnamefont {R.~S.~K.}\ \bibnamefont {Mong}}, \bibinfo {author} {\bibfnamefont {W.}~\bibnamefont {Pfaff}}, \bibinfo {author} {\bibfnamefont {D.}~\bibnamefont {Pekker}}, \emph {et~al.},\ }\bibfield  {title} {\bibinfo {title} {{Realizing all-to-all couplings among detachable quantum modules using a microwave quantum state router}},\ }\href {https://doi.org/10.1038/s41534-023-00723-7} {\bibfield  {journal} {\bibinfo  {journal} {npj Quantum Inf.}\ }\textbf {\bibinfo {volume} {9}},\ \bibinfo {pages} {54} (\bibinfo {year} {2023})}\BibitemShut {NoStop}%
\bibitem [{\citenamefont {Chapman}\ \emph {et~al.}(2023)\citenamefont {Chapman}, \citenamefont {de~Graaf}, \citenamefont {Xue}, \citenamefont {Zhang}, \citenamefont {Teoh}, \citenamefont {Curtis}, \citenamefont {Tsunoda}, \citenamefont {Eickbusch}, \citenamefont {Read}, \citenamefont {Koottandavida} \emph {et~al.}}]{Chapman2022}%
  \BibitemOpen
  \bibfield  {author} {\bibinfo {author} {\bibfnamefont {B.~J.}\ \bibnamefont {Chapman}}, \bibinfo {author} {\bibfnamefont {S.~J.}\ \bibnamefont {de~Graaf}}, \bibinfo {author} {\bibfnamefont {S.~H.}\ \bibnamefont {Xue}}, \bibinfo {author} {\bibfnamefont {Y.}~\bibnamefont {Zhang}}, \bibinfo {author} {\bibfnamefont {J.}~\bibnamefont {Teoh}}, \bibinfo {author} {\bibfnamefont {J.~C.}\ \bibnamefont {Curtis}}, \bibinfo {author} {\bibfnamefont {T.}~\bibnamefont {Tsunoda}}, \bibinfo {author} {\bibfnamefont {A.}~\bibnamefont {Eickbusch}}, \bibinfo {author} {\bibfnamefont {A.~P.}\ \bibnamefont {Read}}, \bibinfo {author} {\bibfnamefont {A.}~\bibnamefont {Koottandavida}}, \emph {et~al.},\ }\bibfield  {title} {\bibinfo {title} {High-on-off-ratio beam-splitter interaction for gates on bosonically encoded qubits},\ }\href {https://doi.org/10.1103/PRXQuantum.4.020355} {\bibfield  {journal} {\bibinfo  {journal} {PRX Quantum}\ }\textbf {\bibinfo {volume} {4}},\ \bibinfo {pages} {020355} (\bibinfo {year} {2023})}\BibitemShut
  {NoStop}%
\bibitem [{\citenamefont {Lu}\ \emph {et~al.}(2023)\citenamefont {Lu}, \citenamefont {Maiti}, \citenamefont {Garmon}, \citenamefont {Ganjam}, \citenamefont {Zhang}, \citenamefont {Claes}, \citenamefont {Frunzio}, \citenamefont {Girvin},\ and\ \citenamefont {Schoelkopf}}]{Lu2023}%
  \BibitemOpen
  \bibfield  {author} {\bibinfo {author} {\bibfnamefont {Y.}~\bibnamefont {Lu}}, \bibinfo {author} {\bibfnamefont {A.}~\bibnamefont {Maiti}}, \bibinfo {author} {\bibfnamefont {J.~W.~O.}\ \bibnamefont {Garmon}}, \bibinfo {author} {\bibfnamefont {S.}~\bibnamefont {Ganjam}}, \bibinfo {author} {\bibfnamefont {Y.}~\bibnamefont {Zhang}}, \bibinfo {author} {\bibfnamefont {J.}~\bibnamefont {Claes}}, \bibinfo {author} {\bibfnamefont {L.}~\bibnamefont {Frunzio}}, \bibinfo {author} {\bibfnamefont {S.~M.}\ \bibnamefont {Girvin}},\ and\ \bibinfo {author} {\bibfnamefont {R.~J.}\ \bibnamefont {Schoelkopf}},\ }\bibfield  {title} {\bibinfo {title} {High-fidelity parametric beamsplitting with a parity-protected converter},\ }\href {https://doi.org/10.1038/s41467-023-41104-0} {\bibfield  {journal} {\bibinfo  {journal} {Nat. Comm.}\ }\textbf {\bibinfo {volume} {14}},\ \bibinfo {pages} {5767} (\bibinfo {year} {2023})}\BibitemShut {NoStop}%
\bibitem [{\citenamefont {Gao}\ \emph {et~al.}(2018)\citenamefont {Gao}, \citenamefont {Lester}, \citenamefont {Zhang}, \citenamefont {Wang}, \citenamefont {Rosenblum}, \citenamefont {Frunzio}, \citenamefont {Jiang}, \citenamefont {Girvin},\ and\ \citenamefont {Schoelkopf}}]{gao2018}%
  \BibitemOpen
  \bibfield  {author} {\bibinfo {author} {\bibfnamefont {Y.~Y.}\ \bibnamefont {Gao}}, \bibinfo {author} {\bibfnamefont {B.~J.}\ \bibnamefont {Lester}}, \bibinfo {author} {\bibfnamefont {Y.}~\bibnamefont {Zhang}}, \bibinfo {author} {\bibfnamefont {C.}~\bibnamefont {Wang}}, \bibinfo {author} {\bibfnamefont {S.}~\bibnamefont {Rosenblum}}, \bibinfo {author} {\bibfnamefont {L.}~\bibnamefont {Frunzio}}, \bibinfo {author} {\bibfnamefont {L.}~\bibnamefont {Jiang}}, \bibinfo {author} {\bibfnamefont {S.~M.}\ \bibnamefont {Girvin}},\ and\ \bibinfo {author} {\bibfnamefont {R.~J.}\ \bibnamefont {Schoelkopf}},\ }\bibfield  {title} {\bibinfo {title} {Programmable interference between two microwave quantum memories},\ }\href {https://doi.org/10.1103/PhysRevX.8.021073} {\bibfield  {journal} {\bibinfo  {journal} {Phys. Rev. X}\ }\textbf {\bibinfo {volume} {8}},\ \bibinfo {pages} {021073} (\bibinfo {year} {2018})}\BibitemShut {NoStop}%
\bibitem [{\citenamefont {Gao}\ \emph {et~al.}(2019)\citenamefont {Gao}, \citenamefont {Lester}, \citenamefont {Chou}, \citenamefont {Frunzio}, \citenamefont {Devoret}, \citenamefont {Jiang}, \citenamefont {Girvin},\ and\ \citenamefont {Schoelkopf}}]{Gao2019}%
  \BibitemOpen
  \bibfield  {author} {\bibinfo {author} {\bibfnamefont {Y.~Y.}\ \bibnamefont {Gao}}, \bibinfo {author} {\bibfnamefont {B.~J.}\ \bibnamefont {Lester}}, \bibinfo {author} {\bibfnamefont {K.~S.}\ \bibnamefont {Chou}}, \bibinfo {author} {\bibfnamefont {L.}~\bibnamefont {Frunzio}}, \bibinfo {author} {\bibfnamefont {M.~H.}\ \bibnamefont {Devoret}}, \bibinfo {author} {\bibfnamefont {L.}~\bibnamefont {Jiang}}, \bibinfo {author} {\bibfnamefont {S.~M.}\ \bibnamefont {Girvin}},\ and\ \bibinfo {author} {\bibfnamefont {R.~J.}\ \bibnamefont {Schoelkopf}},\ }\bibfield  {title} {\bibinfo {title} {Entanglement of bosonic modes through an engineered exchange interaction},\ }\href {https://doi.org/10.1038/s41586-019-0970-4} {\bibfield  {journal} {\bibinfo  {journal} {Nature}\ }\textbf {\bibinfo {volume} {566}},\ \bibinfo {pages} {509} (\bibinfo {year} {2019})}\BibitemShut {NoStop}%
\bibitem [{\citenamefont {Zhang}\ \emph {et~al.}(2019)\citenamefont {Zhang}, \citenamefont {Lester}, \citenamefont {Gao}, \citenamefont {Jiang}, \citenamefont {Schoelkopf},\ and\ \citenamefont {Girvin}}]{yaxing}%
  \BibitemOpen
  \bibfield  {author} {\bibinfo {author} {\bibfnamefont {Y.}~\bibnamefont {Zhang}}, \bibinfo {author} {\bibfnamefont {B.~J.}\ \bibnamefont {Lester}}, \bibinfo {author} {\bibfnamefont {Y.~Y.}\ \bibnamefont {Gao}}, \bibinfo {author} {\bibfnamefont {L.}~\bibnamefont {Jiang}}, \bibinfo {author} {\bibfnamefont {R.~J.}\ \bibnamefont {Schoelkopf}},\ and\ \bibinfo {author} {\bibfnamefont {S.~M.}\ \bibnamefont {Girvin}},\ }\bibfield  {title} {\bibinfo {title} {Engineering bilinear mode coupling in circuit qed: Theory and experiment},\ }\href {https://doi.org/10.1103/PhysRevA.99.012314} {\bibfield  {journal} {\bibinfo  {journal} {Phys. Rev. A}\ }\textbf {\bibinfo {volume} {99}},\ \bibinfo {pages} {012314} (\bibinfo {year} {2019})}\BibitemShut {NoStop}%
\bibitem [{\citenamefont {Pietik{\"{a}}inen}\ \emph {et~al.}(2024)\citenamefont {Pietik{\"{a}}inen}, \citenamefont {{\v{C}}ernot{\'{i}}k}, \citenamefont {Eickbusch}, \citenamefont {Maiti}, \citenamefont {Garmon}, \citenamefont {Filip},\ and\ \citenamefont {Girvin}}]{Pietikainen2024}%
  \BibitemOpen
  \bibfield  {author} {\bibinfo {author} {\bibfnamefont {I.}~\bibnamefont {Pietik{\"{a}}inen}}, \bibinfo {author} {\bibfnamefont {O.}~\bibnamefont {{\v{C}}ernot{\'{i}}k}}, \bibinfo {author} {\bibfnamefont {A.}~\bibnamefont {Eickbusch}}, \bibinfo {author} {\bibfnamefont {A.}~\bibnamefont {Maiti}}, \bibinfo {author} {\bibfnamefont {J.~W.~O.}\ \bibnamefont {Garmon}}, \bibinfo {author} {\bibfnamefont {R.}~\bibnamefont {Filip}},\ and\ \bibinfo {author} {\bibfnamefont {S.~M.}\ \bibnamefont {Girvin}},\ }\bibfield  {title} {\bibinfo {title} {{Strategies and trade-offs for controllability and memory time of ultra-high-quality microwave cavities in circuit QED}},\ }\href {http://arxiv.org/abs/2403.02278} {\bibfield  {journal} {\bibinfo  {journal} {arXiv:2403.02278}\ } (\bibinfo {year} {2024})}\BibitemShut {NoStop}%
\bibitem [{\citenamefont {Huang}\ \emph {et~al.}(2024)\citenamefont {Huang}, \citenamefont {DiNapoli}, \citenamefont {Gupta}, \citenamefont {Patel}, \citenamefont {Bal}, \citenamefont {Crisa}, \citenamefont {Garattoni}, \citenamefont {Lu}, \citenamefont {You}, \citenamefont {Rockwood} \emph {et~al.}}]{jordan}%
  \BibitemOpen
  \bibfield  {author} {\bibinfo {author} {\bibfnamefont {J.}~\bibnamefont {Huang}}, \bibinfo {author} {\bibfnamefont {T.}~\bibnamefont {DiNapoli}}, \bibinfo {author} {\bibfnamefont {E.}~\bibnamefont {Gupta}}, \bibinfo {author} {\bibfnamefont {S.}~\bibnamefont {Patel}}, \bibinfo {author} {\bibfnamefont {M.}~\bibnamefont {Bal}}, \bibinfo {author} {\bibfnamefont {F.}~\bibnamefont {Crisa}}, \bibinfo {author} {\bibfnamefont {S.}~\bibnamefont {Garattoni}}, \bibinfo {author} {\bibfnamefont {Y.}~\bibnamefont {Lu}}, \bibinfo {author} {\bibfnamefont {X.}~\bibnamefont {You}}, \bibinfo {author} {\bibfnamefont {G.}~\bibnamefont {Rockwood}}, \emph {et~al.},\ }\bibfield  {title} {\bibinfo {title} {Control of a long-lived multimode bosonic memory with a weakly coupled transmon ancilla},\ }\href {https://meetings.aps.org/Meeting/MAR24/Session/T47.8} {\bibfield  {journal} {\bibinfo  {journal} {Bull. Am. Phys. Soc.}\ } (\bibinfo {year} {2024})}\BibitemShut {NoStop}%
\bibitem [{\citenamefont {Pechal}\ \emph {et~al.}(2014)\citenamefont {Pechal}, \citenamefont {Huthmacher}, \citenamefont {Eichler}, \citenamefont {Zeytino\ifmmode~\breve{g}\else \u{g}\fi{}lu}, \citenamefont {Abdumalikov}, \citenamefont {Berger}, \citenamefont {Wallraff},\ and\ \citenamefont {Filipp}}]{sideband}%
  \BibitemOpen
  \bibfield  {author} {\bibinfo {author} {\bibfnamefont {M.}~\bibnamefont {Pechal}}, \bibinfo {author} {\bibfnamefont {L.}~\bibnamefont {Huthmacher}}, \bibinfo {author} {\bibfnamefont {C.}~\bibnamefont {Eichler}}, \bibinfo {author} {\bibfnamefont {S.}~\bibnamefont {Zeytino\ifmmode~\breve{g}\else \u{g}\fi{}lu}}, \bibinfo {author} {\bibfnamefont {A.~A.}\ \bibnamefont {Abdumalikov}}, \bibinfo {author} {\bibfnamefont {S.}~\bibnamefont {Berger}}, \bibinfo {author} {\bibfnamefont {A.}~\bibnamefont {Wallraff}},\ and\ \bibinfo {author} {\bibfnamefont {S.}~\bibnamefont {Filipp}},\ }\bibfield  {title} {\bibinfo {title} {Microwave-controlled generation of shaped single photons in circuit quantum electrodynamics},\ }\href {https://doi.org/10.1103/PhysRevX.4.041010} {\bibfield  {journal} {\bibinfo  {journal} {Phys. Rev. X}\ }\textbf {\bibinfo {volume} {4}},\ \bibinfo {pages} {041010} (\bibinfo {year} {2014})}\BibitemShut {NoStop}%
\bibitem [{\citenamefont {Braunstein}\ and\ \citenamefont {van Loock}(2005)}]{Braunstein2005}%
  \BibitemOpen
  \bibfield  {author} {\bibinfo {author} {\bibfnamefont {S.~L.}\ \bibnamefont {Braunstein}}\ and\ \bibinfo {author} {\bibfnamefont {P.}~\bibnamefont {van Loock}},\ }\bibfield  {title} {\bibinfo {title} {Quantum information with continuous variables},\ }\href {https://doi.org/10.1103/RevModPhys.77.513} {\bibfield  {journal} {\bibinfo  {journal} {Rev. Mod. Phys.}\ }\textbf {\bibinfo {volume} {77}},\ \bibinfo {pages} {513} (\bibinfo {year} {2005})}\BibitemShut {NoStop}%
\bibitem [{\citenamefont {Blais}\ \emph {et~al.}(2007)\citenamefont {Blais}, \citenamefont {Gambetta}, \citenamefont {Wallraff}, \citenamefont {Schuster}, \citenamefont {Girvin}, \citenamefont {Devoret},\ and\ \citenamefont {Schoelkopf}}]{blais2007}%
  \BibitemOpen
  \bibfield  {author} {\bibinfo {author} {\bibfnamefont {A.}~\bibnamefont {Blais}}, \bibinfo {author} {\bibfnamefont {J.}~\bibnamefont {Gambetta}}, \bibinfo {author} {\bibfnamefont {A.}~\bibnamefont {Wallraff}}, \bibinfo {author} {\bibfnamefont {D.~I.}\ \bibnamefont {Schuster}}, \bibinfo {author} {\bibfnamefont {S.~M.}\ \bibnamefont {Girvin}}, \bibinfo {author} {\bibfnamefont {M.~H.}\ \bibnamefont {Devoret}},\ and\ \bibinfo {author} {\bibfnamefont {R.~J.}\ \bibnamefont {Schoelkopf}},\ }\bibfield  {title} {\bibinfo {title} {Quantum-information processing with circuit quantum electrodynamics},\ }\href {https://doi.org/10.1103/PhysRevA.75.032329} {\bibfield  {journal} {\bibinfo  {journal} {Phys. Rev. A}\ }\textbf {\bibinfo {volume} {75}},\ \bibinfo {pages} {032329} (\bibinfo {year} {2007})}\BibitemShut {NoStop}%
\bibitem [{\citenamefont {Boissonneault}\ \emph {et~al.}(2009)\citenamefont {Boissonneault}, \citenamefont {Gambetta},\ and\ \citenamefont {Blais}}]{Boissonneault2009a}%
  \BibitemOpen
  \bibfield  {author} {\bibinfo {author} {\bibfnamefont {M.}~\bibnamefont {Boissonneault}}, \bibinfo {author} {\bibfnamefont {J.~M.}\ \bibnamefont {Gambetta}},\ and\ \bibinfo {author} {\bibfnamefont {A.}~\bibnamefont {Blais}},\ }\bibfield  {title} {\bibinfo {title} {{Dispersive regime of circuit QED: Photon-dependent qubit dephasing and relaxation rates}},\ }\href {https://doi.org/10.1103/PhysRevA.79.013819} {\bibfield  {journal} {\bibinfo  {journal} {Phys. Rev. A}\ }\textbf {\bibinfo {volume} {79}},\ \bibinfo {pages} {013819} (\bibinfo {year} {2009})}\BibitemShut {NoStop}%
\bibitem [{\citenamefont {Ithier}\ \emph {et~al.}(2005)\citenamefont {Ithier}, \citenamefont {Collin}, \citenamefont {Joyez}, \citenamefont {Meeson}, \citenamefont {Vion}, \citenamefont {Esteve}, \citenamefont {Chiarello}, \citenamefont {Shnirman}, \citenamefont {Makhlin}, \citenamefont {Schriefl} \emph {et~al.}}]{Ithier2005}%
  \BibitemOpen
  \bibfield  {author} {\bibinfo {author} {\bibfnamefont {G.}~\bibnamefont {Ithier}}, \bibinfo {author} {\bibfnamefont {E.}~\bibnamefont {Collin}}, \bibinfo {author} {\bibfnamefont {P.}~\bibnamefont {Joyez}}, \bibinfo {author} {\bibfnamefont {P.~J.}\ \bibnamefont {Meeson}}, \bibinfo {author} {\bibfnamefont {D.}~\bibnamefont {Vion}}, \bibinfo {author} {\bibfnamefont {D.}~\bibnamefont {Esteve}}, \bibinfo {author} {\bibfnamefont {F.}~\bibnamefont {Chiarello}}, \bibinfo {author} {\bibfnamefont {A.}~\bibnamefont {Shnirman}}, \bibinfo {author} {\bibfnamefont {Y.}~\bibnamefont {Makhlin}}, \bibinfo {author} {\bibfnamefont {J.}~\bibnamefont {Schriefl}}, \emph {et~al.},\ }\bibfield  {title} {\bibinfo {title} {{Decoherence in a superconducting quantum bit circuit}},\ }\href {https://doi.org/10.1103/PhysRevB.72.134519} {\bibfield  {journal} {\bibinfo  {journal} {Phys. Rev. B}\ }\textbf {\bibinfo {volume} {72}},\ \bibinfo {pages} {134519} (\bibinfo {year} {2005})}\BibitemShut {NoStop}%
\bibitem [{\citenamefont {Q-CTRL}(2023)}]{qctrl2022boulder}%
  \BibitemOpen
  \bibfield  {author} {\bibinfo {author} {\bibnamefont {Q-CTRL}},\ }\href {https://q-ctrl.com/boulder-opal} {\bibinfo {title} {Boulder opal}} (\bibinfo {year} {2023})\BibitemShut {NoStop}%
\end{thebibliography}%
\end{document}